\documentclass[fleqn,usenatbib]{mnras}

\usepackage[T1]{fontenc}
\usepackage{ae,aecompl}

\usepackage{graphicx}	
\usepackage{amsmath}	
\usepackage{amssymb}	
\usepackage{txfonts}

\title[Quiescent neutron-star low-mass X-ray binaries in
  47\,Tuc]{Simultaneous {\em Chandra} and {\em HST} observations of
  the quiescent neutron-star low-mass X-ray binaries in 47\,Tucanae}

\author[van den Berg et al.]{M. van den Berg,$^{1}$\thanks{E-mail: mvandenberg@cfa.harvard.edu}
  L.E. Rivera Sandoval,$^{2}$
  C.O. Heinke,$^{3}$
  H.N. Cohn,$^{4}$
  P.M. Lugger,$^{4}$
\newauthor{J.E. Grindlay,$^{1}$
  P.D. Edmonds,$^{1}$ J. Anderson,$^{5}$ A. Catuneanu$^{3,6}$}
\\
\\
$^{1}$Center for Astrophysics $\mid$ Harvard \& Smithsonian,
60 Garden Street,
Cambridge, MA 02138, USA\\
$^{2}$Department of Physics and Astronomy, University of Texas Rio Grande Valley, Brownsville, TX 78520, USA\\
$^{3}$Department of Physics, University of Alberta, CCIS 4-183, Edmonton, AB, T6G 2E1, Canada\\
$^{4}$Department of Astronomy, Indiana University, 727 E.\,Third St, Bloomington, IN 47405, USA\\
$^{5}$Space Telescope Science Institute, 3800 San Martin Drive, Baltimore, MD 21218, USA\\
$^{6}$Dana Canada Corporation, 656 Kerr Street, Oakville, ON L6K 3E4, Canada}

\date{Accepted XXX. Received YYY; in original form ZZZ}

\pubyear{2018}

\usepackage{newtxtext,newtxmath}

\begin{document}
\label{firstpage}
\pagerange{\pageref{firstpage}--\pageref{lastpage}}
\maketitle

\begin{abstract}
We present simultaneous {\em Chandra X-ray Observatory} and {\em
  Hubble Space Telescope} observations of three certain (X5, X7, W37)
and two likely (X4, W17) quiescent neutron-star low-mass X-ray
binaries (qLMXBs) in the globular cluster 47\,Tuc. We study these
systems in the X-ray, optical and near-ultraviolet (NUV) using the
simultaneous data and additional non-contemporaneous {\em HST} data. We have
discovered a blue and variable NUV counterpart to W17. We have not
securely identified the eclipsing qLMXB W37 in the optical or
NUV. Deeper high-resolution imaging is needed to further investigate
the faint NUV excess near the centre of the W37 error circle. We
suggest that a previously identified optical astrometric match to X7
is likely the true counterpart. The Halpha emission and the location
of the counterpart in the colour-magnitude diagram, indicate that the
secondary is probably a non-degenerate, H-rich star. This is
consistent with previous results from fitting X7's X-ray spectrum. In
X4, the simultaneous X-ray and optical behaviour supports the earlier
suggestion that the X-ray variability is driven by changes in
accretion rate. The X-ray eclipses in X5 coincide with minima in the
optical/NUV light curves. Comparison of the 47\,Tuc qLMXBs with the
cataclysmic variables (CVs) in the cluster confirms that overall the
qLMXBs have larger X-ray--to--optical flux ratios. Based on their
optical/NUV colors, we conclude that the accretion disks in the qLMXBs
are less prominent than in CVs. This makes the ratio of X-ray flux to
excess blue optical flux a powerful discriminator between CVs and
qLMXBs.
\end{abstract}




\begin{table*}
  \centering
  \caption{Summary of {\em HST} observations analyzed in this work}
  \label{tab_obs}
  \begin{tabular}{lllllrrl}
    \hline
    Program & Instrument & Start observations & End observations & Filter & $N_{{\rm exp}}$ & \multicolumn{1}{l}{$T_{{\rm exp}}$ (s)} & Comments \\
    \hline
    GO\,9019  & ACS/HRC & 2002--Apr--12 22:48  & 2002--Apr--13 17:51 & F220W & 11 & 1870 &\\
              &         &                      &                     & F330W &  8 &  528 &\\
              &         &                      &                     & F435W & 12 &  720 &\\
              &         &                      &                     & F475W &  5 &  300 &\\
    \hline
    GO\,9443  & ACS/WFC & 2002--Jul--07 22:30  & 2002--Jul--07 22:32 & F435W &  1 & 150 & \\
\hline
    GO\,9281  & ACS/WFC & 2002--Sep--30 01:35  & 2002--Sep--30 06:24 & F435W & 3 &  300 & overlaps with {\em Chandra} ObsID 2735; \\
              &         &                      &                     & F625W & 8 &  465 & analyzed by us but also in HUGS\\
              &         &                      &                     & F658N & 7 & 2630\rule[-1.2ex]{0pt}{0pt} & \\ 
    \cline{3-8}
              &         & 2002--Oct--02 20:47\rule{0pt}{2.6ex}  & 2002--Oct--03 00:43 & F435W & 4 &  310 & overlaps with {\em Chandra} ObsID 2737 \\
              &         &                      &                     & F625W & 7 &  455 & \\
              &         &                      &                     & F658N & 7 & 2630\rule[-1.2ex]{0pt}{0pt} & \\ 
\cline{3-8}
              &         & 2002--Oct--11 02:16\rule{0pt}{2.6ex}  & 2002--Oct--11 10:12 & F435W & 3 &  345 & overlaps with {\em Chandra} ObsID 2738 \\
              &         &                      &                     & F625W & 7 &  400 & \\
              &         &                      &                     & F658N & 6 & 2180 & \\ 
\hline
    GO\,10775 & ACS/WFC   & 2006--Mar--13 01:42  & 2006--Mar--13 05:24 & F606W &  5 & 203 & in HUGS\\
              &           &                      &                     & F814W &  5 & 203 & \\
\hline
    GO\,11729 & WFC3/UVIS & 2010--Sep--28 23:28  & 2010--Sep--29 00:07 & F336W & 3 & 1190 & in HUGS\\
\hline
    GO\,12311 & WFC3/UVIS & 2010--Nov--21 11:59  & 2010--Nov--22 19:34 & F275W & 14 & 4822 & in HUGS\\
\hline
    GO\,12971 & WFC3/UVIS & 2012--Nov--14 10:45  & 2013--Sep--20 10:09 & F336W & 20 & 12050 & in HUGS\\
\hline
    GO\,12950 & WFC3/UVIS & 2013--Aug--13 02:10  & 2013--Aug--13 11:01 & F300X & 24 & 14576 & in HUGS\\
              &           &                      &                     & F390W &  8 &  4624 & \\
    \hline
  \end{tabular}

  {\flushleft
    \begin{flushleft}
The listed start and end times of the observations (in UTC) delineate the time
span that is covered by the specified exposures. For each filter,
$N_{{\rm exp}}$ is the number of images and $T_{{\rm exp}}$ is the
total exposure time.
    \end{flushleft}
  }

\end{table*}

\section{Introduction} \label{sec_intro}

Low-mass X-ray binaries (LMXBs) consist of a neutron star or black
hole that accretes from a low-mass ($\lesssim$1 $M_{\odot}$)
companion. Most known LMXBs are transient systems in which the X-ray
luminosity can change by orders of magnitude when they switch from an
actively accreting state to a quiescent, low-accretion rate state (or
the other way around). The majority of LMXBs are believed to be
quiescent most of the time. In neutron-star systems, quiescence can
offer a glimpse of the accretor: the quiescent X-ray emission may
feature a soft thermal component, thought to be radiated by the
neutron-star surface through the release of heat that is deposited
during episodes of accretion \citep{browea98}. Harder non-thermal
emission can contribute to the X-rays as well; its origin is not well
understood, but suggestions include
synchrotron shock emission or low-level residual accretion (\citealt{campea98}, \citealt{bernardini13}). Since the population of LMXBs
is concentrated towards the Galactic bulge, observations of quiescent
LMXBs (qLMXBs) with typical X-ray luminosities of $L_X\approx
10^{31-33}$ erg s$^{-1}$ are often impeded by high levels of
extinction. As a result, quiescent systems can be difficult to find or
recognize.

It has been known since the early seventies (\citealt{katz75}) that a
good place to look for luminous LMXBs is in globular clusters, where
stellar encounters add new pathways to their formation. Thanks to the
sensitivity and spatial resolution of the {\em Chandra X-ray
  Observatory}, the number of quiescent LMXBs in globular clusters was
also found to be correlated with encounter rate
(e.g.\,\citealt{heinea03b}, \citealt{poolea03},
\citealt{poolhut06}). Due to their dynamical origin, it is not obvious
that globular-cluster (q)LMXBs can serve as a proxy for the field
population. For example, in regard to their orbital periods, it has
been suggested that ultra-compact systems (i.e.~those with periods
$\lesssim$80 min) are more common among cluster LMXBs, possibly a
signature of their different formation scenarios
\citep{deutea00}. Concerning their X-ray emission, it has been found
that the X-ray spectra of many cluster qLMXBs are dominated by the
soft and steady thermal component, and lack the additional harder (and
in many cases variable) non-thermal component that is often seen in
field qLMXBs\footnote{The class of transitional millisecond pulsars is
  an exception, see e.g.~\cite{lina14}.}  (\citealt{walsea15},
\citealt{bahrea15}). Small sample sizes and selection biases may
(partly) explain these apparent differences. We know of fewer than two
dozen (transient or persistently bright) LMXBs in globular clusters,
as well as a similar number of qLMXBs that are bright enough for an
X-ray spectral or variability study. LMXBs are typically discovered
when they go in outburst, but in clusters that are targeted with deep
{\em Chandra} pointings, LMXBs are also first identified in quiescence
by their distinct soft spectra. These separate discovery methods may
select systems with very different outburst histories or duty
cycles. Detailed studies of field and cluster systems can help clarify
to what extent the observed properties really signify different
underlying populations.

Another motivation for studying neutron-star qLMXBs is that fitting
their thermal X-ray emission with a suitable model can provide mass
and radius constraints when combined with a distance estimate
\citep{rutlea02}. Since the mass-radius relation is determined by the
neutron-star equation of state, spectral fitting has become one of the
main methods to place limits on the equation of state of cold dense
matter (see e.g. \citealt{ozelfrei16}). The details of the adopted
model depend on the composition of the neutron-star atmosphere. This
composition in turn is determined by the nature of the mass donor:
accretion supplies fresh material to the outer layers of the neutron
star, which quickly stratifies with the lightest elements settling on
top (\citealt{roma87}, \citealt{alcoilla80}). Main-sequence donors
give rise to Hydrogen (H) rich atmospheres, while material accreted
from white-dwarf donors in ultra-compact systems is rich in Helium,
Carbon, and/or Oxygen. If the atmospheric composition is unknown, this
introduces systematic uncertainties in the mass and radius estimates
(\citealt{servea12}, \citealt{echiea20}). Another complication arises
when the X-ray spectrum is not purely thermal. To investigate whether
accretion could ``contaminate'' the spectrum, variability in X-rays or
at other wavelengths can be used as a diagnostic. Identifying the
nature and properties of the optical or near-ultraviolet (NUV)
counterpart can be crucial for inferring the composition of the
neutron-star atmosphere or to check for ongoing accretion (see
e.g. \citealt{haggea04}, \citealt{heinea14}).

Finally, the non-thermal component of the X-ray emission also provides
an opportunity to study accretion physics at low mass accretion rates.
Accretion has been considered a likely origin of the non-thermal
component \citep{campea98}, supported by the agreement of its spectral
shape with a bremsstrahlung model \citep{chakea14}. Accretion is
likely also responsible for a portion of the thermal spectrum,
evidenced by correlated variability in both components
\citep{campea04,cackea10,wijnea15}.  Different scenarios have been
proposed to explain the emergent spectrum in Cen X-4 and other weakly
magnetic systems. \cite{dangea15} discuss a model where the emission
largely originates in a boundary layer close to the neutron-star
surface, and constrain the properties of this layer and the dominant
radiation mechanism.  Qiao \& Liu consider the radiative coupling
between the thermal (from the neutron-star surface) and non-thermal
(from the accretion flow) emission, and find that only a small
fraction of the energy transferred to the neutron star needs to be
thermalized and scattered back to the—radiatively
inefficient—accretion flow to explain the observations
(\citealt{qiaoliu2018}, \citealt{qiaoliu2020},
\citealt{qiaoliu2021}). Knowing the conditions in and around the
accretion region is also important to understand the observed
mass-donor properties (e.g.~\citealt{shahea22}).

The massive globular cluster 47\,Tuc (NGC\,104) hosts at least three
certain and two probable neutron-star qLMXBs, none of which have ever
been caught in outburst. These are, respectively, X5 (also known as
W58), X7 (also known as W46) and W37; and the candidates X4 (also known
as W125) and W17. The ROSAT sources X5 and X7 were already identified
as likely qLMXBs (also called soft X-ray transients) by
\cite{hasiea94}, and their nature was confirmed by the first {\em
  Chandra} studies of the cluster (\citealt{grinea01a},
\citealt{heinea03a}). The {\em Chandra} sources W37 and W17, and the
ROSAT source X4 were added as (candidate) qLMXBs by
\cite{heinea05a}. X4 and W17 are notable because, unlike in many other
cluster qLMXBs, their X-ray spectra feature prominent non-thermal
emission, suggesting that accretion may be happening at a very low
rate. X5 and especially the extremely stable source X7 are among the
best-studied cluster qLMXBs, for several reasons. They are bright in
X-rays, and their spectra contain no discernible non-thermal
components (\citealt{heinea06,bogdea16}). Moreover, 47\,Tuc suffers
from little foreground extinction and its distance is precisely
determined. From the known orbital periods of X5 (8.667 hr) and W37
(3.087 hr), we know that their neutron-star atmospheres must be
H-rich.

In this paper we study these five (candidate) qLMXBs in the X-ray,
optical and NUV. We investigate the known optical counterparts to X5,
which was found and studied in detail by \cite{edmoea02a}, and to X4,
which was briefly reported in \cite{beccea14}. We look for
counterparts to the three others, which so far have not been
identified in the optical or NUV. Our study is based on data from
several {\em Hubble Space Telescope} ({\em HST}) programs, including
images from the Advanced Camera for Surveys (ACS) that were taken
simultaneously with deep {\em Chandra} observations of 47\,Tuc.  This
allows us to study correlated X-ray/optical behavior in our
sources. There is a sixth source in 47\,Tuc that is also believed to
be an LMXB, viz.\,X9 (W42). It has been suggested that the accretor in
this ultra-compact system is a black hole, although a neutron-star
primary cannot be excluded. Details can be found in \cite{millea15},
\cite{bahrea17}, \cite{churea17}, and \cite{tudoea18}.

For the cluster reddening, we assume $E$($B-V$)=$0.04\pm0.02$
\citep{salaea07}. Following \cite{bogdea16} who calculated the
weighted mean of published distances to 47\,Tuc, we adopt a distance
to the cluster of 4.53$^{{\text +}0.08}_{-0.04}$ kpc.  This average
does not include the Gaia DR2 estimate \citep{chenea18}, but at
$4.45\pm0.01\pm0.12$ kpc (random and systematic errors, respectively)
the latter is consistent with our adopted value. We start with an
overview of the observations in Section~\ref{sec_obs}, followed by a
description of the data reduction and analysis in
Section~\ref{sec_ana}. In Section~\ref{sec_oid} we present our
findings, which we discuss in the broader context of other
globular-cluster qLMXBs, and other faint cluster X-ray sources in
Section~\ref{sec_dis}.

\section{Observations} \label{sec_obs}

\subsection{Simultaneous X-ray and optical observations} \label{sec_obs_sim}

We observed 47\,Tuc simultaneously with {\em Chandra} and {\em HST} on
2002 September 30, October 2--3, and October 11. This temporal spacing
was chosen to sample variability on timescales from hours to
weeks. The {\em HST} data were taken with the Wide Field Channel (WFC)
on ACS under program GO\,9281. The two CCD detectors of the WFC give a
combined field of view of 3.4\arcmin~$\times$ 3.4\arcmin~with a
2.5\arcsec~gap in the middle. We obtained images through a blue
(F435W), red (F625W) and narrow-band H$\alpha$ (F658N) filter for
total exposure times of 955 s, 1320 s, and 7440 s,
respectively. During each of the three visits, we obtained exposures
in all three filters; this way, we were able to construct
quasi-instantaneous colours that represent the stars' actual colours
as closely as possible, minimizing the effects of variability. In the
crowded fields of globular clusters, this filter choice has proven to
be efficient in separating the true optical counterparts of {\em
  Chandra} sources from optical sources that happen to lie in the
X-ray error circles by chance (see e.g.~\citealt{cohnea10},
\citealt{luggea17}).  Accreting binaries with low-mass donors
typically have strong H$\alpha$ emission lines superposed on a blue
continuum coming from an accretion disk or stream (or, in the case of
cataclysmic variables (CVs), from the white dwarf as well), while the
stars in magnetically active binaries with enhanced coronal X-ray
emission can show chromospheric H$\alpha$ emission lines.  For each
filter, we employed a dither pattern with fractional-pixel offsets so
that the native detector resolution of 0.05\arcsec~per pixel could be
improved when combining the individual images. A summary of the
GO\,9281 exposures can be found in Table~\ref{tab_obs}.

The {\em Chandra} exposures that overlap with the {\em HST}
observations are part of a larger data set that consists of four pairs
of ACIS-S exposures, each pair made up of a long ($\sim$65 ks)
full-frame exposure (ObsIDs 2735--2738) and a short ($\sim$5 ks)
sub-array exposure (1/4 of an ACIS chip; ObsIDs 3384--3387). The {\em
  HST} data were taken during the first (ObsID 2735), third (ObsID
2737) and fourth (ObsID 2738) of the long exposures. Details of the
entire 2002 {\em Chandra} data set of 47\,Tuc and its analysis are
given in \cite{heinea05b}.

\subsection{Additional {\em Chandra} data}

We take the positions of the qLMXBs and the sources we use for
registering the {\em Chandra} and {\em HST} astrometry (see
Section~\ref{sec_astro}) from the catalog of 47\,Tuc {\em Chandra}
sources by \cite{bhatea17}. This deep source list is based on all
ACIS-S data from 2002 described in the previous section, as well as
ACIS-I data from 2000 and ACIS-S data from 2014--2015. The combined
data represent 540 ks of exposure time on the cluster, with some parts
receiving more exposure than others as a result of the use of
different imaging modes (full-frame, 1/4 or 1/8
sub-array). \cite{bhatea17} applied the EDSER algorithm \citep{liea04}
to the event files, improving the spatial resolution and the ability
to separate sources in the crowded cluster core. The 95\% confidence
radii on the positions in the catalog were calculated using the
prescription by \cite{hongvandea05}, based on {\em wavdetect} runs on
mock ACIS-I data.

\begin{table}
  \begin{tabular}{l@{\hskip0.3cm}l@{\hskip0.3cm}l@{\hskip0.18cm}l@{\hskip0.3cm}l}
    \hline
    ID    & $f_X$ ($\times10^{-14}$) & $\log f_X$/$f_{U_{300}}$ & $\log f_X$/$f_{B_{435}}$ & epoch $f_X$\\
          & erg s$^{-1}$ cm$^{-2}$   &                        &                        & \\
    \hline
    X4  & 4.3$\pm$1.5             &  2.1(2)                      & 1.6(2)$^{*}$                 & ObsID 2735--high\rule[-1.5ex]{0pt}{0pt}\\
          & 2.4\textsuperscript{\raisebox{1pt}{\rlap{+0.5}}}\textsubscript{\raisebox{-1pt}{\rlap{--1.1}}}       &  1.8(2)                      & 1.7(2)$^{*}$                & other 2002 ObsIDs\rule[-1.5ex]{0pt}{0pt}\\
    X5  & 58$\pm$3            & 2.50(4)                & 2.13(3)                & 2000\rule[-1.5ex]{0pt}{0pt}\\
    X7  & 78$\pm$4            & 2.63(4)                & 2.04(3)                & 2000\rule[-1.5ex]{0pt}{0pt}\\
    W17 & 1.7\textsuperscript{\raisebox{1pt}{\rlap{+0.3}}}\textsubscript{\raisebox{-1pt}{\rlap{--0.5}}}   & 0.9(1)                & $\gtrsim$1.5           & 2000 and 2002 data\rule[-1.5ex]{0pt}{0pt} \\
    W37 & 7.4\textsuperscript{\raisebox{1pt}{\rlap{+0.8}}}\textsubscript{\raisebox{-1pt}{\rlap{--0.5}}}      & $\gtrsim$0.9                       &  $\gtrsim$$0.5^{*}$                      & ObsID 2735--high\rule[-1.5ex]{0pt}{0pt}\\
    \hline    
  \end{tabular}
  
  \caption{X-ray fluxes $f_X$ and magnitudes $U_{\rm 300}$ and $B_{\rm
      435}$ for the counterparts discussed in Sec.~\ref{sec_oid} are
    corrected for absorption. For X7, we assumed N1 as the
    counterpart. For W37, we used the magnitudes of star 4, the
    faintest neighbor, to compute limits on the flux ratio's. The last
    column gives the epoch of the X-ray fluxes, where `ObsID
    2735--high' refers to the high-flux part of ObsID 2735 (different
    for X4 and W37). Flux ratios based on X-ray and optical
    measurements from the same epoch are marked with an asterisk
    (*). X-ray fluxes for X4 and W17 are given for the 0.5--10 keV
    band. For X5, X7 and W37 X-ray fluxes are given in the 0.5--2.5
    keV band but due to their very soft spectra, the values match the
    0.5--10 keV flux very well. A systematic error on the optical
    magnitude calibration of 2\% \citep[upper limit quoted
      in][]{siriea05} would add an error of $\sim$0.2 to the flux
    ratios.} \label{tab_xflux}

  \end{table}

\begin{table*}
\centering
\caption{Properties of the (likely) optical and NUV counterparts of the 47\,Tuc qLMXBs}
\label{tab_avgphot}
\begin{tabular}{lllllllll}
\hline
ID & $\alpha_{2000}$ & $\delta_{2000}$ & $d$ & \multicolumn{1}{l}{$U_{300}$} & \multicolumn{1}{l}{$B_{390}$} & \multicolumn{1}{l}{$B_{435}$} & \multicolumn{1}{l}{$R_{625}$} & \multicolumn{1}{l}{H$\alpha_{\rm 658}$} \\
   &  ($^{\rm h}$ $^{\rm m}$ $^{\rm s}$) & ($^{\circ}$ $\arcmin$ $\arcsec$) & ($\arcsec$) \\
\hline
X4$_{\rm opt}$  &  00:23:53.986 & $-$72:03:50.13  & 0.06 & 24.93(3)   & 24.599(18) & 24.20(4)    & 22.116(15) & 21.343(17)  \\
X5$_{\rm opt}$  &  00:24:00.958 & $-$72:04:53.22  & 0.08 & 22.997(14)  & 22.318(9) & 21.847(17)  & 20.334(11) & 19.732(15) \\
X7$_{\rm opt}$  &  00:24:03.501 & $-$72:04:51.96  & 0.07 & 23.06(2)   & 21.74(2)   & 21.289(17)  & 19.836(17) & 19.42(2) \\
W17$_{\rm NUV}$ &  00:24:08.307 & $-$72:04:31.44  & 0.05    & 22.883(18) & 23.17(3)   & $\gtrsim24$ & \ldots & \ldots \\
W37            &  \ldots       & \ldots          & \ldots  & $\gtrsim21.3$        & $\gtrsim20.1$    & $\gtrsim20.1$         & \ldots  & \ldots     \\
\hline
\end{tabular}

{\flushleft
\begin{flushleft}  
The columns $\alpha_{2000}$ and $\delta_{2000}$ give the coordinates
of the (likely) optical/NUV counterpart in the stacked GO\,9281 images
(X4$_{\rm opt}$, X5$_{\rm opt}$, X7$_{\rm opt}$), or in the stacked
GO\,12950 images (W17$_{\rm NUV}$). The column $d$ is the angular
offset between the X-ray and optical/NUV positions after applying the
boresight correction. The counterparts lie well within the 95\% error
circle around the boresighted {\em Chandra} positions
($\sim$0.36\arcsec).  Magnitudes were measured from the stacked
images, and thus represent averages. Errors on the magnitudes (in
parentheses) are formal DAOPHOT 1 $\sigma$ errors on the last
significant digit(s). Systematic errors on the absolute magnitude
calibration can amount up to 2\% \citep{siriea05}. We have not
identified a counterpart to W37 and provide lower limits to its
magnitudes.
\end{flushleft}
}

\end{table*}

\subsection{Additional NUV and optical imaging with {\em HST}}

We have analyzed NUV images of 47\,Tuc from two {\em HST} programs. On
2013 August 13 we observed the cluster with the UVIS channel on the
Wide Field Camera 3 (WFC3) as part of program GO\,12950. UVIS covers a
field of view of 2.7\arcmin~$\times$ 2.7\arcmin~with a scale of
0.04\arcsec~per pixel. During two consecutive visits, 24 images were
obtained through the F300X filter for a total exposure time of 14\,576
s, and eight images were taken through the F390W filter for a total
exposure of 4624 s. The dithering pattern included fractional-pixel
offsets. More details can be found in \cite{riveea15}.

We also analyzed archival images of 47\,Tuc from program GO\,9019 that
were obtained with the ACS High Resolution Camera (HRC) on 2002 April
12--13. The HRC offers a relatively small field of view of
29\arcsec$\times$25\arcsec, but a superb detector scale of
$\sim$0.025\arcsec~per pixel. This data set comprises images taken
through ten filters, which are laid out in a mosaic pattern covering
the cluster centre. We only analyzed the exposures in the NUV and blue
filters F220W, F330W, F435W (only the 60 s exposures) and F475W that
include X7 and/or W37 in their field of view. The other three systems
are not covered by this data set. Basic information for the GO\,12950
and GO\,9019 data that we used can be found in Table~\ref{tab_obs}.

Finally, we use the star catalogs of 47\,Tuc from the {\em HST UV
  Globular cluster Survey} (HUGS; \citealt{nardea18}). These are based
on ACS/WFC F606W and F814W images from program GO\,10775
(P.I.\,Sarajedini; \citealt{saraea07}) and ACS/WFC and WFC3/UVIS
images in the filters F275W (GO\,12311, P.I.\,Piotto), F336W
(GO\,11729, P.I.\,Holtzman; GO\,12971, P.I.\,Richer), and F435W
(GO\,9443, P.I.\,King; GO\,9281---which is our data set as described
in Sect.~\ref{sec_obs_sim}). The HUGS images in the blue filters were
taken over a time span of more than eleven years: F275W images are
from 2010, F336W images from 2010--2013 and F435W images from
2002. This could impact the blue colours of stars if they are
(long-term) variables. The F606W and F814W images were taken within a
time span of $<$4 hours, so the derived red colour is only sensitive
to short-term brightness variations. We refer to Table~\ref{tab_obs}
for the start and end times of all 47\,Tuc HUGS data. We have used the
magnitudes extracted with the HUGS ``method 3'' photometry, which is
appropriate for very crowded environments.

\section{Data Reduction and Analysis} \label{sec_ana}

\subsection{X-ray data} \label{ssec_ana_xray}

We refer to \cite{heinea05a,heinea05b} and \cite{bhatea17} for details
on the reduction of the {\em Chandra} data. We extracted light curves
from the four long ACIS-S exposures from 2002 with time bins of 3000 s
(X4, W17, W37) or 500 s (for X5 and X7, which are much
brighter). Event times were converted to solar-system barycentre times
using the CIAO tools. X-ray luminosities from earlier works
\citep{heinea05a,heinea05b} that assumed a cluster distance of 4.85
kpc, were adjusted for the distance we adopt here (4.53 kpc).

Neither X5 nor X7 show evidence for intrinsic X-ray luminosity
variability. We use the unabsorbed X-ray fluxes that were derived from
hydrogen-atmosphere spectral fits to the 2000 data in
\cite{heinea03a}. The count rate for X5 is variable, but (apart from
the obvious eclipses) the variability appears to reflect changes in
the amount of obscuration in front of the X-ray source, instead of
changes in the intrinsic flux, as evidenced by the increased $N_H$ fit
in periods of lower countrate \citep{heinea03a,bogdea16}.  The X5
observation from 2000 appears to be the least affected by
obscuration. The adopted flux for W37 was taken from Table 3 of
\citet{heinea05a}, and reflects the results of fitting the spectrum
extracted from the segment of ObsID 2735 that is least affected by
absorption. The model used is a combination of an absorbed hydrogen
atmosphere plus a non-thermal (power-law) component.  The flux values
for X5, X7 and W37 from \cite{heinea03a} and \cite{heinea05a} were
computed for the 0.5--2.5 keV band, but the spectra are so soft, with
no (X5, X7) or almost no (W37; $\lesssim2$\% flux contribution)
power-law component, that these are consistent for the 0.5--10 keV
band as well. W17 is not significantly variable. The spectrum from the
2000 and 2002 data can be fit with a combination of a
hydrogen-atmosphere plus power-law model, with the latter the dominant
component. We convert the total intrinsic luminosity in the 0.5--10
keV band from Table 2 in \cite{heinea05a} to an unabsorbed flux. The
flux of X4, on the other hand, is thought to vary intrinsically,
likely due to changes in the power-law component. \cite{heinea05a}
identified a period of high flux in ObsID 2735, between 6.7 and 12.8
hr after the start of the observation. We use 0.5--10 keV X-ray fluxes
separately for this interval and the other three ObsIDs in 2002, and
paired each value with the overlapping GO\,9281 F435W flux to compute
quasi-instantaneous X-ray--to--optical flux ratios.

We list the X-ray flux values and resulting X-ray--to--(optical or
NUV) flux ratios in Table~\ref{tab_xflux}. Estimated systematic errors
in the quoted X-ray fluxes are at least 5\%, and arise from the use of
different energy bands, and from uncertainties in modelling the
spectra and the line-of-sight absorption column density $N_H$. Details
of the spectral modelling can be found in \cite{heinea03a} and
\cite{heinea05a}.

\subsection{Optical and NUV {\em HST} images}

We started the reduction of the GO\,9281 WFC data with the images that
were flat-fielded and corrected for charge transfer efficiency (CTE)
losses by the ACS calibration pipeline (CALACS version 8.2.0). These
{\tt flc} images still suffer from significant geometric distortion,
which we corrected for by subsequent processing with the DrizzlePac
software \citep{gonzea12}. First, the small relative offsets of the
{\tt flc} images were corrected for with the routine {\tt
  tweakreg}. The aligned individual images were then used as input for
{\tt astrodrizzle} to create twice-oversampled (to
0.025\arcsec\,pixel$^{-1}$) stacked images in each filter that are
corrected for geometric distortion. Three short 10 s exposures, which
were taken to obtain photometry of bright cluster stars, were excluded
from the stacks. The stacked images are cosmic-ray cleaned with the
effects of bad columns and hot pixels removed as much as possible.

We reduced the GO\,9019 HRC data in a similar way, starting from the
pipeline-processed (CALACS version 9.2.0) flat-fielded images. This
program was executed about a month after the installation of ACS on
{\em HST} in 2002 March. The effects of CTE degradation are therefore
negligible, and we have not corrected for it. As for the GO\,9281
data, the GO\,9019 {\tt flt} frames were aligned with {\tt tweakreg}
and stacked with {\tt astrodrizzle} with a factor of two oversampling
to produce an image scale of 0.0125\arcsec\,pixel$^{-1}$ in the final
stacked images.

The starting point for the reduction of the GO\,12950 UVIS data was
the flat-fielded images from the UVIS pipeline (CALWF3 version
3.1.2). We ran the dedicated STScI
software\footnote{\url{http://www.stsci.edu/hst/wfc3/tools/cte\_tools}}
to correct for CTE losses, before aligning and stacking the images
with DrizzlePac. The scale of the image stacks is
0.02\arcsec\,pixel$^{-1}$. Details can be found in \cite{riveea15}.

\begin{figure*}
  \centerline{
      \includegraphics[width=8cm,angle=0]{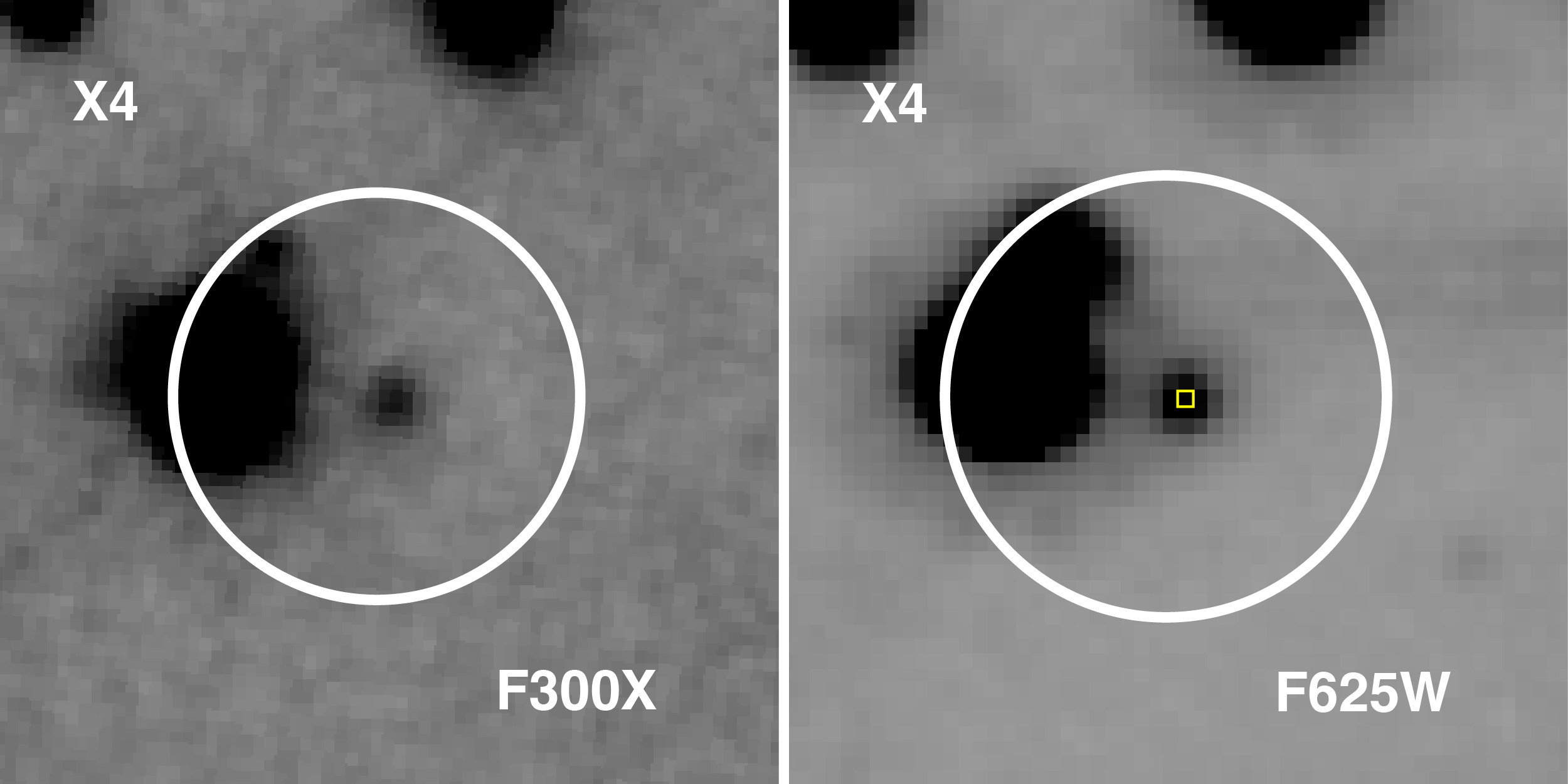} \hspace{0.3cm}
      \includegraphics[width=8cm,angle=0]{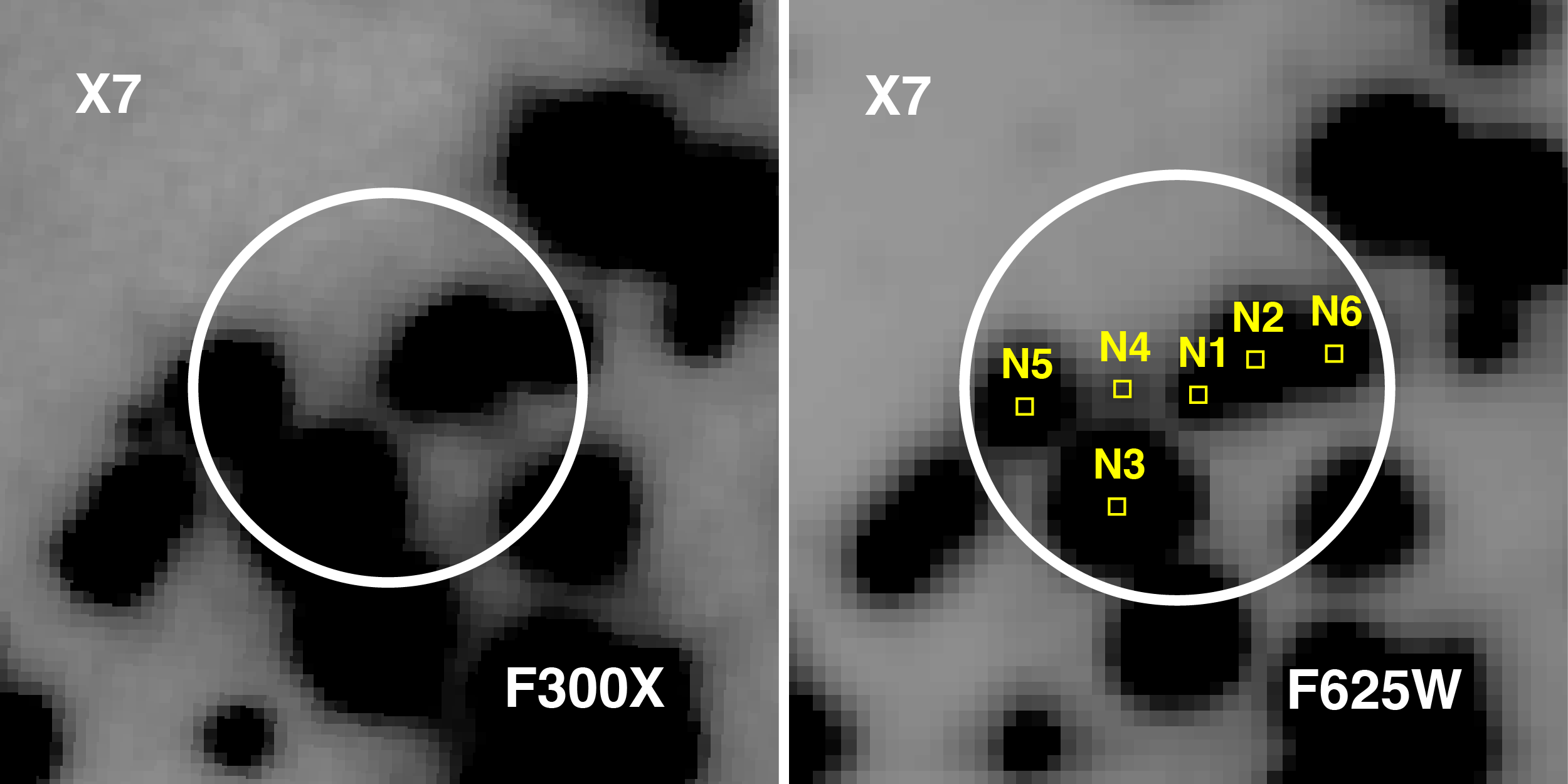}
  }
  \vspace{0.3cm}
  \centerline{
     \includegraphics[width=8cm,angle=0]{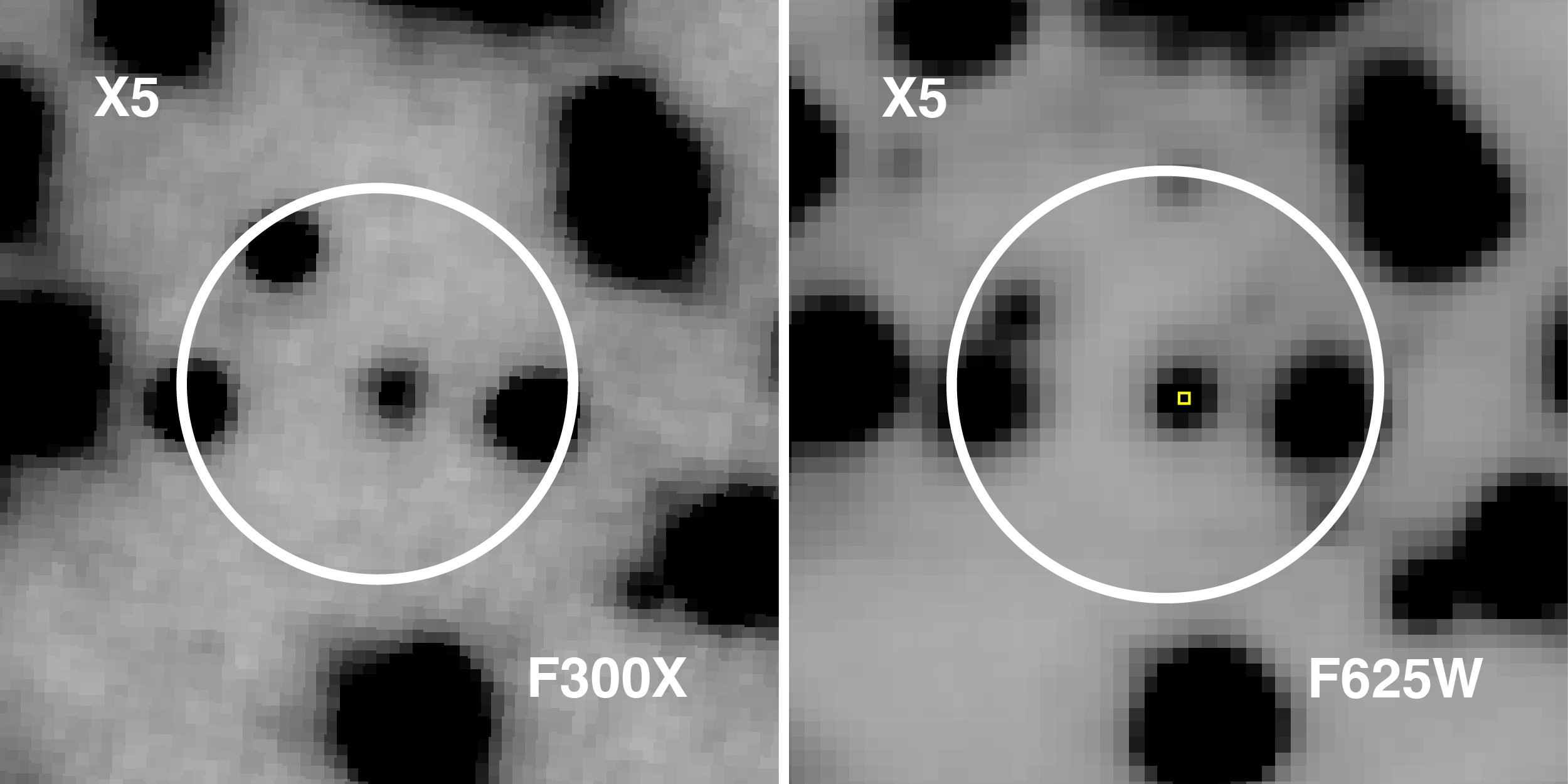} \hspace{0.3cm}
     \includegraphics[width=8cm,angle=0]{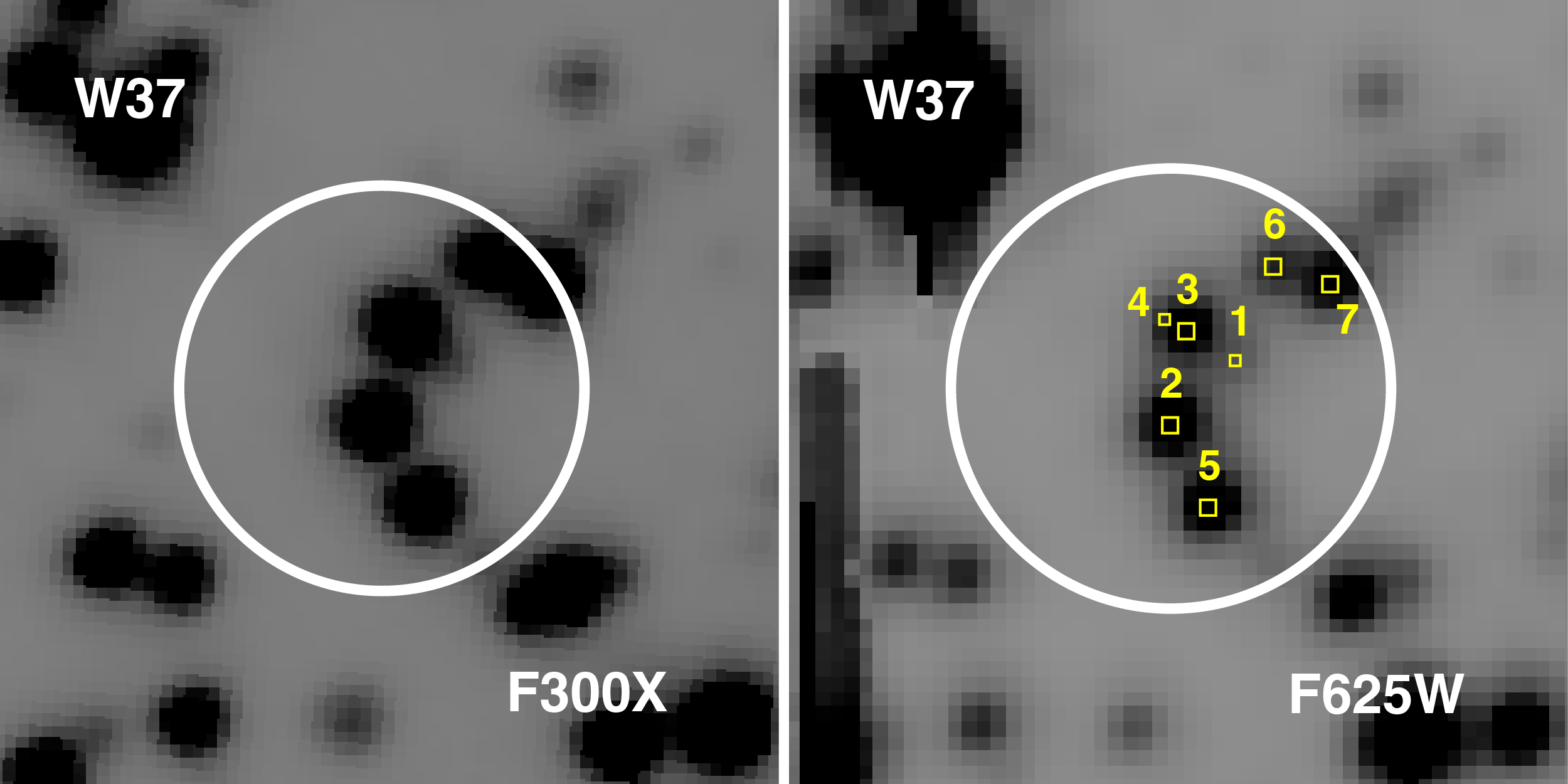}
  }
  \vspace{0.3cm}
  \centerline{
     \includegraphics[width=8cm,angle=0]{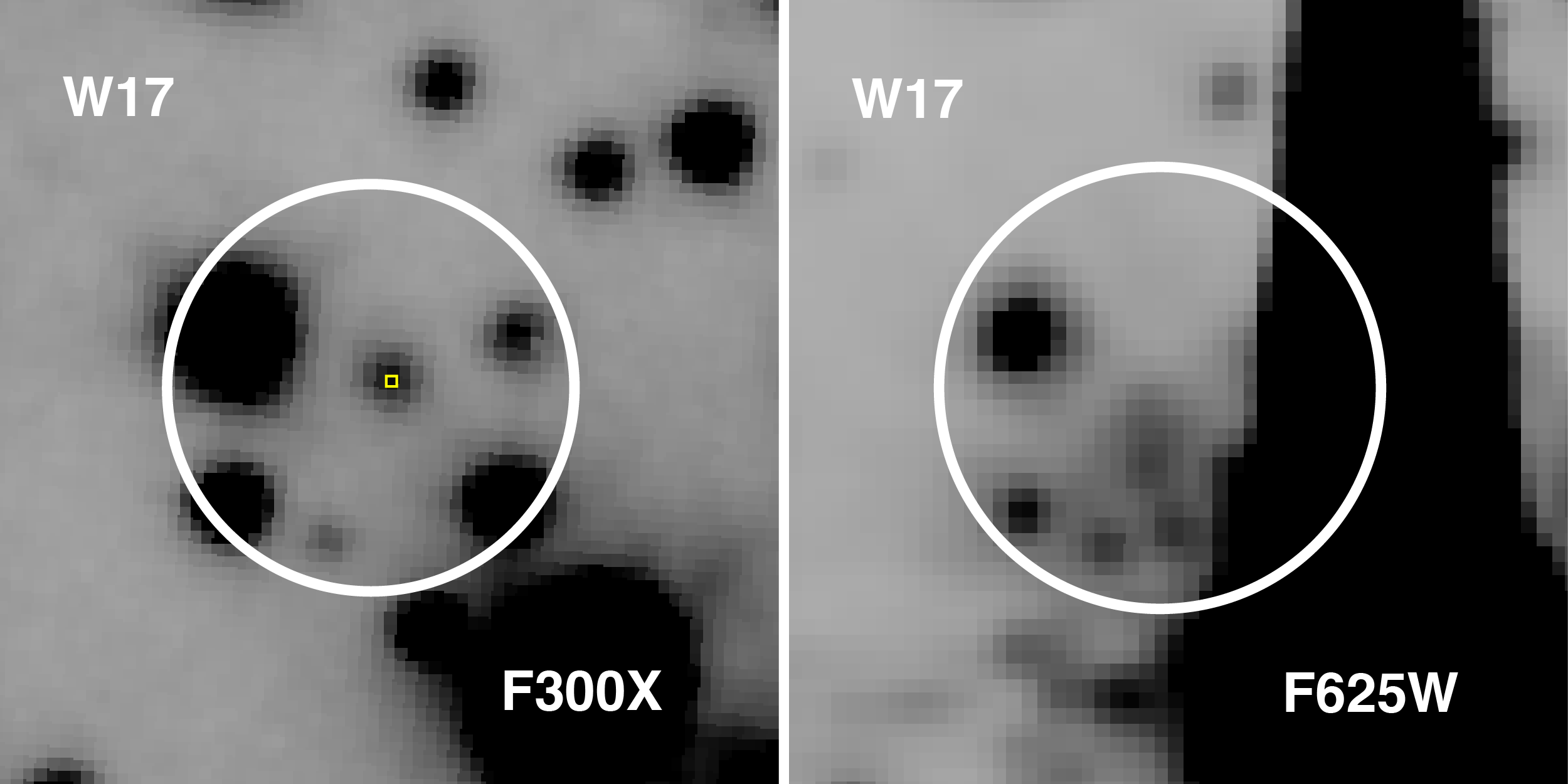} \hspace{8.4cm}
  }
\caption{Finding charts for the five qLMXBs discussed in this
  paper, showing 1.3\arcsec~$\times$ 1.3\arcsec~sections from the
  GO\,12950 F300X and GO\,9281 F625W stacked images. White circles are
  the 95\% error circles that combine the error in the X-ray position,
  the absolute astrometry of the {\em HST} images, and the
  boresight. The counterparts for X4, X5 and W17 are marked with small unlabelled
  yellow squares in the image where it is best visible. For X7 and
  W37, sources that are discussed in the text are labeled. We propose
  N1 as the counterpart for X7. North is up, east to the left.}
  \label{fig_fcs}
  \end{figure*}

\begin{figure}
   \centerline{\includegraphics[width=8cm,angle=0]{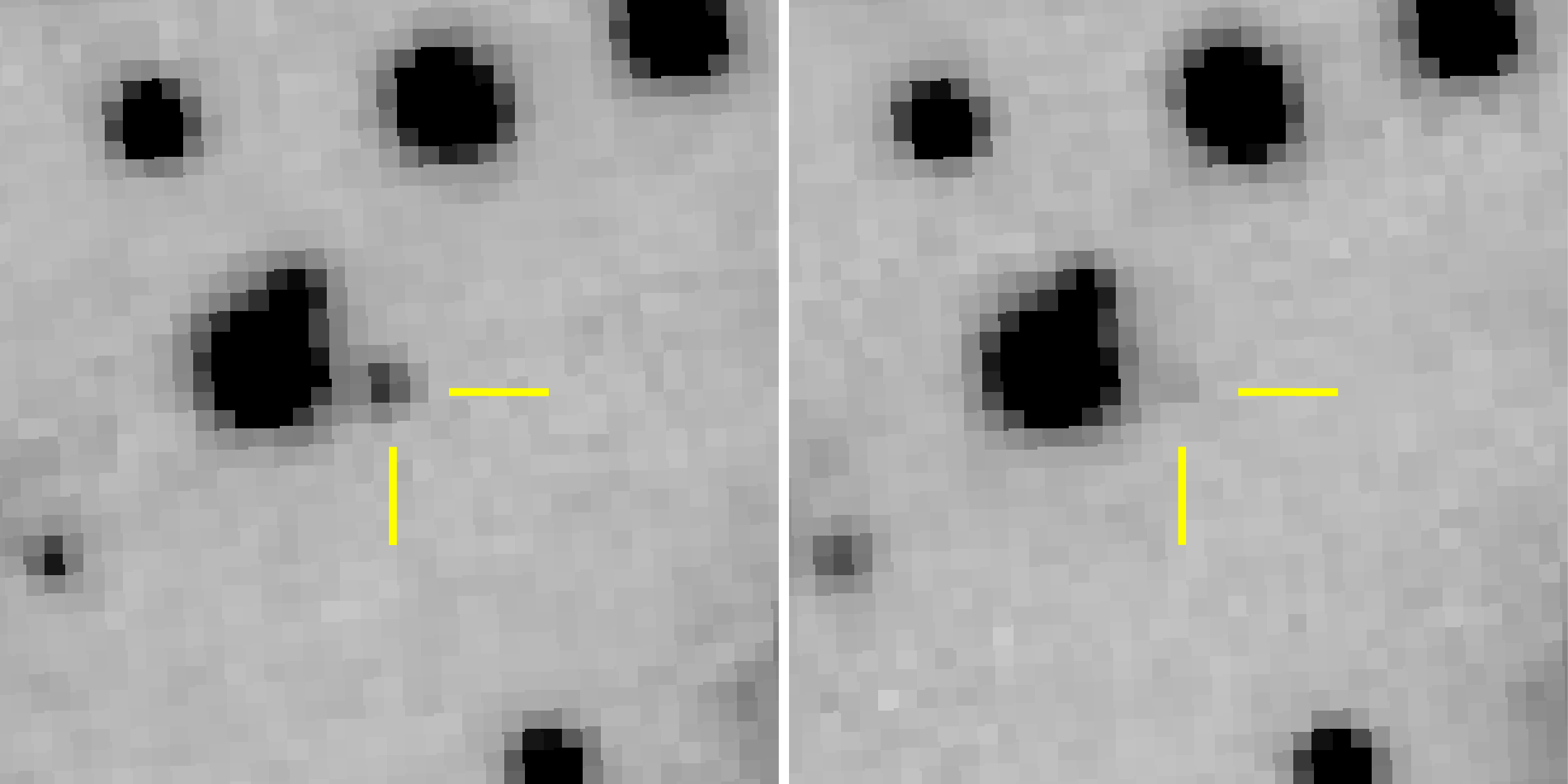}}
\caption{Two GO\,9281 F435W images (2\arcsec $\times$ 2\arcsec) that
  illustrate the variability of X4$_{\rm opt}$ (indicated with yellow
  tick marks). The exposures were taken 2.8 days apart (left: 2002
  September 30 3:22 UT, right: 2002 October 2 22:39 UT). At
  $B_{435}\approx24.1$, the counterpart is 1.2 mag fainter in the
  second epoch than in the first epoch. North is up, east is to the
  left. \label{fig_x4var}}
\end{figure}

\begin{figure*}
   \centerline{\includegraphics[width=17.2cm,angle=0]{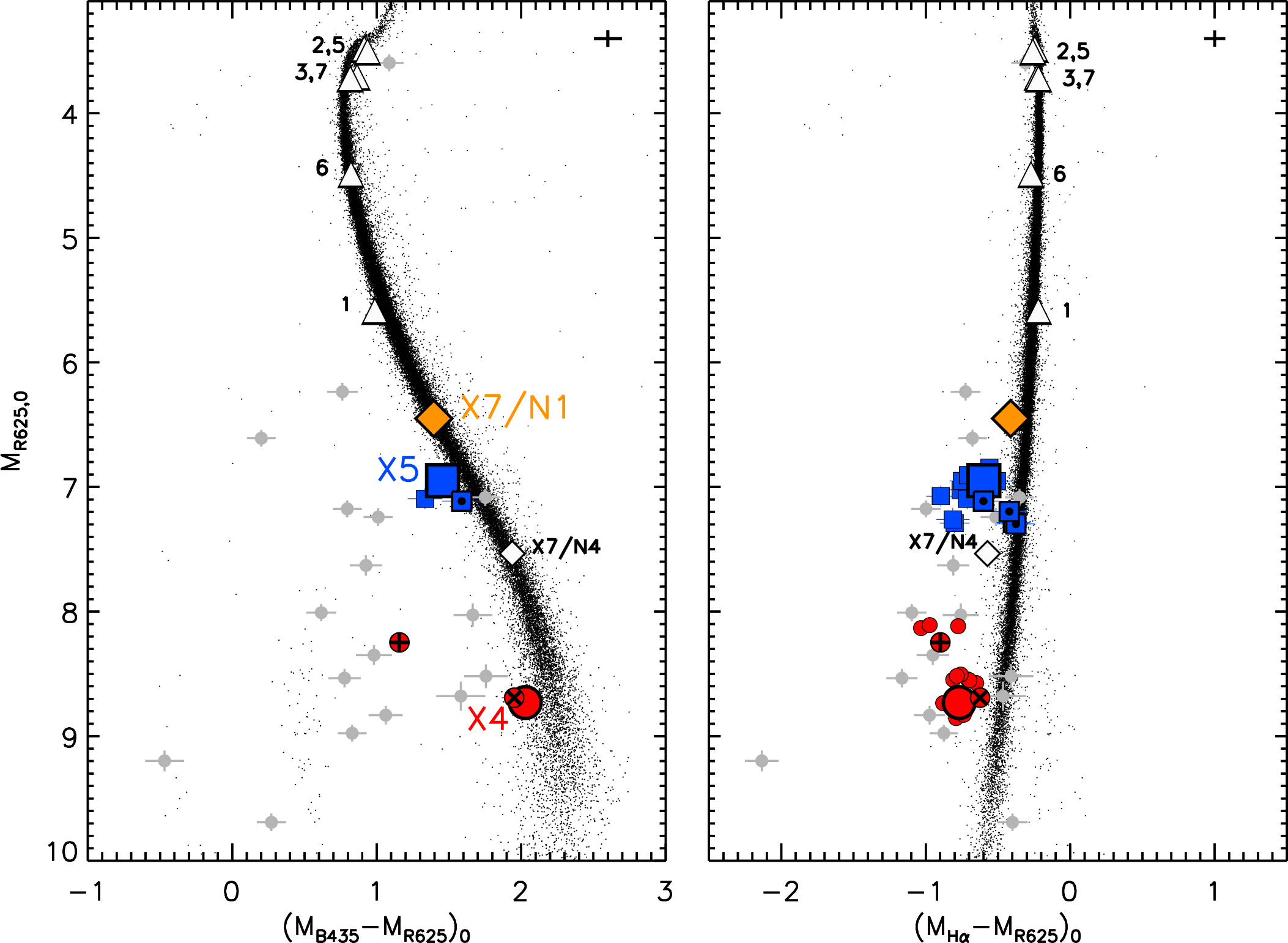}}
\caption{GO\,9281 $R_{\rm 625}$ versus $B_{435}-R_{625}$ and $R_{\rm
    625}$ versus H$\alpha_{\rm 658}-R_{625}$ colour-magnitude diagrams
  showing the average photometry of X4$_{\rm opt}$ (large red circle),
  X5$_{\rm opt}$ (large blue square), and the likely counterpart for
  X7 (N1, orange diamond). Stars close to the {\em Chandra} position
  of W37 (1--3 and 5--7 in Fig.~\ref{fig_fcs}; star 4 is only detected
  in $B_{435}$) are plotted as upward triangles, N4 near X7 as an open
  diamond. For X4$_{\rm opt}$ and X5$_{\rm opt}$, the smaller symbols
  show the quasi-instantaneous colours. For X4$_{\rm opt}$, the +
  ($\times$) signs mark measurements taken when the X-ray flux was high
  (low); see the light curve in Fig.~\ref{fig_xlcx4} (bottom) for the
  corresponding data points. For X5$_{\rm opt}$, black dots mark
  colours from visit 2 around the time of X-ray eclipse (see light
  curves in Figs.~\ref{fig_xlcx5} and \ref{fig_color}). Magnitudes are
  converted to dereddened absolute magnitudes using a distance
  $d$=4.53 kpc and reddening $E$($B-V$)=0.04. For comparison, colours
  of securely identified CVs in 47\,Tuc are shown as grey circles
  \citep{riveea18}. In the top right, we indicate the uncertainty as a
  result of errors on the reddening and the distance
  modulus. \label{fig_optcmd}}
\end{figure*}

\begin{figure}
 \centerline{\includegraphics[width=8cm,angle=0]{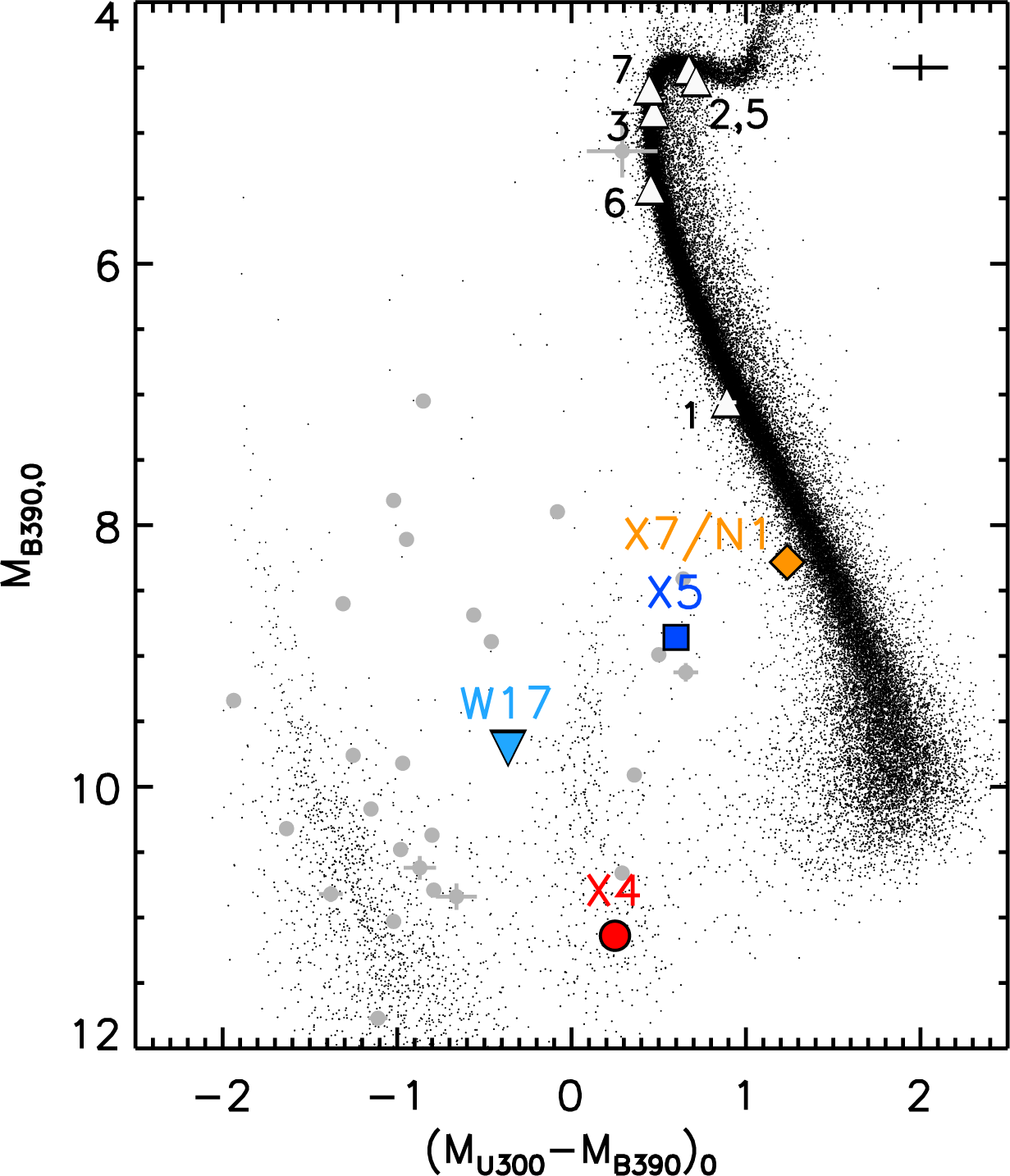}}
\caption{GO\,12950 $B_{{\rm 390}}$ versus $U_{{\rm 300}}-B_{{\rm
      390}}$ colour-magnitude diagram. The (likely) counterparts to
  X4, X5, X7 and W17 are plotted with coloured symbols. Photometry is
  extracted from the stacked images, so variability (see
  Fig.~\ref{fig_nuvlc}) is "averaged out''. Stars close to the {\em
    Chandra} position of W37 are plotted as open upward triangles;
  star 4 is omitted since its photometry is very uncertain (see
  text). Magnitudes are converted to dereddened absolute magnitudes
  using $d$=4.53 kpc and $E$($B-V$)=0.04. The uncertainty as a result
  of errors on the reddening and distance modulus is indicated in the
  top right. For comparison, colours of securely identified CVs in
  47\,Tuc are shown as grey circles \citep{riveea18}.}
\label{fig_nuvcmd}
\end{figure}

\subsubsection{Astrometry} \label{sec_astro}

The {\em Chandra} ACIS-S absolute astrometry is tied to the
International Celestial Reference System (ICRS) with an accuracy of
$\sim$1.1\arcsec\,(90\%
confidence)\footnote{\url{http://cxc.cfa.harvard.edu/cal/ASPECT/celmon/}}. The
absolute astrometry of our distortion-corrected {\em HST} images on
the other hand, can be off by $\sim$1\arcsec~or more \citep{koekea05}.
To make the optical identification of the {\em Chandra} sources
easier, we first tie the {\em HST} astrometry to the ICRS using UCAC2
stars, which have positional errors
$\lesssim$0.070\arcsec\ \citep{zachea04}. However, the few UCAC2
standards in the GO\,9281 images are often saturated even in the 10 s
exposures. Therefore, we do the astrometric calibration in several
steps. First, we astrometrically calibrated a ground-based image of
47\,Tuc. From the ESO archive we obtained a 30 s $V$ image taken with
the 2.2m/Wide Field Imager (WFI) at La Silla, Chile on 2002 October
29. An astrometric solution was derived based on 225 UCAC2 stars;
fitting for relative shift and rotation, pixel scale, and distortions,
gives rms residuals of $\sim$0.035\arcsec\ in right ascension and in
declination. Using 126 stars that are unsaturated and relatively
isolated in the WFI and the short GO\,9281 F435W exposure, we derived
an astrometric solution for the latter, which has rms residuals of
$\sim$0.030\arcsec~in each direction. Finally, this solution was
transferred to the stacked GO\,9281 images with negligible errors
using $>$10\,000 stars. We estimate the final 1\,$\sigma$ accuracy of
the WFC absolute astrometry as the quadratic sum of the random errors
in the UCAC2 positions, the UCAC2--WFI tie and the WFI--{\em HST} tie,
augmented with the estimated systematic error in UCAC2 positions (10
mas, \citealt{zachea04}). This results in an error of
$\sim$0.095\arcsec. We checked for a systematic offset between the
{\em Chandra} and {\em HST} astrometry by computing the weighted
average of the differences in {\em Chandra} and WFC positions of 29
known matches between X-ray sources and their optical counterparts
(see e.g.~the finding charts in
\citealt{edmogillea03a,edmogillea03b}). The resulting offset (X-ray
minus optical) is $\Delta\alpha$=$-$0.027\arcsec$\pm$0.006\arcsec and
$\Delta\delta$=+0.060\arcsec$\pm$0.009\arcsec.

The GO\,12950 images were calibrated in a similar way, yielding a
final error in the astrometry of 0.074\arcsec, and a small boresight
correction of $\Delta\alpha$=$-$0.023\arcsec$\pm$0.008\arcsec and
$\Delta\delta$=$-$0.003\arcsec$\pm$0.009\arcsec.

We apply the boresight correction to the {\em Chandra} positions and
looked for optical and NUV counterparts in the 95\% error circles,
computed by adding the appropriate errors in the optical (or NUV)
positions, the X-ray positions, and the boresight errors (in
quadrature). The GO\,9019 and HUGS images were not explicitly aligned
to the X-ray positions, but it was easy to cross-reference them to the
aligned GO\,9281 and GO\,12950 images by eye.

\subsubsection{Photometry} \label{ssec_phot}

We used the IRAF {\sc DAOPHOT} package to extract magnitudes from the
stacked GO\,9281 and GO\,12950 images with point spread function
photometry. For the GO\,12950 images, we also carried out photometry
with the KS2 package \citep{andea08} as an independent check. In
general, KS2 and {\sc DAOPHOT} produce consistent results. Unless
mentioned otherwise, we report the {\sc DAOPHOT} magnitudes in this
paper. All photometry is calibrated to the Vega-mag system using the
STScI zeropoints. We denote calibrated magnitudes in the F300X, F390W,
F435W, F625W, and F658N filters by $U_{{\rm 300}}$, $B_{{\rm 390}}$,
$B_{{\rm 435}}$, $R_{{\rm 625}}$, and H$\alpha_{{\rm 658}}$,
respectively. To compute dereddened magnitudes, we converted the
cluster reddening of $E$($B-V$)=0.04$\pm$0.02 to filter-specific
extinction values. For the NUV filters, we obtained $A$($U_{{\rm
    300}}$)=$0.26\pm0.13$ and $A$($B_{{\rm 390}}$)=$0.18\pm0.08$ using
the UVIS Exposure Time Calculator. For the optical filters, the
conversions given in Table 14 in \cite{siriea05} yield $A$($B_{{\rm
    435}}$)=$0.16\pm0.08$, $A$($R_{{\rm 625}}$)=$0.11\pm0.05$, and
$A$(H$\alpha_{{\rm 658}}$)=$0.10\pm0.05$. Optical and NUV fluxes used
to estimate X-ray--to--optical or X-ray--to--NUV flux ratio's
(Table~\ref{tab_xflux}) were computed from the dereddened magnitudes
using the zeropoint fluxes provided by
STScI.\footnote{\url{https://acszeropoints.stsci.edu}}$^,$\footnote{
  \url{https://www.stsci.edu/hst/instrumentation/wfc3/data-analysis/photometric-calibration/uvis-photometric-calibration}}

In order to create optical light curves for the (candidate)
counterparts, we ran aperture photometry on the individual images
(except for the short exposures). Observation times were converted to
barycentric times. We created NUV light curves with the {\sc DOLPHOT}
software \citep{dolp00}. {\sc DOLPHOT} also produces average
photometry, which we find to be in general agreement with the {\sc
  DAOPHOT} and KS2 photometry. More details about the photometric
analysis can be found in \cite{riveea15,riveea18}.

We carried out aperture photometry on the twice-oversampled, drizzled
GO\,9019 frames using the DAOPHOT {\tt FIND} and {\tt PHOT}
utilities. We used a 2 pixel radius (0.025 arcsec) aperture and
corrected the flux to an infinite aperture using the encircled energy
fractions from Table 4 in \cite{siriea05}. The magnitudes were
calibrated to the Vega-mag system using the zeropoints from Table 11
in \cite{siriea05}. In order to filter out outliers, such as blended
stars and PSF artifacts, we only included objects in the CMD for which
the positions across frames agreed to within 0.7 pixels.

\section{Results} \label{sec_oid}

\begin{figure*}
   \centerline{\includegraphics[width=16.5cm,angle=0]{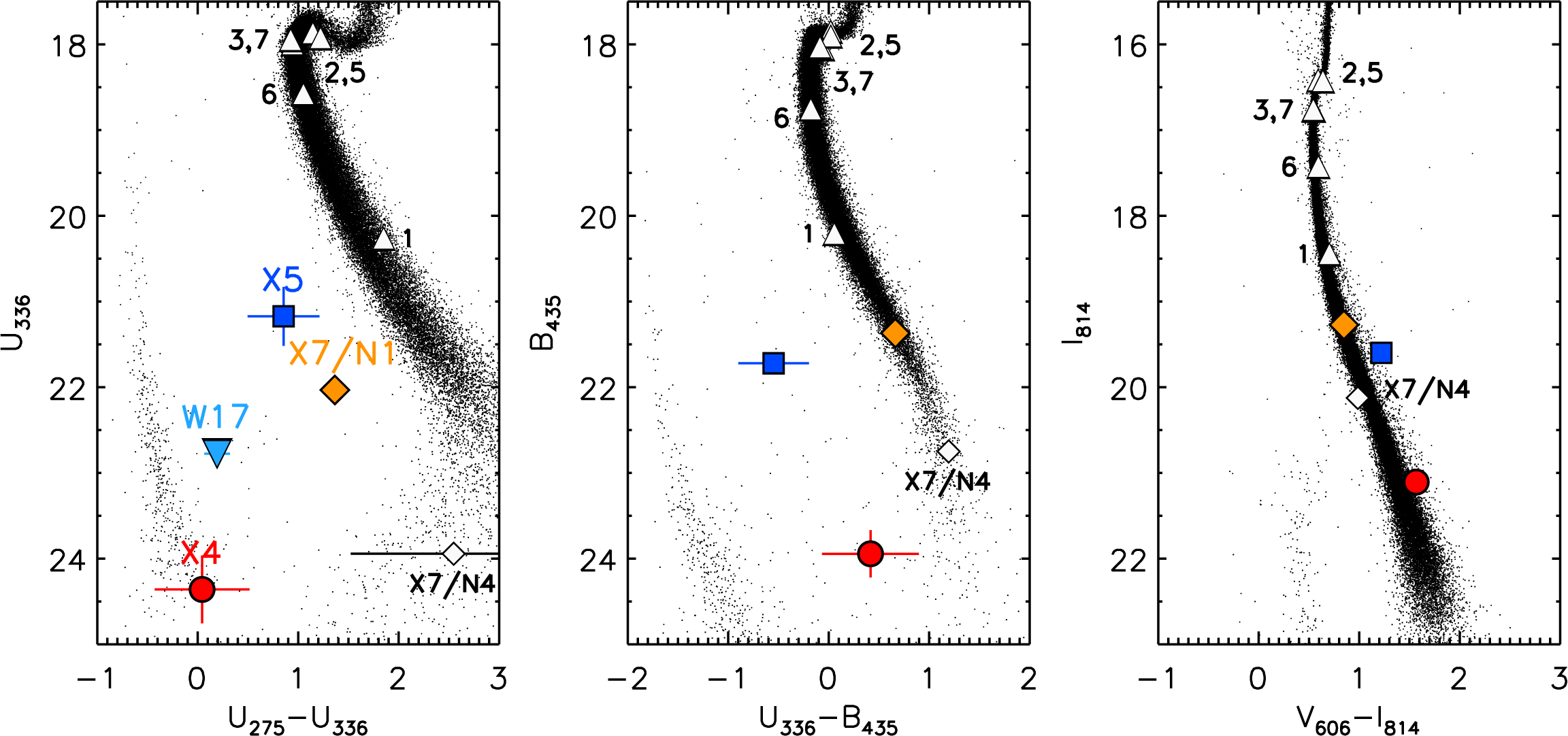}}
\caption{HUGS colour-magnitude diagrams with the (likely) counterparts
  marked with coloured symbols. The HUGS photometry in F275W, F336W
  and F435W is not contemporaneous (see Table~\ref{tab_obs}), which
  could affect the blue colours of objects in the HUGS catalogue if
  they are (long-term) variables. Unlikely counterparts near X7 and
  W37 are marked with the same (open) symbols as in
  Fig.~\ref{fig_optcmd}. Errors on the photometry of the (proposed or
  unlikely) counterparts are computed with the HUGS-reported rms
  values divided by square-root of the number of measurements. Error
  bars are plotted but in some case they are smaller than the symbol
  sizes. The relatively large (for its brightness) errors of X5$_{\rm
    opt}$ in the left and middle panels are dominated by the large rms
  in the F336W magnitudes.}
  \label{fig_hugscmd}
\end{figure*}

\begin{figure*}
  \centerline{
    \includegraphics[width=17.5cm,angle=0]{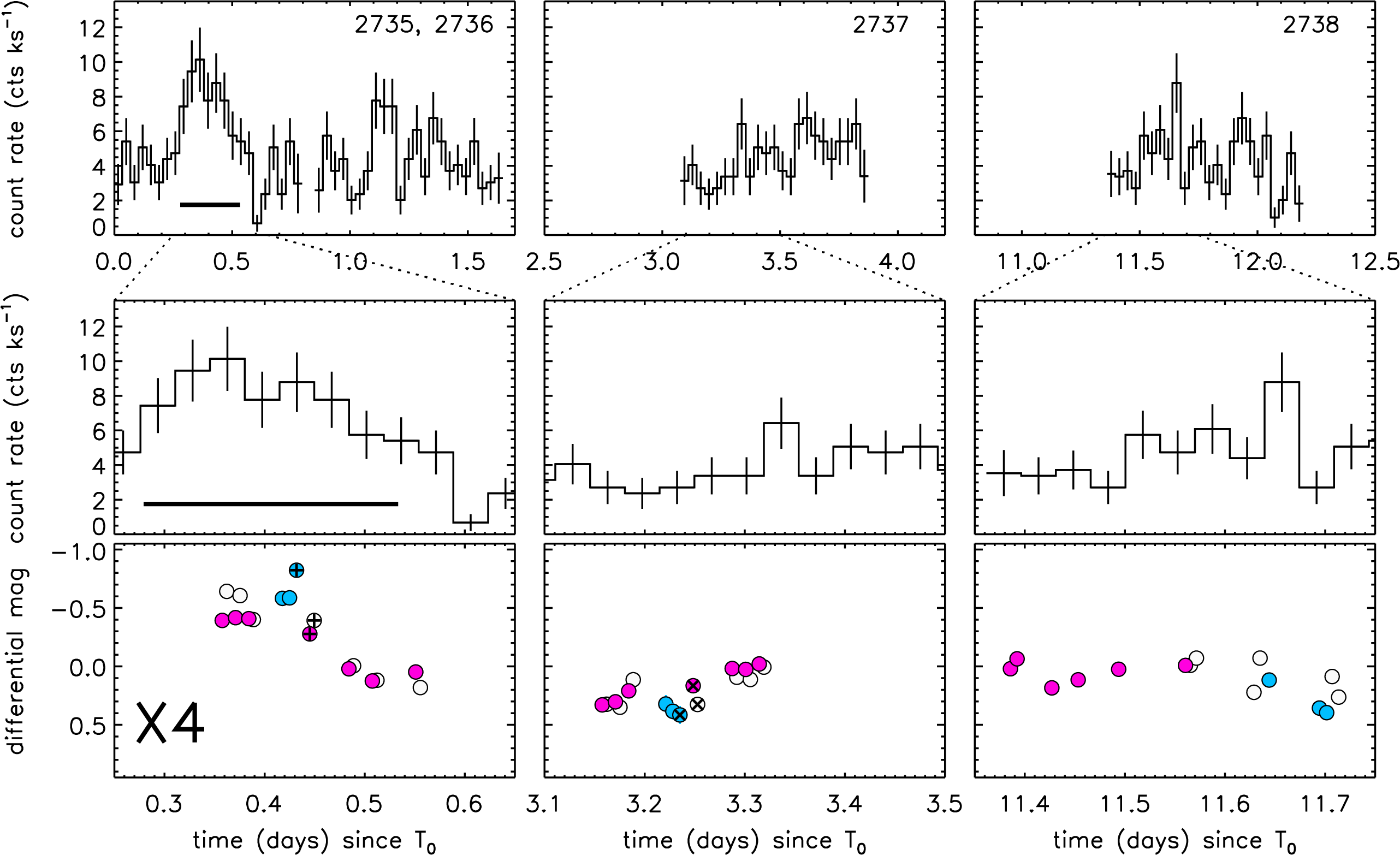}}
\caption{Simultaneous {\em Chandra} and GO\,9281 light curves of
  X4. The top row shows the X-ray light curves (0.3--8 keV) from the
  four long 2002 {\em Chandra} observations (labeled by ObsID). The
  middle row zooms in on the sections with simultaneous {\em HST}
  coverage. The thick horizontal lines in the top-left and middle-left
  panels indicate the high X-ray flux interval between 6.7 hr and 12.8
  hr after the start of the observation. In the bottom panel we show
  the optical light curves (F435W: blue; F625W: pink; F658N: white)
  after subtracting the average magnitudes. The three circles with +
  signs (bottom-left) mark the $B_{\rm 435}$, $R_{\rm 625}$ and
  H$\alpha_{\rm 658}$ magnitudes that were combined to compute colours
  when the X-ray flux is high, and the three circles with $\times$
  signs (bottom-middle) mark magnitudes that were combined to compute
  colours during the low X-ray--flux interval. Time on the x-axis is
  in units of days since the start of the first 2002 {\em Chandra}
  observation (2002 Sep 29 17:01 UTC).  \label{fig_xlcx4}}
\end{figure*}

\begin{figure}
      \centerline{\includegraphics[width=8cm,angle=0]{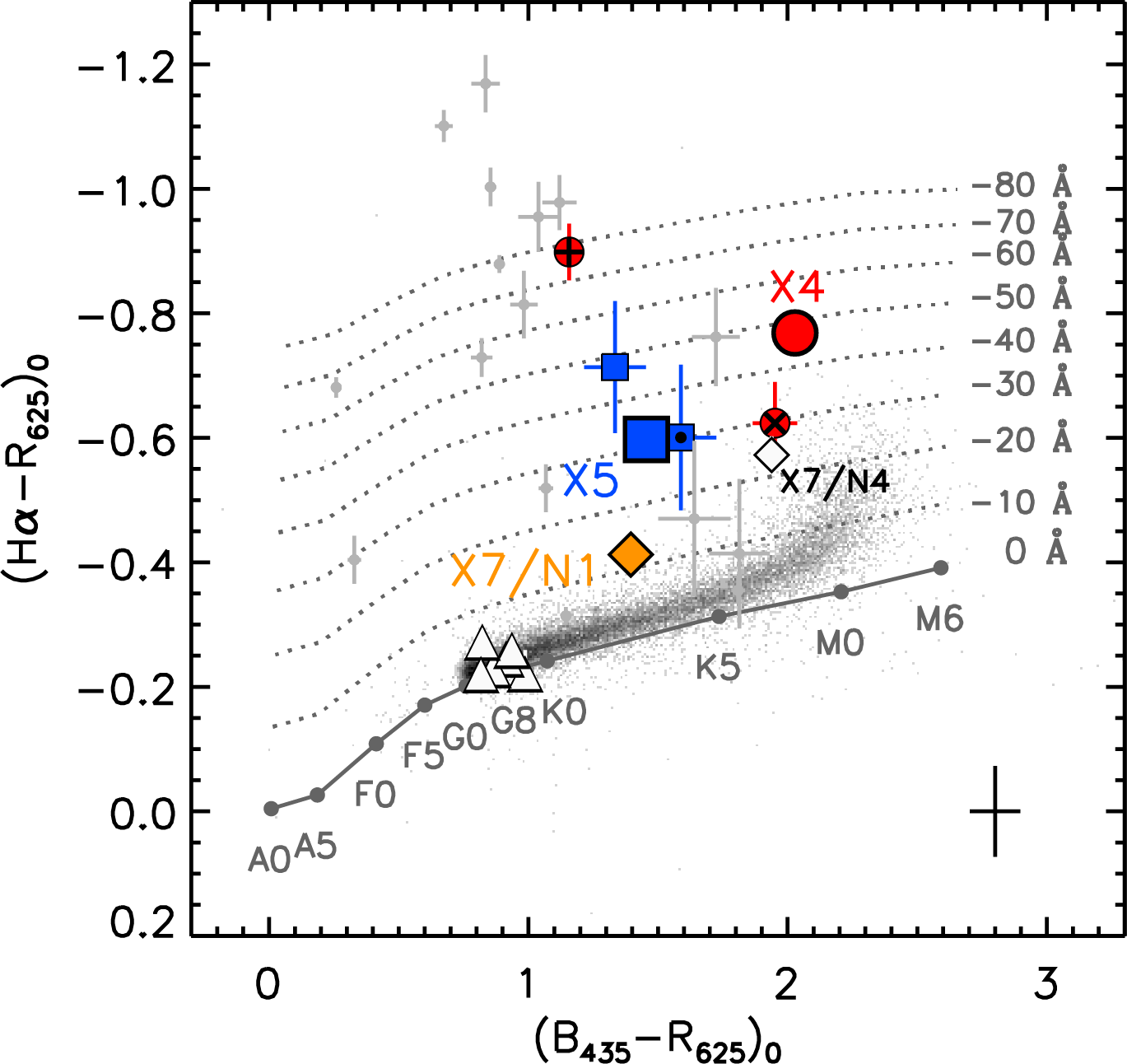}}
  \caption{GO\,9281 colour-colour diagram with the average colours of
    the (likely) counterparts marked with the larger coloured
    symbols. The small symbols show the quasi-instantaneous colours
    for X4$_{\rm opt}$ and X5$_{\rm opt}$.  Unlikely counterparts near
    X7 and W37 are marked with the same (open) symbols as in
    Fig.~\ref{fig_optcmd}. The photometry has been dereddened using
    $E$($B-V$)=0.04$\pm$0.02. The grey solid line connects the
    synthetic colours of stars as computed from Castelli-Kurucz model
    spectra of $\log Z$=$-2.5$ with the {\tt Synphot} package. The
    dotted grey lines represent synthetic colours that result from
    adding a Gaussian emission line with FWHM 25\AA\,to the model
    spectra, with the line strength increasing in increments of
    $\Delta$EW(H$\alpha$)=$-10$\AA. The grey circles are the colours
    of securely identified CVs in 47\,Tuc. In the bottom right, we
    show the uncertainty as a result of the error on the
    reddening.  \label{fig_optccd}}
\end{figure}

\subsection{X4} \label{ssec_w125}

X4 was first detected in ROSAT High Resolution Imager (HRI)
observations of 47\,Tuc \citep{hasiea94}. The source was identified
with W125 in the catalog of sources inside the cluster's half-mass
radius detected in the {\em Chandra} data from 2000 and 2002
\citep{heinea05b}. The model for X4's X-ray spectrum requires a soft
component in addition to a dominant harder component to yield an
acceptable fit. Consequently, X4 was classified as a likely qLMXB
based on its spectral resemblance to other such sources. During the
2002 {\em Chandra} observations, X4 displayed variability on time
scales of $\lesssim1$ day. As X4 brightened in ObsID 2735, its
spectrum became harder, indicating that a reduction in $N_H$ was not
responsible for the increase in count rate. \cite{heinea05a} argued in
favor of changes in the power-law component driving the variability,
possibly a sign of ongoing accretion.

X4 lies $\sim$1.4\arcmin\ to the north-west of the centre of 47\,Tuc,
and outside the field of view of the {\em HST} images analyzed by
\cite{edmogillea03a,edmogillea03b} to look for optical counterparts to
     {\em Chandra} sources. \cite{beccea14} examined the GO\,9281
     F658N (but not the F435W and F625W) images that we also analyze
     here, and reported a likely counterpart with excess H$\alpha$
     emission. In order to look for H$\alpha$-excess sources, they
     combined the H$\alpha_{\rm 658} $ magnitudes from GO\,9281 with
     $V$ (F606W) and $I$ (F814W) magnitudes from ACS/WFC program
     GO\,10775 (also included in HUGS), taken $\sim$3.5 yr after the
     GO\,9281 data. As we show below, X4 is considerably variable in
     the optical as well. As a result, colours constructed from
     non-contemporaneous measurements can be affected by
     variability. We revisit X4's H$\alpha_{\rm 658}$ colour below.
     
In the GO\,9281 data, we readily found a variable object at only
0.06\arcsec\, from the boresighted {\em Chandra} position of X4
(Fig.~\ref{fig_fcs}, top-left). Comparison of its coordinates with
those of the counterpart found by \cite{beccea14} shows that these are
the same object. X4 lies in an uncrowded part of the
cluster. Examination of the individual GO\,9281 exposures, in which
X4's likely counterpart (X4$_{\rm opt}$ henceforth) is easily
resolved, reveals that this star is clearly distinguishable in some
images but almost fades away in others. In Fig.~\ref{fig_x4var} we
compare images from 2002 September 30 and October 2. The counterpart
dimmed by $\Delta B_{435}\approx1.2$ mag in $\sim$2.8 days. Its
average magnitudes as derived from the stacked images, place it to the
blue and H$\alpha$-bright (i.e.~left) side of the bottom of the main
sequence in the optical colour-magnitude diagram (CMD) of GO\,9281
(Fig.~\ref{fig_optcmd}; large red filled circle). At increasingly
shorter wavelengths, we see that X4$_{\rm opt}$ becomes relatively
bluer. In the NUV CMD of GO\,12950, the object lies between the
47\,Tuc main sequence and the white-dwarf sequence
(Fig.~\ref{fig_nuvcmd}), a region also occupied by background stars in
the Small Magellanic Cloud. The position and optical/NUV magnitudes of
X4$_{\rm opt}$ can be found in Table~\ref{tab_avgphot}.

In the HUGS $B_{\rm 435}$ versus $U_{\rm 336}-B_{\rm 435}$ CMD
X4$_{\rm opt}$ lies in a similar region, and for the bluest HUGS CMD,
the counterpart falls right on the white-dwarf sequence
(Fig.~\ref{fig_hugscmd}, middle and left CMDs). Long-term brightness
variations add uncertainty to the blue HUGS colours of X4 since the
combined data sets are non-contemporaneous. The relative location of
X4$_{\rm opt}$ with respect to the main sequence is qualitatively
different in the reddest ($I_{\rm 814}$ versus $V_{\rm 606}-I_{\rm
  814}$) CMD, where it lies slightly to the red or above the main
sequence (Fig.~\ref{fig_hugscmd}, right CMD).  The HUGS catalogue does
not provide a proper-motion membership probability for X4$_{\rm opt}$.

In order to establish to what extent variability impacts the colours
of X4$_{\rm opt}$, we also examine its quasi-instantaneous GO\,9281
colours. We created light curves in the $B_{\rm 435}$, $R_{\rm 625}$
and H$\alpha_{658}$ filters using aperture photometry on the
individual exposures, adopting a small aperture radius of 1.5 pixels
to minimise contamination by the nearby bright star to the east (see
Fig.~\ref{fig_xlcx4}, bottom row).  By combining pairs of $R_{\rm
  625}$ and H$\alpha_{\rm 658}$ measurements that are closest in time
($\sim$6--7 min apart), we constructed fifteen H$\alpha_{\rm
  658}-R_{\rm 625}$ measurements; these are plotted as small red
circles in the right panel of Fig.~\ref{fig_optcmd}. The two small red
circles in the left panel represent $B_{\rm 435}-R_{\rm 625}$ colours
from images taken $\sim$19 min apart (the time separation is different
here because of the observing sequence). With plus-signs inside the
red circles we have marked one $B_{\rm 435}-R_{\rm 625}$ and one
H$\alpha_{\rm 658}-R_{\rm 625}$ measurement taken during a $\sim$31
min time span during the high X-ray flux interval in ObsID 2735
(individual data points are marked in the bottom-left panel of
Fig.~\ref{fig_xlcx4}). Similarly, we marked a pair of colours taken
during the low X-ray flux interval with x-shaped crosses (see also the
bottom-middle panel of Fig.~\ref{fig_xlcx4}). We find that X4$_{\rm
  opt}$ shows large optical brightness and colour variations that
appear to be correlated with its behaviour in X-rays. When the X-ray
flux is high, X4$_{\rm opt}$ is brighter, bluer, and has a larger
H$\alpha_{658}$ excess than during the remainder of GO\,9281 (see
middle and bottom rows of Fig.~\ref{fig_xlcx4}). The observed
X-ray/optical correlation supports the interpretation that the rise in
X-ray emission is caused by a temporary higher accretion rate. The
average colour of X4$_{\rm opt}$, as well as the colours during the
high and low X-ray flux intervals, are also plotted in the
colour-colour diagram (CCD) of Fig.~\ref{fig_optccd}. We infer that
the strength of the H$\alpha_{\rm 658}$ excess emission in terms of
equivalent width (EW) can vary at least between about $-$30 \AA~and
$-$80 \AA. The average EW in the GO\,9281 images is about $-$50 \AA\,,
larger than the EW estimate by \cite{beccea14} of $-$28 \AA.

\begin{figure}
   \centerline{\includegraphics[width=8cm,angle=0]{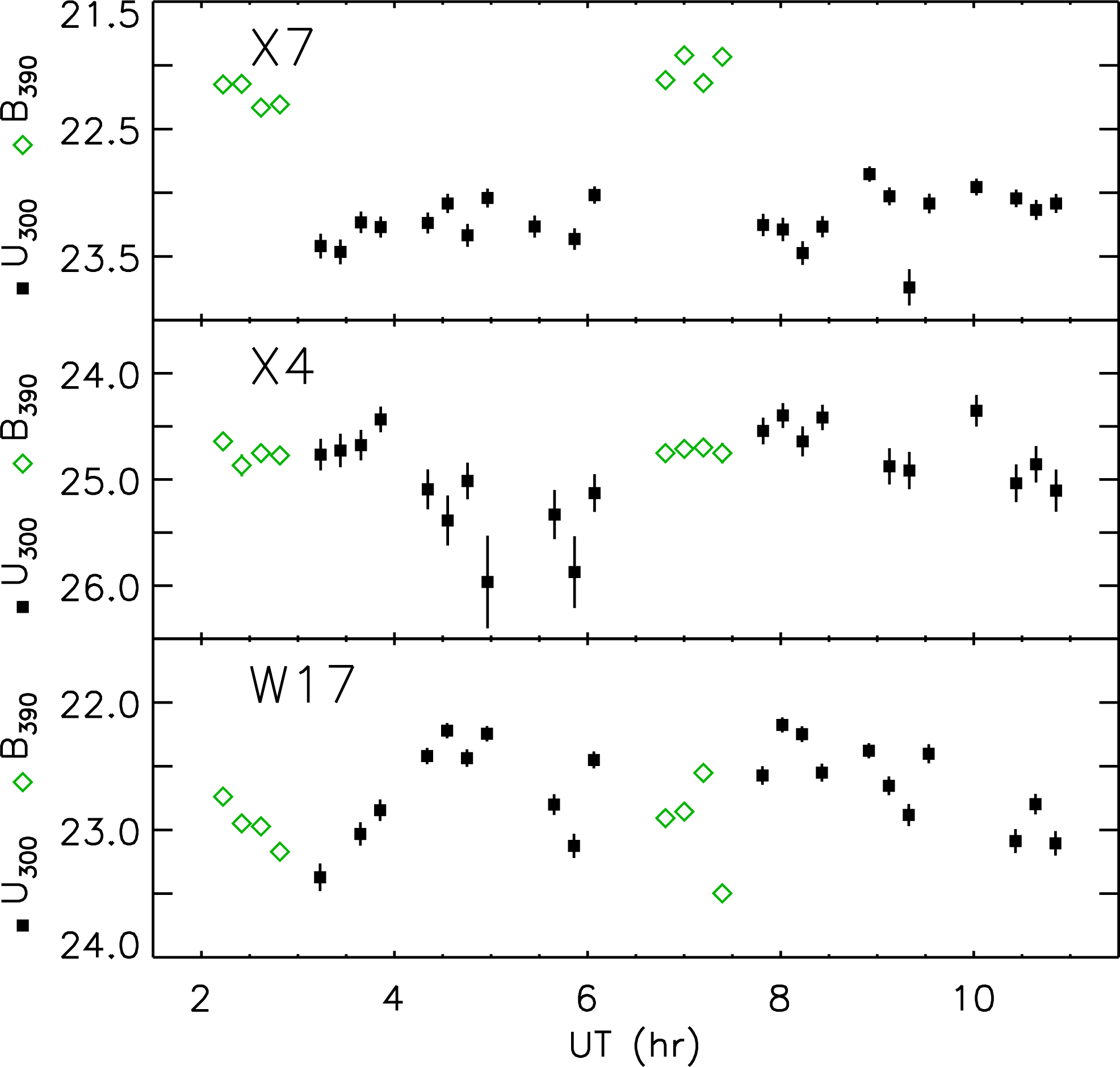}}
       \caption{GO\,12950 light curves in $U_{\rm 300}$ (filled squares) and
  $B_{\rm 390}$ (open green diamonds) for the counterparts to X7, X4 and
  W17. Magnitudes that could be affected by CRs are excluded. Time on
  the x-axis is UT on 2013 August 13.
\label{fig_nuvlc}}
\end{figure}

In the GO\,12950 images, the object is significantly variable in
$U_{\rm 300}$ as well, by $\sim$1.6 mag (although at the faint end
magnitude errors are $\sim$0.4 mag). In $B_{\rm 390}$, the source is
more stable (Fig.~\ref{fig_nuvlc}).

\begin{figure*}
  \centerline{
     \includegraphics[width=17.5cm,angle=0]{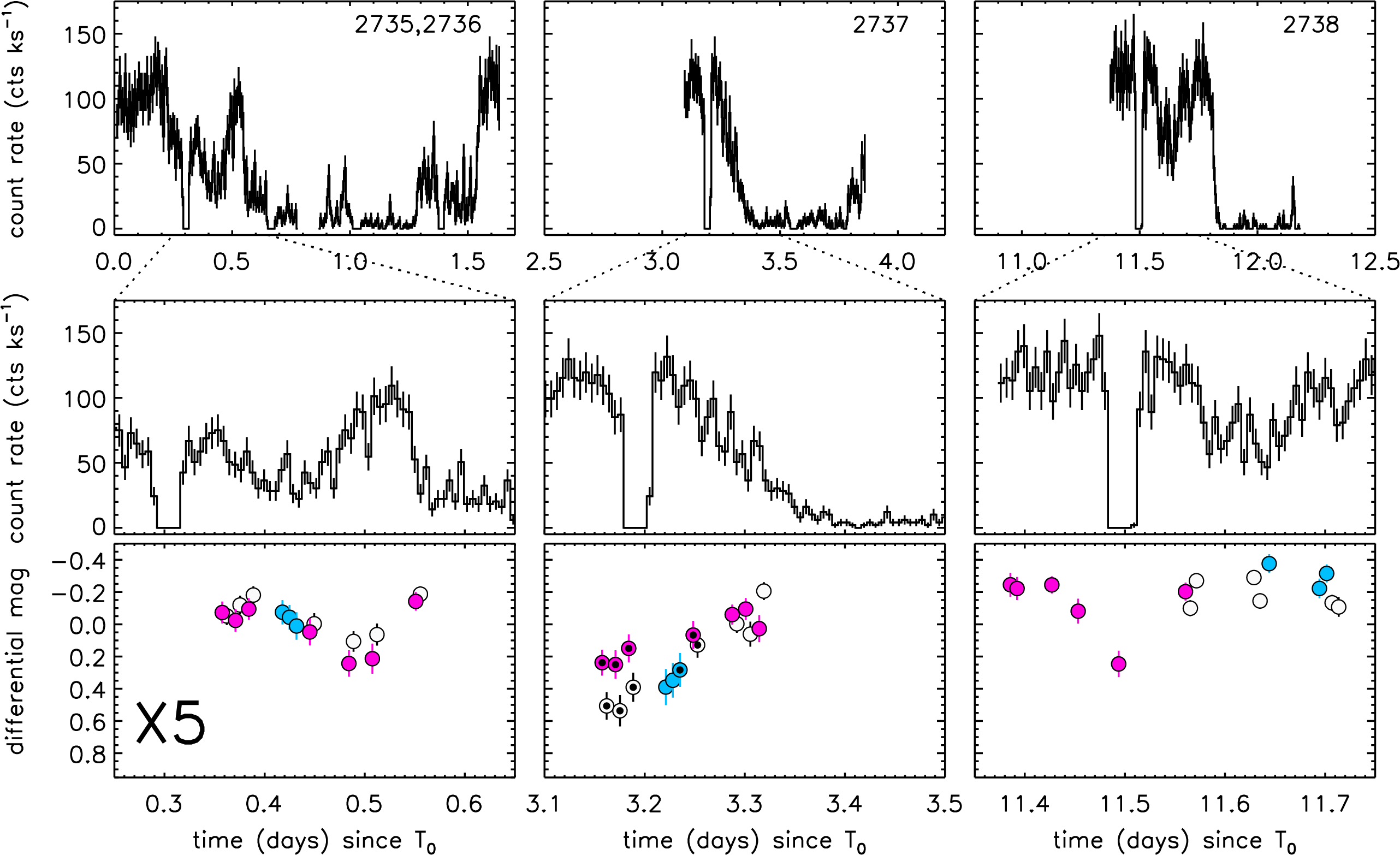}}
\caption{Simultaneous {\em Chandra} and GO\,9281 light curves of
  X5. The top row shows the X-ray light curves (0.3--8 keV) from the
  four long 2002 {\em Chandra} observations (labeled by ObsID). The
  middle row zooms in on the sections with simultaneous {\em HST}
  coverage. In the bottom panel we show the optical light curves
  (F435W: blue; F625W: pink; F658N: white) after subtracting the
  average magnitudes. Data points with the black filled dots (in
  bottom-middle panel) are the magnitudes that were combined to
  compute the colours of X5 around the time of the X-ray eclipse. Time
  on the x-axis is in units of days since the start of the first 2002
  {\em Chandra} observation (2002 Sep 29 17:01
  UTC). \label{fig_xlcx5}}
\end{figure*}

\begin{figure}
   \centerline{\includegraphics[width=8cm,angle=0]{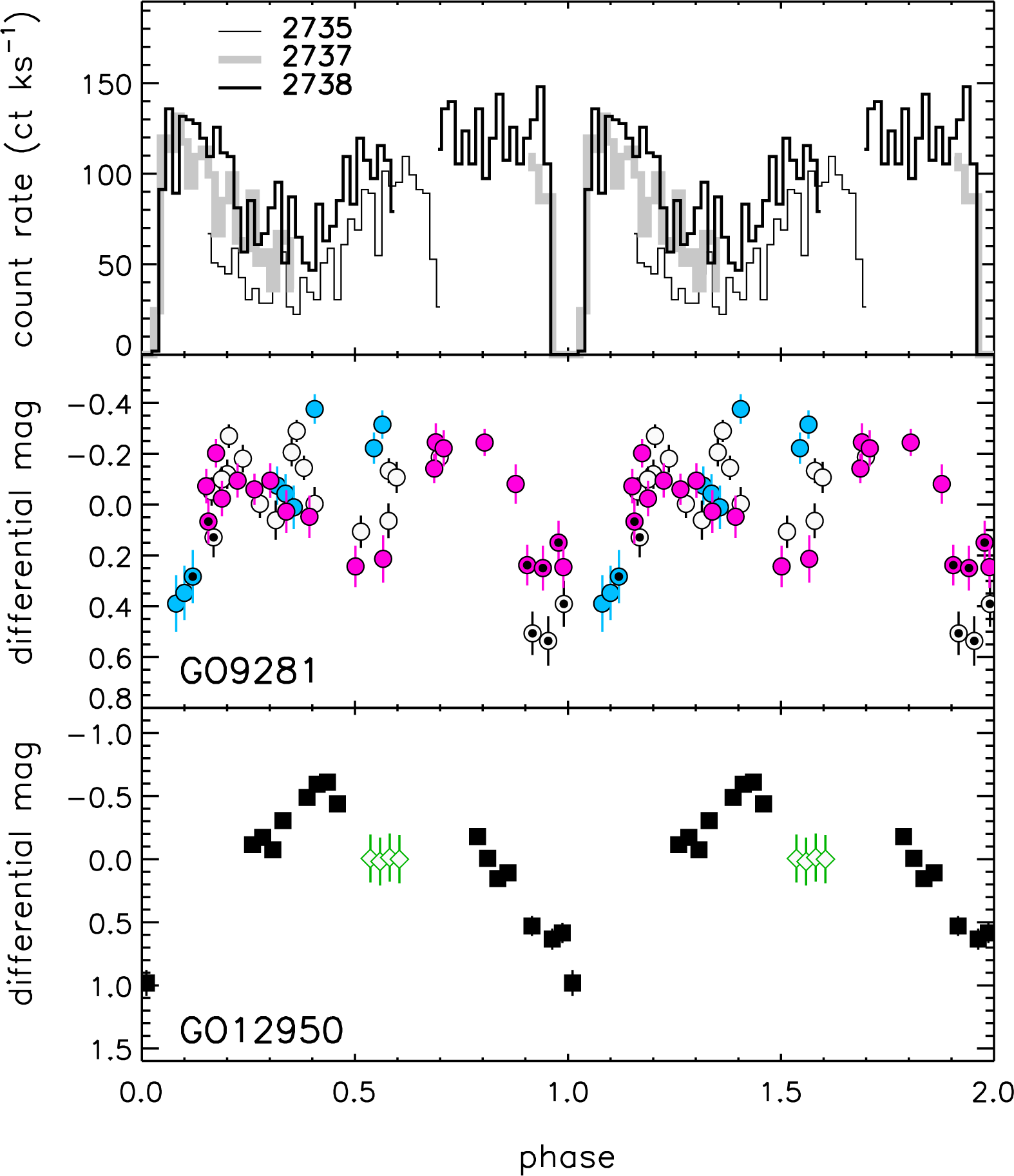}}
\caption{{\em Chandra} and {\em HST} light curves for X5, folded on
  the orbital period of 8.667 hr; phasing was chosen such that the
  middle of the X-ray eclipse is phase=0. {\em Top:} The 0.3--8 keV
  light curves only show the sections that were simultaneous with the
  GO\,9281 observations. The line thickness is different for each
  epoch.  {\em Middle:} GO\,9281 F435W (blue), F625W (pink), and F658N
  (white) differential light curves. Black dots mark the data points
  from visit 2 that were combined to create colours around the time of
  X-ray eclipse. {\em Bottom:} GO\,12950 lightcurves in F300X (filled squares), F390W
  (open green diamonds). \label{fig_color}}
\end{figure}

\begin{figure}
  \centerline{\includegraphics[width=8cm]{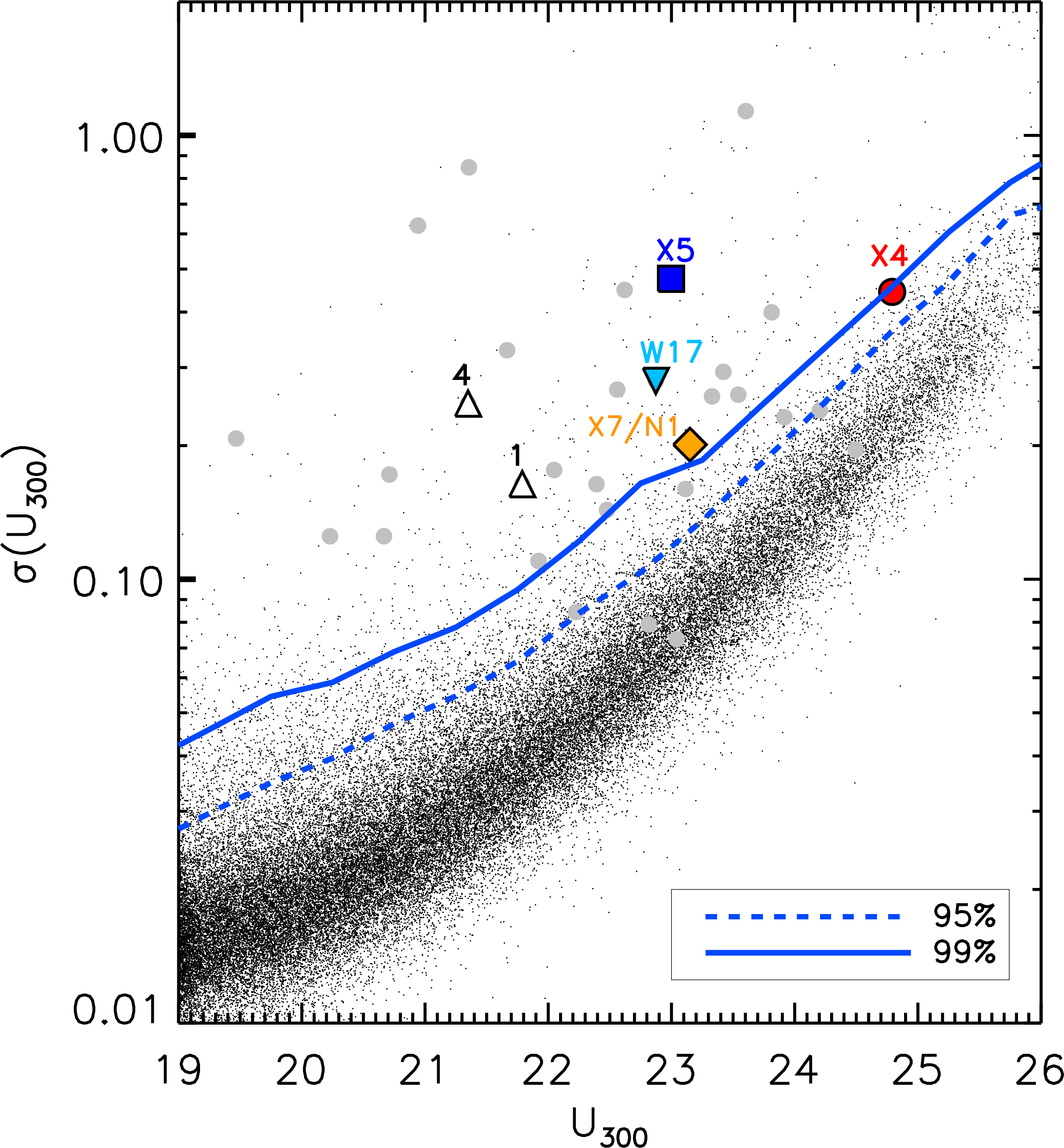}}
  \caption{Rms variability in the GO\,12950 $U_{\rm 300}$ light curves
    versus average $U_{\rm 300}$ magnitude. Outliers were filtered
    using a 3$\sigma$-clipping algorithm. (Likely) counterparts and
    two stars near the {\em Chandra} position of W37 are marked with
    the same symbols as in Fig.~\ref{fig_nuvcmd}. For comparison,
    securely identified CVs from \citep{riveea18} are shown as grey
    filled circles; the CV W36 lies outside the plot ranges, and the
    CV W23 is omitted as it overlaps with X5$_{\rm opt}$. The blue
    dashed lines represent the 95\% and 99\% limits of the percentile
    distribution (computed in bins with a 0.5 mag width) below which
    95\% and 99\%, respectively of the stars are found. Other stars in
    the cluster are represented with black dots.}
\label{fig_nuvrms}
\end{figure}

\begin{figure}
\centerline{\includegraphics[width=8cm]{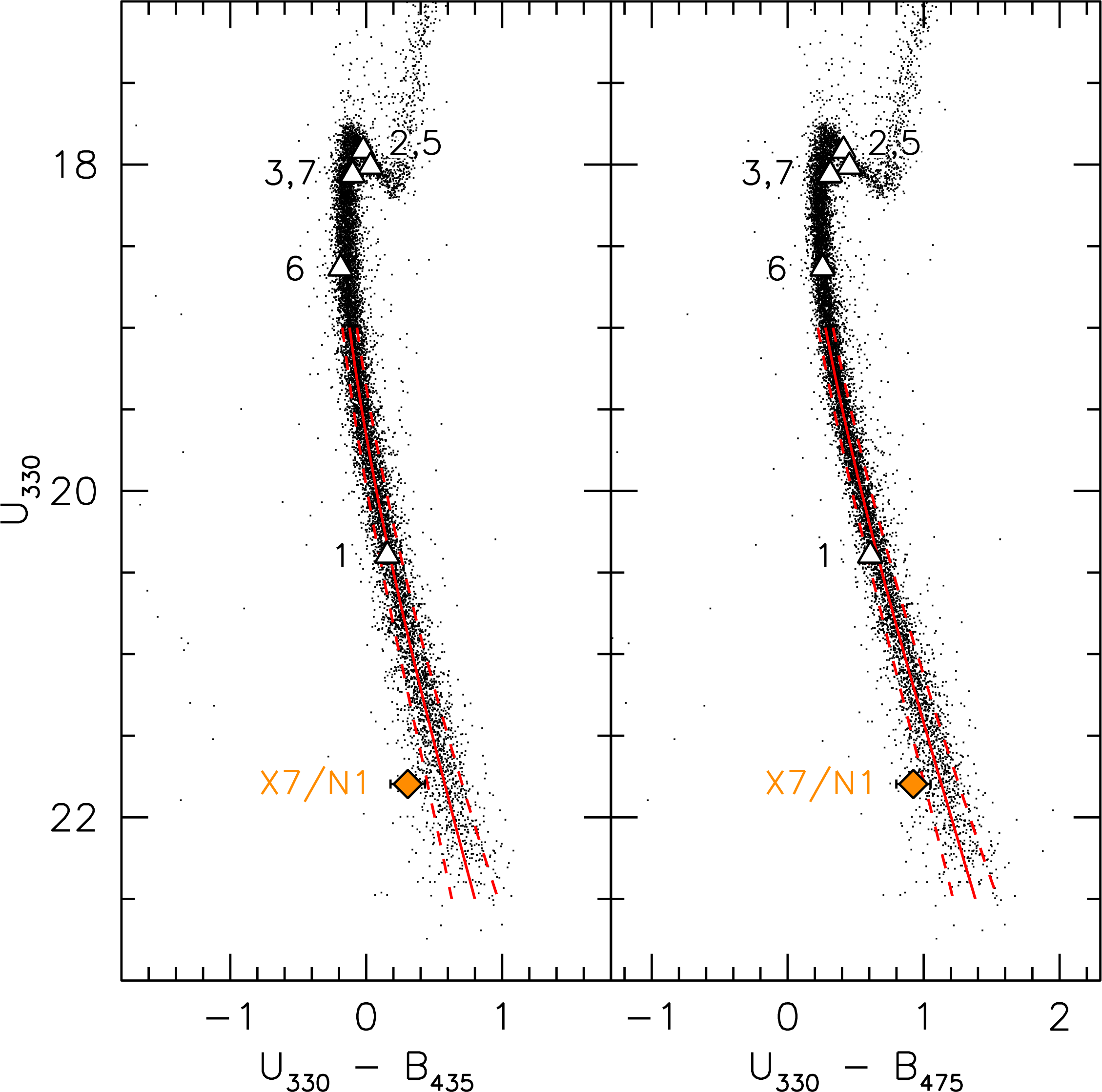}}
  \caption{GO\,9019 $U_{\rm 330}$ versus $U_{\rm 330}-B_{\rm 435}$ and
    $U_{\rm 330}$ versus $U_{\rm 330}-B_{\rm 475}$ colour-magnitude
    diagrams with the likely counterpart to X7 (N1) indicated with an
    orange diamond.  Upward triangles are stars close the {\em
      Chandra} position of W37. Solid red lines represent the mean
    colour as function of magnitude, while the dashed red lines are
    the 1 $\sigma$ deviations about the mean, calculated using an
    iterated 3 $\sigma$ clip.}
  \label{fig_9019cmd}
\end{figure}

\subsection{X5} \label{ssec_x5}

The {\em Chandra} light curve of X5 (or W58) shows eclipses at an
$\sim$8.7 hr period as well as ``dipping'' activity due to rapid
changes in the column density obscuring the regions close to the
neutron star (\citealt{heinea03a}; see top panel of
Fig.~\ref{fig_xlcx5}). These changes are likely caused by
inhomogeneities in the outer parts of an accretion disk seen (almost)
edge on. Using the precise {\em Chandra} position, \cite{edmoea02a}
identified the counterpart (X5$_{\rm opt}$) as a blue object in WFPC2
images that shows variability on long and short time scales (see
Fig.~1 in \citealt{edmoea02a} and our Fig.~\ref{fig_fcs} for a finding
chart). In $U$ (F300W), the short-term variations resemble eclipses
and ellipsoidal variations; no such variability was seen by Edmonds et
al.\,in $V$ (F555W). The average $U$ magnitude was found to change by
$\sim$1 mag between epochs separated by years.  While the dipping in
X-rays, long-term $U$ variability, and blue $U-V$ colour point to the
presence of an accretion disk, \cite{heinea03a} (and later
\citealt{bogdea16}, with even more stringent constraints) found that
any contribution from non-thermal emission to the {\em Chandra}
spectrum is negligible. As suggested by \cite{heinea03a}, gas from the
disk may not currently be reaching the neutron star. Possibly, gas is
building up in the disk, or it could be swept away by a pulsar wind or
propeller effect before arriving at the neutron-star surface.

In the GO\,9281 photometry, X5$_{\rm opt}$ is also blue (although not
as pronounced as in $U-V$) and shows H$\alpha$ excess emission
(Fig.~\ref{fig_optcmd}). The simultaneous X-ray and optical light
curves of X5$_{\rm opt}$ show correlated variability in all three
filters, in the sense that the optical brightness also decreases
around the time of the X-ray eclipses (Fig.~\ref{fig_xlcx5}, middle
and bottom rows). This is illustrated more clearly in the folded light
curves of Fig.~\ref{fig_color}, where we adopted an orbital period of
31200.197(1) s (see Appendix \ref{app_x5period}). The folded X-ray
light curve in the top panel of Fig.~\ref{fig_color} only shows those
time intervals that are simultaneous with the optical data
(middle). Double maxima and minima per orbital phase are seen most
clearly in the $R_{625}$ light curve. These features could be the
combined effect of ellipsoidal variations and eclipses, or a signature
of multiple shocks that give rise to hot spots in the disk (see
e.g. \citealt{konoea17} for the cataclysmic variable V455 And in
quiescence, and \citealt{davaea05} for the neutron-star LMXB Cen X-4
in quiescence). Moreover, as the hot disk should be more extended than
the region from which the X-rays originate, the optical/NUV eclipse
should also last longer. This likely explains why the optical minimum
around phase 0 is broader than the X-ray eclipses, as was also noted
by \cite{edmoea02a} for the $U$ band. Parts of the hot disk could be
obscured by the stream impact point, as well, which would shift the
midpoint of the optical/NUV eclipse to earlier phases. The sampling of
our light curve is not good enough to see this potential effect.

The quasi-instantaneous colours, computed in the same way as for
X4$_{\rm opt}$, show that X5$_{\rm opt}$ moves closer to the main
sequence in H$\alpha_{\rm 658}-R_{\rm 625}$ around phase 0 (defined to
be the centre of the X-ray eclipse; Fig.~\ref{fig_optcmd}). In the
bottom-middle panel of Fig.~\ref{fig_xlcx5}, the data points that were
combined to compute the colours around the X-ray eclipse are marked
with filled dots, which makes them easier to recognise in the folded
light curve of Fig.~\ref{fig_color}, the CMD of Fig.~\ref{fig_optcmd},
and the CCD of Fig.~\ref{fig_optccd}. Towards the end of this optical
dip, around 3.25 days after the start of GO\,9281, the H$\alpha_{\rm
  658}-R_{\rm 625}$ excess is again at the average level for X5$_{\rm
  opt}$ (see the left-most ``dotted'' blue square in the $R_{\rm 625}$
versus H$\alpha_{\rm 658}-R_{\rm 625}$ CMD in
Fig.~\ref{fig_optcmd}). The $B_{\rm 435}-R_{\rm 625}$ colour at this
time also lies closer to the main sequence compared to the average,
but the corresponding individual $B_{\rm 435}$ and $R_{\rm 625}$ data
points lie in a part of the light curve where the brightness may be
changing relatively quickly, which could affect the colour. These
changes suggest that part of the region where the H$\alpha$ (and
perhaps also blue) emission is relatively strong, is obscured together
with the X-rays.  Unlike \cite{edmoea02a}, who noted that X5$_{\rm
  opt}$ shows no signs of eclipses or ellipsoidal variability in $V$,
we do see variations at the redder wavelengths. This suggests
long-term qualitative changes in the light curve. The minima around
phase 0.5 in $R_{\rm 625}$ and H$\alpha_{\rm 658}$ could be
ellipsoidal variations. On the other hand, there is no obvious dip
around that phase in $B_{\rm 435}$; given the poor sampling, we cannot
draw firm conclusions. The light curve looks especially noisy around
phases 0.2--0.4 in H$\alpha_{\rm 658}$ (Fig.~\ref{fig_color}). This
interval overlaps with the part of the X-ray light curve most affected
by dipping.

In the GO\,12950 data, the counterpart to X5 is much bluer relative to
the main sequence than in the optical (Fig.~\ref{fig_nuvcmd}). With
$U_{\rm 300}\approx22.3$ at maximum brightness, the source is about as
bright as in 1997 November and 1999 October, but almost a magnitude
fainter than in 1995 October \citep{edmoea02a}.  The uncertainty in
the orbital period is small enough that we can phase-connect the
X-ray/optical light curves from 2002 with the NUV light curve from
GO\,12950 taken about 11 years later (see Appendix A). In the bottom
panel of Fig.~\ref{fig_color}, we show the phased NUV light curve
taken over a time span of $\sim$6.5 hr. The $U_{\rm 300}$ light curve
looks smooth, and shows a broad dip centred around the X-ray eclipse.
The four $B_{\rm 390}$ data points show a lack of
variability. Unfortunately, in the other four F390W exposures, X5
falls in a chip gap.

In the HUGS CMDs, X5$_{\rm opt}$ is blue except in the reddest
filters, where it lies $\sim$0.35 mag to the red of the main sequence
in $V_{\rm 606} - I_{\rm 814}$. This excess may be explained by
periodic changes in the light curve on a time scale of hours affecting
the colours, or intrinsically anomalous colours of the donor.

\subsection{X7} \label{ssec_x7}

X7 (or W46) is one of the best-studied X-ray sources in 47\,Tuc. A
neutron-star atmosphere model gives an excellent fit to the {\em
  Chandra} spectrum. The lack of a harder power-law spectral
component, and the high level of stability in X-rays over the years,
suggest there is no accretion occurring in the system. This makes X7
an attractive source for deriving constraints on the neutron-star
equation of state \citep{heinea06,bogdea16}. Some systematic
uncertainties in X7's radius and mass measurements remain, because the
composition of the neutron-star atmosphere is unknown. Although an
H-rich atmosphere is deemed more likely than a He-rich atmosphere (see
the discussion in \cite{bogdea16} and \cite{steiea18}), the nature of
the optical or NUV counterpart would shed more light on this
issue. However, no counterpart has been identified
yet. \cite{edmoea02a} found no variable or blue source in the X-ray
error circle.

We examined the colours of the three nearby stars that were also
considered by \cite{edmoea02a}, N1 to N3 (see Fig.~\ref{fig_fcs} top
right for a finding chart). Like these authors, we find that N2 and N3
have normal main-sequence colours in the GO\,9281 and GO\,12950
images, so we agree that these stars are not viable counterparts;
moreover, their angular separations from X7's {\em Chandra} position
are relatively large. At an offset of $\sim$0.07\arcsec, N1 is an
excellent astrometric match to X7. However, its $U-V$ colour resembles
that of a normal low-mass main-sequence star, and in the high-cadence
GO\,8267 WFPC2 time series spanning 8.3 days N1 shows no variability
(see Fig.~3d in \citealt{edmoea02a}). We find that the $B_{\rm
  435}-R_{\rm 625}$ colour of N1 is also unremarkable, and consistent
with that of a $\sim$0.62 $M_{\odot}$ main-sequence star (based on
comparison with the isochrones from \citealt{bresea12}). On the other
hand, its H$\alpha_{\rm 658}-R_{\rm 625}$ colour puts it slightly to
the H$\alpha$-excess side of the main sequence (see
Fig.~\ref{fig_optcmd}). Interestingly, in the $B_{\rm 390}$ versus
$U_{\rm 300}-B_{\rm 390}$ CMD, N1 lies a little bit to the blue of the
main sequence (Fig.~\ref{fig_nuvcmd}). This modest blue excess is also
found independently in the KS2 photometry of the same data, where the
apparent deviation from the main sequence is in fact a bit
larger. (Below, we quantify N1's blue excess in more detail for the
GO\,9019 CMDs, where it is found in a similar location with respect to
the main sequence). Its $U_{\rm 300}$ light curve
(Fig.~\ref{fig_nuvlc}, top) shows scatter at a level that is higher
than $\gtrsim$99\% of stars with a similar magnitude
(Fig.~\ref{fig_nuvrms}). We also identify a faint star at an offset of
$\sim$0.15\arcsec~from X7 that is visible (but unnamed) in the $V$
finding chart in \cite{edmoea02a}, as well. We name this star N4 and
mark it in the finding chart of Fig.~\ref{fig_fcs}. N4 has a small
H$\alpha$ excess that is more pronounced than that of N1, but lies on
the main sequence in the $B_{\rm 435}-R_{\rm 625}$ CMD. From the NUV
images, we find that $B_{\rm 390}$=24.18(4) but since this star is
undetected in $U_{\rm 300}$, we do not know its NUV colour. However,
the limit on $U_{\rm 300}$ suggests that $U_{\rm 300}-B_{\rm
  390}$$\gtrsim$1.8.  There are two more stars inside the error circle
(N5 and N6); both lie on the main sequence, and therefore we do not
consider them plausible counterparts. These findings spurred us to
focus on N1 and N4, and look for coverage in other {\em HST} images.

In Fig.~\ref{fig_hugscmd}, we mark N1 and N4 in the HUGS CMD with the
orange and white diamond, respectively. N1 lies well to the blue of
the main sequence in the $U_{\rm 275}$ versus $U_{\rm 275}-U_{\rm
  336}$ CMD, whereas the location of N4 is poorly constrained due to
the large errors in $U_{\rm 275}$. In the redder CMDs, both stars lie
on, or close to, the main sequence. Taking the HUGS colours at face
value, we conclude that N1 is the more likely counterpart given its
blue NUV colour and smaller offset from X7.  We caution that the value
of the ``sharp'' parameter, which indicates how well a star's profile
matches that of the PSF, is outside the optimal range in all HUGS
measurements of N1 and N4. This may be due to the crowding near X7 and
could indicate that their photometry is affected.

We examined N1 and N4 in the GO\,9019 HRC data as well. N4 is not
detected in the GO\,9019 F220W and F330W images, only in the F435W and
F475W images.  N1 is too faint to be detected in F220W. The average
$U_{\rm 330}-B_{\rm 435}$ and $U_{\rm 330}-B_{\rm 475}$ colours put N1
to the blue of the main sequence. To quantify this blue offset, we
first computed the width of the main sequence at the magnitude of N1
in the CMDs of Fig.~\ref{fig_9019cmd}. The solid red lines in the
panels represent the mean colour as a function of magnitude, while the
dashed red lines are the 1 $\sigma$ deviations about the mean,
calculated using an iterated 3 $\sigma$ clip. The curves were smoothed
by fitting third-order polynomials. We took half the separation
between the dashed lines as the 1 $\sigma$ uncertainty on the colour
index, which for N1 is plotted as the error bar on the orange diamond
in each panel. N1 lies more than 1 $\sigma$ to the blue side of the
main sequence in both CMDs, especially in the $U_{\rm 330}-B_{\rm
  435}$ CMD, suggesting a blue (NUV) excess.

We also considered indications for variability in the individual
GO\,9019 images. For $B_{\rm 435}$ and $B_{\rm 475}$, the DAOPHOT
errors are $\sim$0.08--0.1 mag, consistent with the scatter in the
magnitudes from the individual images (three in each filter). In
$U_{\rm 330}$, N1 shows a 0.35 mag variation between the three
images. Going from the first to the third exposure, N1 varies from
21.71(25) mag, to 21.36(18) mag (4 min later) and finally to 21.57(20)
mag (19 min after the first image). For all three filters, we compared
the standard deviation in the N1 data points with those of other stars
in the images, and found that N1 does not stand out. We conclude that
N1 is not significantly variable in the GO\,9019 images that we
analyzed. Comparison between the HUGS magnitude for N1 in F336W from
2010 Sep ($U_{\rm 336}$=22.03(07)) with the value reported by
\cite{edmoea02a} from 1999 Jul ($U_{\rm 336}$=21.97 with an absolute
error $\sim$0.2 mag) also shows good agreement.

In conclusion, we consider it very likely that N1 is the true
optical/NUV counterpart of X7. The variability in F300X
(Fig.~\ref{fig_nuvrms}), the blue colours in the NUV filters and the
weak H$\alpha$ excess emission suggest the presence of an accretion
disk. However, in X-rays X7 has not shown any signs of ongoing
accretion. Possibly, accretion from the secondary feeds the disk but
either the gas has not reached the neutron star just yet, or is
expelled from the system by a pulsar or propeller wind. Alternatively,
low-level accretion {\em is} taking place and generates faint X-ray
emission of the order $\sim$10$^{31}$ erg s$^{-1}$, but the
corresponding weak power-law component in the X-ray spectrum remains
hidden under the $\sim$10$^{33}$ erg s$^{-1}$ soft X-ray emission.
Another implication of N1 being the counterpart, is that the
atmosphere of the neutron star in X7 would be H-rich as N1 looks like
a main-sequence star with a small H$\alpha$ excess. This is consistent
with the conclusions of \cite{bogdea16} and \cite{steiea18} from X-ray
spectral fitting.

\subsection{W17}

W17 is among the new sources that were discovered in the {\em Chandra}
observations of 2000. \cite{edmogillea03b} identified this source as a
candidate qLMXB based on its relatively soft X-ray colours, and the
high X-ray--to--optical flux ratio ($\gtrsim$0.9) that is inferred
from its X-ray luminosity ($L_X\approx 2 \times 10^{31}$ erg s$^{-1}$
(0.5--6 keV) for the distance adopted here) and the high upper limit
to the optical magnitude ($U\gtrsim24$). The proximity of a very
bright, saturated star at only $\sim$0.62\arcsec~to the south-west
hampered their search for an optical counterpart in the deep GO\,8267
WFPC2 $V$ and $I$ data.  Fits to the 2000 and 2002 {\em Chandra}
spectra led \cite{heinea05a} to the conclusion that W17 is indeed a
qLMXB: its spectrum is dominated by a $\Gamma \approx 1.9$ power-law
component, but also shows a distinct soft component that can be fitted
with a neutron-star atmosphere model.  W17 does not show X-ray
variability on short time scales (days to weeks) or between the 2000
and 2002 observations.

In our GO\,9281 data, the light of the nearby bright star also
inhibits a sensitive search for a counterpart, so we have no
simultaneous X-ray/optical measurements of this source. We estimate an
upper limit of $B_{435}\gtrsim24$, with an uncertainty of $\sim$0.5
mag given that the light of the saturated star makes it difficult to
account for the local background. The situation is much better in the
NUV. In the GO\,12950 images we readily found a very blue star near
the centre of the {\em Chandra} error circle (see Fig.~\ref{fig_fcs}).
This star is highly variable with a full amplitude of $\Delta U_{{\rm
    300}}$$\gtrsim$ 1 mag in the $\sim$8 hr time span of the GO\,12950
observations (Fig.~\ref{fig_nuvlc}). Its time-averaged $U_{\rm
  300}-B_{\rm 390}$ colour puts it far to the blue of the main
sequence, in between the white dwarfs and the SMC sequence
(Fig.~\ref{fig_nuvcmd}). The HUGS $U_{\rm 275}-B_{\rm 336}$ colour
puts it even closer to the white-dwarf sequence. In the F435W, F606W,
F814W filters, we do not consider the photometry reliable given the
proximity of the bright star.

The GO\,12950 NUV colours of the counterpart are bluer than those of
X4$_{\rm opt}$ and X5$_{\rm opt}$. A larger contribution from an
accretion disk could be the reason for this. However, the X-ray
luminosity of the power-law component in W17 is about the same as the
average power-law luminosity in X4, suggesting that on average the
level of accretion in these two systems is similar. The orbital period
would shed light on the nature of the companion. The shape of the
GO\,12950 light curves show a hint of repetition at a time scale of
$\sim$3.5--4 hr but do not display significant periodicity. If the
secondary is a white dwarf rather than a main-sequence star, it could
contribute to the blue colours as well.

\subsection{W37} \label{ssec_w37}

W37 is also one of the sources discovered in the 2000 {\em Chandra}
observation.  Its spectrum is soft and can be fitted with a
neutron-star atmosphere model. W37's X-ray light curve shows eclipses
(at a period of 3.087 hr) as well as drastic changes in the count rate
due to variable obscuration. During the eclipses, a small amount of
residual soft emission of unknown origin was seen by
\cite{heinea05a}. W37 has no known optical or NUV counterpart. No
nearby blue or variable object is seen in the {\em HST} data analyzed
by \cite{edmogillea03a}, although the presence of two bright
main-sequence stars close to the X-ray position reduces the
sensitivity to finding faint counterparts. If the companion is a
main-sequence star, the orbital period indicates a secondary mass of
$\lesssim$0.34 $M_{\odot}$.

In the GO\,9281 and GO\,12950 images we find seven optical/NUV sources
inside the 95\% error circle, marked 1 to 7 in the panel for W37 in
Fig.~\ref{fig_fcs}. Five of these (2, 3, 5--7) look like regular
main-sequence or main-sequence turnoff stars, with some lying slightly
offset to the red side of the main sequence in $B_{\rm 435}-R_{\rm
  625}$. We do not consider these stars plausible counterparts. Star 1
lies close to the main sequence in the $R_{\rm 625}$ versus
H$\alpha_{\rm 658}-R_{625}$ CMD (near the expected location of a
$\sim$0.7 $M_{\odot}$ star) and the $U_{336}$ versus $U_{275}-U_{336}$
and $I_{814}$ versus $V_{606}-I_{814}$ CMDs, but in the $R_{\rm 625}$
versus $B_{\rm 435}-R_{\rm 625}$, $B_{\rm 390}$ versus $U_{\rm
  300}-B_{\rm 390}$, and $B_{\rm 435}$ versus $U_{\rm 336}-B_{\rm
  435}$ CMDs of Figs.~\ref{fig_optcmd}, \ref{fig_nuvcmd} and
\ref{fig_hugscmd}, respectively, it appears to be slightly offset to
the blue. Given the estimated upper mass limit of a main-sequence
donor (see previous paragraph), it is not likely that star 1 is the
true counterpart. It could be blended with an undetected fainter and
bluer object that draws its photometry to the blue side of the main
sequence, or the PSF wings of the nearby star 3 may not be properly
accounted for. The photometry for star 4 is ambiguous. In the optical,
it is only convincingly detected in the F435W images, at $B_{\rm
  435}\approx20.1$. Forcing DAOPHOT to extract $R_{\rm 625}$ and
H$\alpha_{\rm 658}$ magnitudes from the position of this F435W
detection, does not produce credible results. In the NUV, the
photometry of star 4 puts it either to the red (DAOPHOT) or the blue
(KS2, Dolphot) of the main sequence. Given the uncertain photometry,
we have omitted this star from Fig.~\ref{fig_nuvcmd}. Perhaps the only
thing that can be concluded is that, at the time of the GO\,9281
observations, this object was blue for it to be detected in the F435W,
but not in the F625W and F658N images. Stars 1 and 4 are highly
variable in the $U_{\rm 300}$ images of the GO\,12950 data set; both
lie above the 99\% percentile limit in Fig.~\ref{fig_nuvrms}. Star 1
was clearly detected in the NUV photometry and the variability seems
to be intrinsic, while star 4 seems to be closer to star 3 and
variability could be due to contamination from that star. In the HUGS
catalog and the GO\,9019 images, star 4 is not detected.

\begin{figure}
   \centerline{\includegraphics[width=8.5cm]{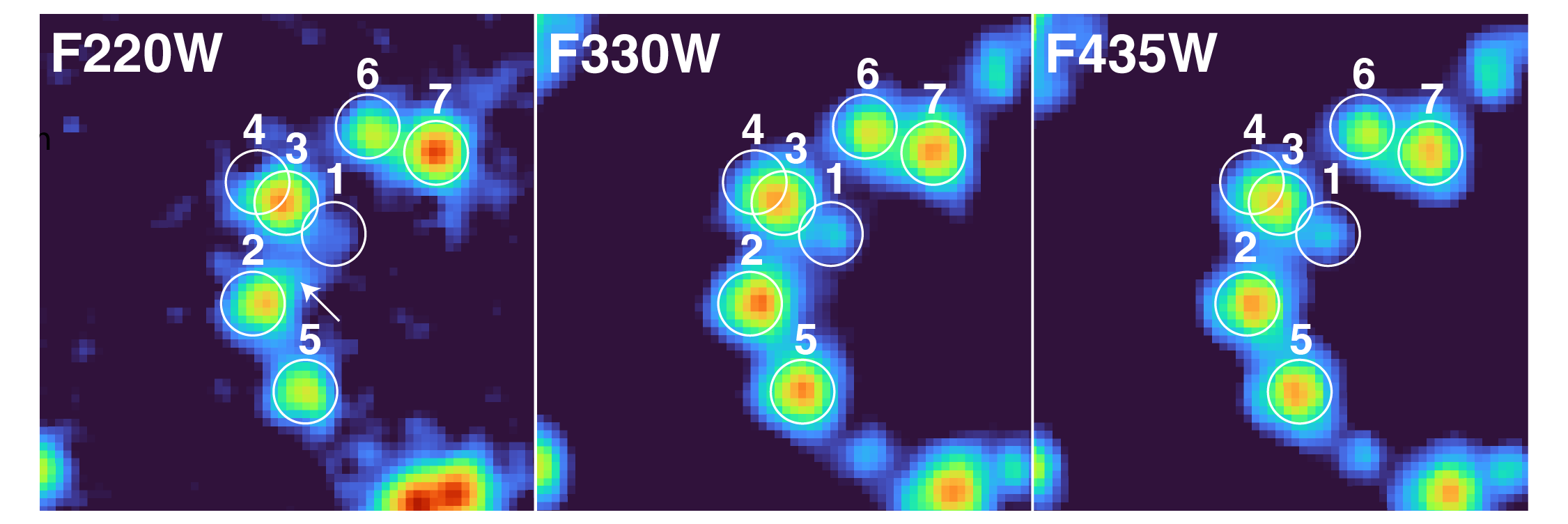}}
\caption{Three-panel GO\,9019 ACS HRC finding charts for a
  $0.8\arcsec\times0.8$\arcsec\,region centred on the position of
  W37. North is up and east is to the left. The excess flux to the
  north-west of star 2 in the F220 image is indicated with a white
  arrow. The frames for each filter have been drizzle-combined with
  $2\,\times$ oversampling and are rectified to the \emph{Gaia}
  reference frame. The drizzled pixel size is 0.0125\arcsec\,on a
  side. The stars are labelled as in the W37 panel of
  Fig.~\ref{fig_fcs}.}
\label{fig_w37}
\end{figure}

Examination of the GO\,9019 F220W image suggests the presence of
excess flux at a position about 0.06\arcsec\,to the north-west of star
2 and about 0.04\arcsec\,from the location of W37. This excess flux is
not apparent at longer wavelengths, which is consistent with the
presence of a blue star. This is illustrated in Fig.~\ref{fig_w37},
which shows the region around W37 in the F220W, F330W, and F435W
filters. These drizzled ACS HRC frames have been aligned with the HUGS
coordinate system, which itself is aligned with the \emph{Gaia}
reference system. While the extension of the image of star 2 in F220W
is suggestive of the presence of a blue star near the position of W37,
it is by no means definitive. Deeper high-resolution imaging is
necessary to further investigate this issue.

In order to identify the counterpart of W37 through optical
variability on the known orbital period, we have performed a series of
light curve analyses on the GO\,8267 WFPC2 data. The F555W and F814W
light curves of the four stars closest to the centre of the X-ray
error circle (stars 1 to 4) were kindly provided to us by
R.\,Gilliland. The light curves contain 636 (F555W) and 653 (F814W)
data points spread over 8.3 days in 1999 Jul. We refer to
\cite{gillea00} and \cite{albrea01} for more details about the data
and light curve extraction.  First, we phase-folded the $V_{\rm 555}$
and $I_{\rm 814}$ light curves on the known period from the X-ray
eclipses, but no periodicity was observed in either of the two
bands. To reduce the scatter in these light curves, we averaged the
measurements in phase bins of 0.001 and 0.003 days, well below the
duration of the X-ray eclipse. Still, no periodicity was observed for
any of the four stars.  Another attempt was carried out by using a
sigma-clipping approach on each of the previously defined phase bins.
We removed all photometric measurements that were smaller or larger
than $m \pm 3 \sigma$, where $m$ is the median value in each bin and
$\sigma$ is the standard deviation. This approach was taken to remove
outliers, but did not reveal any periodicity.  Finally, we performed a
Lomb-Scargle \citep{lomb76,scar} analysis to look for the X-ray period
in the optical data. The analysis was performed separately in both
filters. In multiple cases a period corresponding to the {\em HST}
orbit was observed ($\sim$96.5 min, so very close to half W37's
orbital period, or 92.6 min), but none consistent with the X-ray
period of W37 or half of it.

In conclusion, we have not securely identified the counterpart to
W37. To compute limits on the X-ray--to--optical flux ratio (see
Table~\ref{tab_xflux}) we use the magnitudes of star 4, which is the
faintest of the stars near W37.

\section{Discussion} \label{sec_dis}

\begin{figure}
  \centerline{\includegraphics[width=8cm]{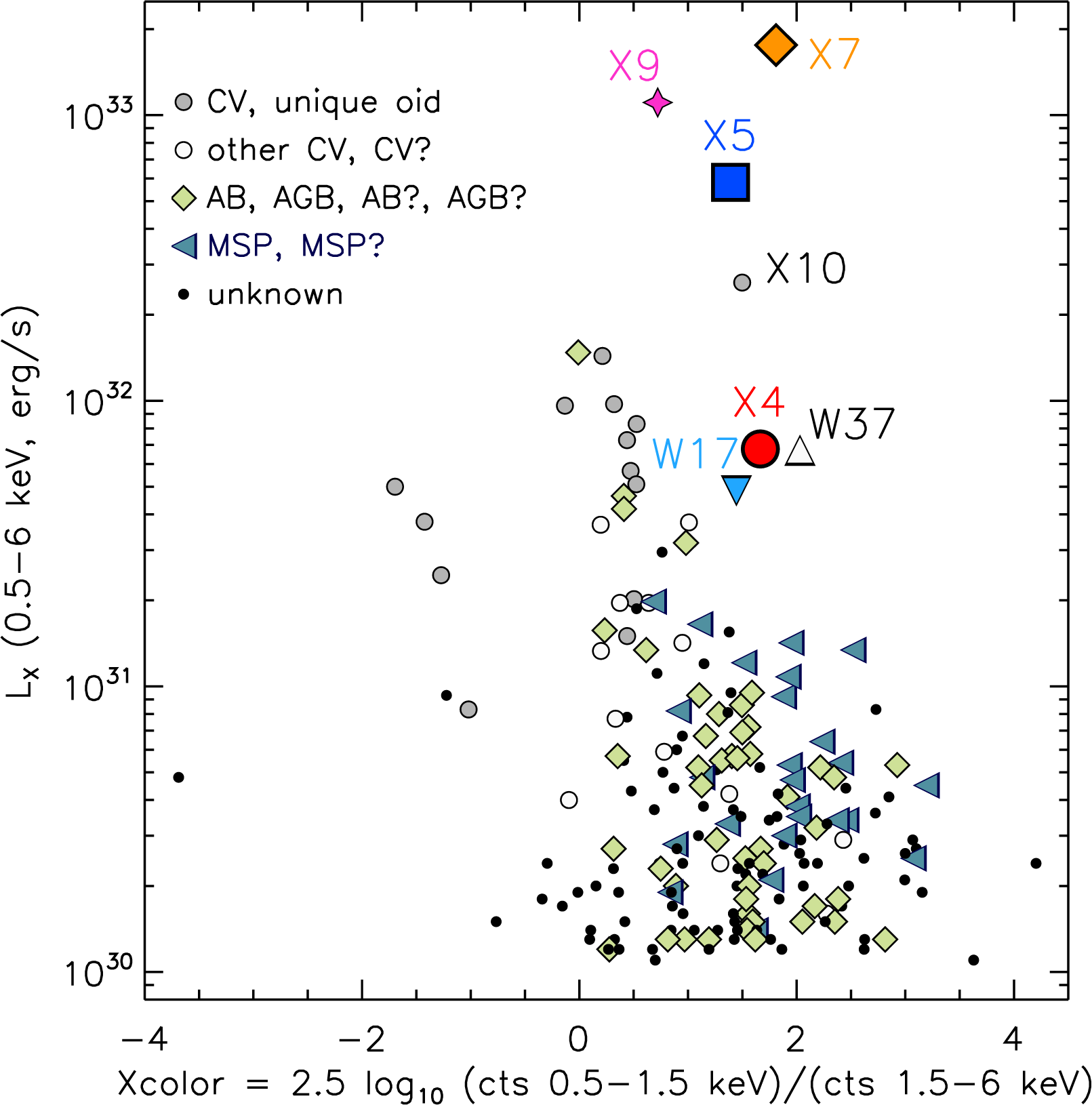}}
\caption{X-ray luminosity versus (0.5--1.5 keV)/(1.5--6 keV) X-ray
  colour for 47\,Tuc sources in the 2002 {\em Chandra} observations
  \citep{heinea05b}. Only sources with more than 20 counts (0.5--6
  keV) are included. The qLMXBs and the CV X10 are labelled. Other
  source classes are plotted with symbols according to the legend in
  the top left. Classifications were taken from \citet{heinea05b}
  unless the classification was revised by more recent studies. We
  distinguish between the securely identified CVs with unique optical
  counterparts (\citealt{riveea18}, filled grey circles) and other CVs
  and CV candidates (open circles).}
\label{fig_xcmd}
\end{figure}

\begin{figure}
\vspace{1cm}
\centerline{\includegraphics[width=8.cm]{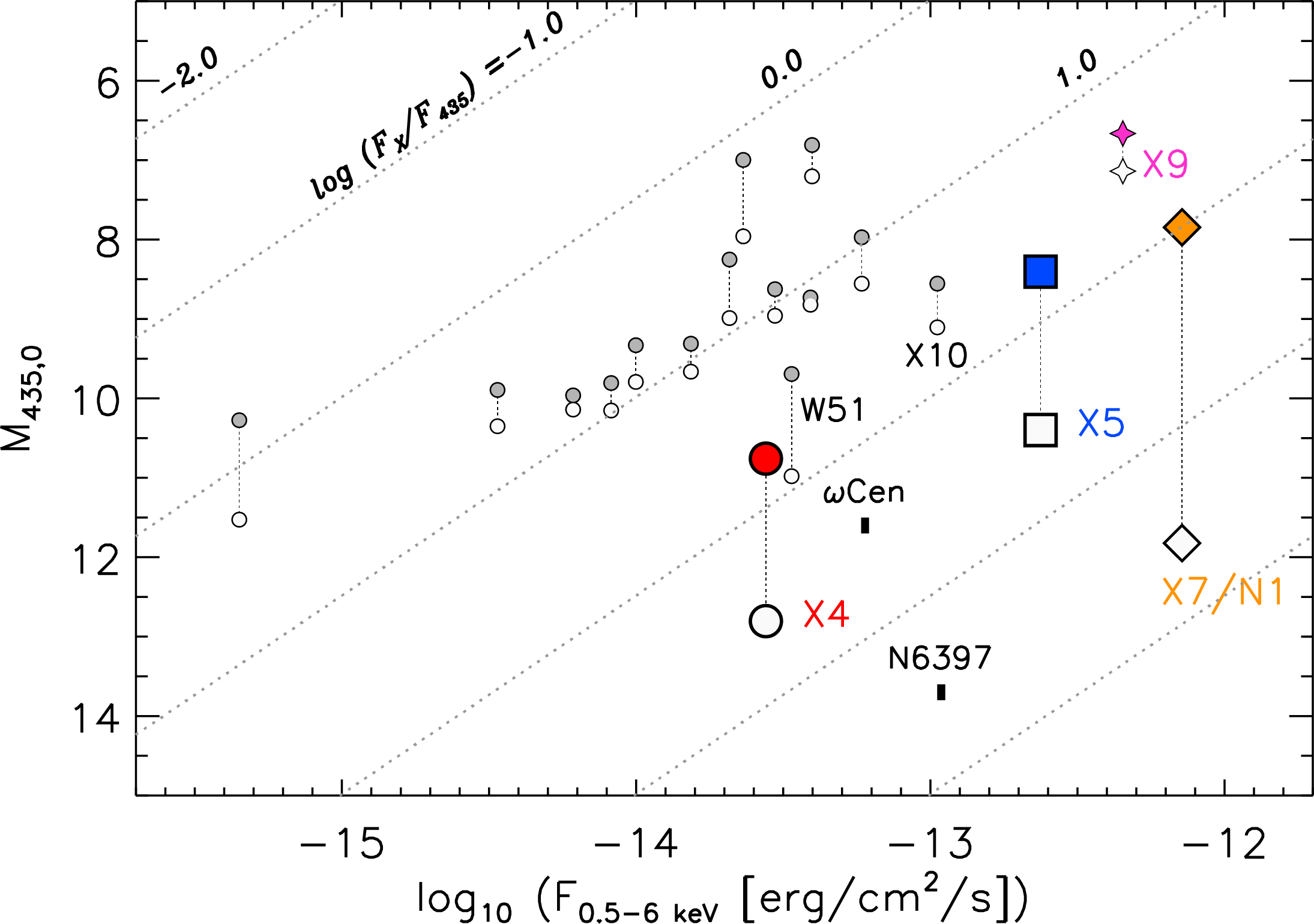}}
\caption{Filled symbols show the absolute $B_{\rm 435}$ magnitudes of
  the qLMXBs (coloured) and CVs (grey) versus the X-ray fluxes (0.5--6
  keV, from Table~2 in \citet{heinea05b}) from the simultaneous 2002
  {\em HST} and {\em Chandra} data. Open circles mark the lower limit
  on the excess blue fluxes (see text). The vertical dotted lines mark
  the range where the actual contribution from the accretion flow to
  the blue light is expected. The diagonal dotted lines mark lines of
  constant X-ray to $B_{\rm 435}$ flux ratio. For comparison, we also
  show the two other globular-cluster qLMXBs with optical counterparts
  and X9, the ultra-compact qLMXB in 47\,Tuc.}
\label{fig_fxfopt}
\end{figure}

Our search for optical/NUV counterparts to the qLMXBs in 47\,Tuc has
yielded the discovery of a blue and variable counterpart to W17. We
also find compelling indications that a known astrometric match to X7
is indeed a likely counterpart. Of the five (likely) neutron-star
qLMXBs in the cluster, only one (W37) remains unidentified as the
proximity of several stars near the centre of the X-ray error circle
complicates a sensitive search for a counterpart. Optical counterparts
have been identified for only two other globular-cluster
qLMXBs\footnote{and for a handful of globular-cluster transient LMXBs
  in quiescence, including IGR\,J18245$-$2452 in M\,28
  \citep{pallea13}; Terzan\,5 X-1 and Terzan\,5 X-2
  \citep{ferrea15,testea12}; and SAX J1748.9$-$2021 in NGC\,6440
  \citep{verbea00,intzea01}}: the source in $\omega$\,Cen (discovered
by \cite{haggea04}, later named 44e by \cite{haggard09}) and U24 in
NGC\,6397 \citep{heinea14}. With $R_{\rm 625}$=25.2 and $R_{\rm
  625}$=26.2, respectively, these systems are much fainter than the
47\,Tuc qLMXBs (with $R_{\rm 625}$ between 19.8 and 22.1, see
Table~\ref{tab_avgphot}), making them unsuitable for detailed
follow-up studies. The $\omega$\,Cen and NGC\,6397 systems are also
intrinsically much fainter in the optical ($M_{\rm 625}$=$11.6$ and
$M_{\rm 625}$=13.7, respectively) than the 47\,Tuc neutron-star qLMXBs
($M_{\rm 625}$ between $\sim$6.5 and $\sim$8.8, see
Fig.~\ref{fig_optcmd}). It is unlikely that similar
systems\footnote{with X-ray spectra dominated by a thermal component;
  note that a number of qLMXBs show a dominant non-thermal component
  \citep{Campana02,Jonker04,Wijnands05}; such systems could be
  confused with CVs even in 47 Tuc \citep[cf.][]{millea15}.} are also
present in 47\,Tuc but have been missed; while in the optical they lie
below the sensitivity of the {\em HST} data, they would have been
easily detected by {\em Chandra}. For uncrowded sources, the
\cite{heinea05b} 47\,Tuc source catalogue is complete down to
$L_X\approx8\times10^{29}$ erg s$^{-1}$, whereas
$L_X\approx2\times10^{32}$ erg s$^{-1}$ for 44e \citep{henlea18} and
$L_X\approx8\times10^{31}$ erg s$^{-1}$ for U24 \citep{bogdea10}).

In the optical, the counterparts of X4 and X5 and the likely
counterpart to X7 lie relatively close to the main sequence and show
H$\alpha$ excess emission. This implies that the mass donors in these
systems are non-degenerate, H-rich main-sequence stars. For X5, the
non-degenerate nature of the secondary already followed directly from
the known orbital period and eclipse light curve. For X7, the
properties of the suggested counterpart are consistent with earlier
conclusions from fitting X7's X-ray spectrum.  Unfortunately, the
counterpart of W17 is not detected in H$\alpha$ and the nature of the
donor remains unknown. Based on the large value of the lower limit on
the X-ray--to--optical flux ratio, \cite{edmogillea03b} identified W17
as a likely qLMXB. Our limit on the flux ratio from the simultaneous
data (see Table~\ref{tab_xflux} and discussion below) confirms this
view. While we still cannot say with certainty whether X4 is a qLMXB
or not, the properties of the counterpart to X4 (modest blue optical
colour like X5 and X7, high flux ratio) make the case for X4 being a
qLMXB stronger.

In a diagram of X-ray luminosity versus X-ray colours, X5 and X7 stand
out as soft luminous sources in the cluster (see Fig.~\ref{fig_xcmd},
adapted from \cite{heinea05b}). The more actively accreting
ultra-compact LMXB X9 is comparably bright, but has a harder
spectrum. Classification based on X-ray luminosity and colours alone
becomes more challenging for the less luminous sources X4, W17 and W37
(which could be classified as a qLMXB thanks to its X-ray light curve)
because members of other source classes can reach similar
luminosities. A case in point is X10 (also known as V3 or W27), which
is also a bright soft source.  X10 was classified as a probable
magnetic CV based on its X-ray spectrum and periodic (4.67 hr) X-ray
light curve \citep{heinea05b,baoea23} that are both typical for its
source class.

Taking into account the properties of the optical or NUV counterparts
helps to tell source classes apart. The known 47\,Tuc CVs and qLMXBs
span similar ranges in absolute magnitude (see Figs.~\ref{fig_optcmd}
and \ref{fig_nuvcmd}) and excess H$\alpha$ emission (see
Fig.~\ref{fig_optccd}). The CVs, however, have counterparts that tend
to be bluer than the qLMXB counterparts. In particular in the optical
$R_{\rm 625}$ versus $B_{\rm 435}-R_{\rm 625}$ CMD, the counterparts
of X4, X5 and X7 are not far offset from the blue of the main
sequence. In other words, the relative contribution of an accretion
disk or stream (and in addition, for some CVs, the emission of the
white dwarf itself) to the combined light is more pronounced for CVs.

An often-used criterion for source classification is the ratio of the
X-ray flux to optical flux, which is highest for qLMXBs, followed in
order by CVs and ABs \citep{verbea08}. This can be understood based on
the different origins of the X-rays: emission from the heated surface
of the compact object and accretion in qLMXBs and CVs (and from a
shock region in magnetic systems) versus coronal activity of the
rapidly rotating stars in a close binary of main-sequence and/or
sub-giant stars. Neutron stars and black holes have the deepest
potential wells and give rise to the highest accretion-powered X-ray
luminosities. \cite{edmoea02b} demonstrated the diagnostic power of
the X-ray--to--optical flux ratio for the classified {\em Chandra}
sources in 47\,Tuc. Since our 2002 {\em Chandra} and {\em HST}
GO\,9281 observations were taken simultaneously, the resulting flux
ratios address concerns stemming from the X-ray and optical data being
taken at different epochs. For example, \cite{edmoea02b} found that X5
and X10 have similar flux ratios but cautioned that since X10 is
highly variable both in X-rays and in the optical, its actual $F_{\rm
  X}$/$F_{\rm opt}$ value could be lower. The filled symbols in
Fig.~\ref{fig_fxfopt} show the X-ray fluxes from the 2002 {\em
  Chandra} observation against the GO\,9281 F435W fluxes $F_{\rm 435}$
(converted to unabsorbed absolute magnitudes)\footnote{Note that in
this figure, X-ray fluxes in the 0.5--6 keV band from the 2002 {\em
  Chandra} data are plotted (based on Table~2 in
\citealt{heinea05b}). The corresponding X-ray--to--optical flux ratios
may be different from those in Table~\ref{tab_xflux}, which were
computed for specific (partial) ObsIDs or for a different epoch.}. We
include the securely identified CVs from \cite{riveea18} but omitted
W23 and W299 that have other candidate counterparts in the error
circle, turning their reported X-ray luminosities \citep{heinea05b}
into upper limits. We find that at $F_{\rm X}$/$F_{\rm
  435}\approx1.4$, the flux ratio for X10 is somewhat lower than that
of X5 ($\sim$1.7), but still not far off, even for the simultaneous
fluxes. The high flux ratio for X10 can be explained by its nature as
a magnetic CV: due to the high magnetic field of the white dwarf,
there is no accretion disk in the system to contribute to the blue
optical flux.  Furthermore, we find that W51 is also relatively bright
in X-rays compared to its optical flux.  This may be due to the large
flare produced by W51 in the 2002 {\it Chandra} observations
\citep[][Fig. 6]{heinea05b}, which substantially increased its average
flux; in the 2000 {\it Chandra} observations, its average flux was 5.7
times lower (which would bring W51 more in line with other CVs in 47
Tuc). We do not understand the nature of this flare.

In order to try to further separate the qLMXBs and CVs, we make use of
the finding that in the optical the qLMXB counterparts tend to lie
closer to the main sequence than the CV counterparts. We turn this
offset into an {\em excess} blue flux $F_{\rm 435,exc}$ in the
following way. We assume that all the light in the F625W band is
contributed by a main-sequence secondary. In an isochrone (age 11 Gyr,
Z=0.004) generated with the PARSEC code \citep{bresea12}\footnote{see
  also \url{http://stev.oapd.inaf.it/cgi-bin/cmd}}, we look up the
corresponding $B_{\rm 435}$ magnitude for such a star. The excess blue
flux $F_{\rm 435,exc}$ is found by the difference between the observed
$B_{\rm 435}$ magnitude of the counterpart and the (theoretical)
secondary-star flux. This is likely a lower limit to the excess blue
flux, as the accretion flow may also contribute light in the redder
filters. But it does give us an upper limit on $F_{\rm X}$/$F_{\rm
  435,exc}$ and combined with the value of $F_{\rm X}$/$F_{\rm 435}$
this delineates the range inside which the true value of $F_{\rm
  X}$/$F_{\rm blue,exc}$ should lie. As can be seen in
Fig.~\ref{fig_fxfopt}, the distinction between the CVs and qLMXBs
becomes more pronounced when considering the excess blue flux.

Using the excess blue fluxes versus X-ray fluxes, we see three groups
in Fig.~\ref{fig_fxfopt}. At log$F_{\rm X}$/$F_{\rm 435,exc} > 2.0$,
only qLMXBs are seen. Between 1.5 $<$log$F_{\rm X}$/$F_{\rm 435,exc}
<2.0$, we see a qLMXB without NS surface emission (X9, likely a black
hole qLMXB) and a CV without an optically bright accretion disk (X10,
a polar CV). We also find W51 here, which is classified as a CV, but
has a higher $F_{\rm X}$/$F_{\rm 435,exc}$ than either X9 or X10,
spurring us to speculate that W51 might be a magnetic CV or a hidden
qLMXB (with a dominant nonthermal X-ray spectral component). Then at
log$F_{\rm X}$/$F_{\rm 435,exc} < 1.5$, we see only CVs.

As a final note, we point out that the (likely) optical/NUV
counterparts to X4, X5, X7 and W17 lie well within the error circles
(Fig.~\ref{fig_fcs}). Since the radii of these error circles are
dominated by the X-ray errors, this suggests that the prescription by
\cite{hongvandea05} for the 95\% positional uncertainty on the {\em
  Chandra} positions overestimates the actual uncertainties. For the
five sources that we consider here, we have also computed the X-ray
errors using the formula in \cite{kimkimea07}. We find that for these
bright sources the resulting 95\% X-ray errors are about 3--5 times
smaller than the Hong errors. The optical/NUV counterparts lie very
close to, or within, the Kim error radii. We suggest that for future
studies of bright {\em Chandra} sources, the use of the Kim errors is
more appropriate as they eliminate many astrometric counterparts that
would otherwise fall inside the (larger) error circles.

\section{Conclusions} \label{sec_con}

We have performed a detailed study of three certain (X5, X7 and W37)
and two likely (X4 and W17) qLMXBs in 47\,Tuc using simultaneous {\em
  Chandra} and optical {\em HST} data, complemented with
non-contemporaneous optical and NUV {\em HST} data. We have discovered
a blue and variable NUV counterpart to W17. The emission of a nearby
bright star overwhelms this star in the optical. For X7, we suggest
that the star N1, a previously identified astrometric match to this
source, is the likely true counterpart. Since N1 looks like a
main-sequence star, the atmosphere of the neutron star is probably
H-rich. N1 has blue colours (in particular in the NUV), shows a modest
H$\alpha$ excess and enhanced variability in $U_{\rm 300}$. These
properties point at the presence of an accretion disk, but at the same
time X7 has not shown any signs of accretion in X-rays. We suggest
that either gas from the disk has not reached the neutron star (yet),
or that the power-law component associated with accretion is
overwhelmed by the soft X-ray emission. We find a possible NUV
counterpart for W37 but deeper higher-resolution imaging is needed to
investigate the association with the X-ray source.

We have investigated the properties of the known optical/NUV
counterparts to X4 and X5. For X4, we find that the simultaneous X-ray
and optical light curves confirm that the X-ray variability is driven
by changes in the accretion rate. In X5, minima in the optical and NUV
light curves accompany the X-ray eclipses. The simultaneous X-ray and
optical data suggest that the regions in the binary system that give
rise to the blue and H$\alpha$ excess emission are obscured together
with the X-rays. We find evidence for long-term variations in the
optical from comparing our light curves with those taken about three
years earlier.

Overall, based on the smaller blue (F435W) excess fluxes for X4$_{\rm
  opt}$, X5$_{\rm opt}$, and X7$_{\rm opt}$ compared to the CV optical
counterparts, we conclude that the accretion disks in the 47\,Tuc
qLMXBs are less prominent than in CVs. This makes the ratio of X-ray
flux to excess blue optical flux a powerful discriminator between CVs
and qLMXBs.

\section*{Acknowledgements}
  The authors thank R.~Gilliland for providing GO\,8267 light
  curves. This work is supported by Chandra grant GO3-4033A. This work
  made use of data from ESO telescopes obtained from the ESO/ST-ECF
  Science Archive Facility.
  CH is supported by NSERC Discovery Grants RGPIN-2016-04602 and RGPIN-2023-04264.

\section*{Data availability}
The {\em Chandra} data used in this work are available in the {\em Chandra} Data Archive at \url{https://cxc.harvard.edu/cda}. The {\em HST} data used in this work can be retrieved from the Mikulski Archive for Space Telescopes (MAST) Portal at \url{https://mast.stsci.edu/search/ui/#/hst}.





\begin{thebibliography}{}
\makeatletter
\relax
\def\mn@urlcharsother{\let\do\@makeother \do\$\do\&\do\#\do\^\do\_\do\%\do\~}
\def\mn@doi{\begingroup\mn@urlcharsother \@ifnextchar [ {\mn@doi@}
  {\mn@doi@[]}}
\def\mn@doi@[#1]#2{\def\@tempa{#1}\ifx\@tempa\@empty \href
  {http://dx.doi.org/#2} {doi:#2}\else \href {http://dx.doi.org/#2} {#1}\fi
  \endgroup}
\def\mn@eprint#1#2{\mn@eprint@#1:#2::\@nil}
\def\mn@eprint@arXiv#1{\href {http://arxiv.org/abs/#1} {{\tt arXiv:#1}}}
\def\mn@eprint@dblp#1{\href {http://dblp.uni-trier.de/rec/bibtex/#1.xml}
  {dblp:#1}}
\def\mn@eprint@#1:#2:#3:#4\@nil{\def\@tempa {#1}\def\@tempb {#2}\def\@tempc
  {#3}\ifx \@tempc \@empty \let \@tempc \@tempb \let \@tempb \@tempa \fi \ifx
  \@tempb \@empty \def\@tempb {arXiv}\fi \@ifundefined
  {mn@eprint@\@tempb}{\@tempb:\@tempc}{\expandafter \expandafter \csname
  mn@eprint@\@tempb\endcsname \expandafter{\@tempc}}}

\bibitem[\protect\citeauthoryear{{Albrow}, {Gilliland}, {Brown}, {Edmonds},
  {Guhathakurta}  \& {Sarajedini}}{{Albrow} et~al.}{2001}]{albrea01}
{Albrow} M.~D.,  {Gilliland} R.~L.,  {Brown} T.~M.,  {Edmonds} P.~D.,
  {Guhathakurta} P.,   {Sarajedini} A.,  2001, \mn@doi [\apj] {10.1086/322353},
  \href {http://adsabs.harvard.edu/abs/2001ApJ...559.1060A} {559, 1060}

\bibitem[\protect\citeauthoryear{{Alcock} \& {Illarionov}}{{Alcock} \&
  {Illarionov}}{1980}]{alcoilla80}
{Alcock} C.,  {Illarionov} A.,  1980, \mn@doi [\apj] {10.1086/157656}, \href
  {https://ui.adsabs.harvard.edu/abs/1980ApJ...235..534A} {235, 534}

\bibitem[\protect\citeauthoryear{{Anderson} et~al.,}{{Anderson}
  et~al.}{2008}]{andea08}
{Anderson} J.,  et~al., 2008, \mn@doi [\aj] {10.1088/0004-6256/135/6/2055},
  \href {http://adsabs.harvard.edu/abs/2008AJ....135.2055A} {135, 2055}

\bibitem[\protect\citeauthoryear{{Bahramian}, {Heinke}, {Degenaar}, {Chomiuk},
  {Wijnands}, {Strader}, {Ho}  \& {Pooley}}{{Bahramian}
  et~al.}{2015}]{bahrea15}
{Bahramian} A.,  {Heinke} C.~O.,  {Degenaar} N.,  {Chomiuk} L.,  {Wijnands} R.,
   {Strader} J.,  {Ho} W. C.~G.,   {Pooley} D.,  2015, \mn@doi [\mnras]
  {10.1093/mnras/stv1585}, \href
  {https://ui-adsabs-harvard-edu.ezp-prod1.hul.harvard.edu/abs/2015MNRAS.452.3475B}
  {452, 3475}

\bibitem[\protect\citeauthoryear{{Bahramian} et~al.,}{{Bahramian}
  et~al.}{2017}]{bahrea17}
{Bahramian} A.,  et~al., 2017, \mn@doi [\mnras] {10.1093/mnras/stx166}, \href
  {http://adsabs.harvard.edu/abs/2017MNRAS.467.2199B} {467, 2199}

\bibitem[\protect\citeauthoryear{{Bao}, {Li}  \& {Cheng}}{{Bao}
  et~al.}{2023}]{baoea23}
{Bao} T.,  {Li} Z.,   {Cheng} Z.,  2023, \mn@doi [\mnras]
  {10.1093/mnras/stad836}, \href
  {https://ui.adsabs.harvard.edu/abs/2023MNRAS.521.4257B} {521, 4257}

\bibitem[\protect\citeauthoryear{{Beccari}, {De Marchi}, {Panagia}  \&
  {Pasquini}}{{Beccari} et~al.}{2014}]{beccea14}
{Beccari} G.,  {De Marchi} G.,  {Panagia} N.,   {Pasquini} L.,  2014, \mn@doi
  [\mnras] {10.1093/mnras/stt2074}, \href
  {http://adsabs.harvard.edu/abs/2014MNRAS.437.2621B} {437, 2621}

\bibitem[\protect\citeauthoryear{{Bernardini}, {Cackett}, {Brown}, {D'Angelo},
  {Degenaar}, {Miller}, {Reynolds}  \& {Wijnands}}{{Bernardini}
  et~al.}{2013}]{bernardini13}
{Bernardini} F.,  {Cackett} E.~M.,  {Brown} E.~F.,  {D'Angelo} C.,  {Degenaar}
  N.,  {Miller} J.~M.,  {Reynolds} M.,   {Wijnands} R.,  2013, \mn@doi [\mnras]
  {10.1093/mnras/stt1741}, \href
  {https://ui.adsabs.harvard.edu/abs/2013MNRAS.436.2465B} {436, 2465}

\bibitem[\protect\citeauthoryear{{Bhattacharya}, {Heinke}, {Chugunov},
  {Freire}, {Ridolfi}  \& {Bogdanov}}{{Bhattacharya} et~al.}{2017}]{bhatea17}
{Bhattacharya} S.,  {Heinke} C.~O.,  {Chugunov} A.~I.,  {Freire} P.~C.~C.,
  {Ridolfi} A.,   {Bogdanov} S.,  2017, \mn@doi [\mnras]
  {10.1093/mnras/stx2241}, \href
  {http://adsabs.harvard.edu/abs/2017MNRAS.472.3706B} {472, 3706}

\bibitem[\protect\citeauthoryear{{Bogdanov}, {van den Berg}, {Heinke}, {Cohn},
  {Lugger}  \& {Grindlay}}{{Bogdanov} et~al.}{2010}]{bogdea10}
{Bogdanov} S.,  {van den Berg} M.,  {Heinke} C.~O.,  {Cohn} H.~N.,  {Lugger}
  P.~M.,   {Grindlay} J.~E.,  2010, \mn@doi [\apj]
  {10.1088/0004-637X/709/1/241}, \href
  {http://adsabs.harvard.edu/abs/2010ApJ...709..241B} {709, 241}

\bibitem[\protect\citeauthoryear{{Bogdanov}, {Heinke}, {{\"O}zel}  \&
  {G{\"u}ver}}{{Bogdanov} et~al.}{2016}]{bogdea16}
{Bogdanov} S.,  {Heinke} C.~O.,  {{\"O}zel} F.,   {G{\"u}ver} T.,  2016,
  \mn@doi [\apj] {10.3847/0004-637X/831/2/184}, \href
  {http://adsabs.harvard.edu/abs/2016ApJ...831..184B} {831, 184}

\bibitem[\protect\citeauthoryear{{Bressan}, {Marigo}, {Girardi}, {Salasnich},
  {Dal Cero}, {Rubele}  \& {Nanni}}{{Bressan} et~al.}{2012}]{bresea12}
{Bressan} A.,  {Marigo} P.,  {Girardi} L.,  {Salasnich} B.,  {Dal Cero} C.,
  {Rubele} S.,   {Nanni} A.,  2012, \mn@doi [\mnras]
  {10.1111/j.1365-2966.2012.21948.x}, \href
  {http://adsabs.harvard.edu/abs/2012MNRAS.427..127B} {427, 127}

\bibitem[\protect\citeauthoryear{{Brown}, {Bildsten}  \& {Rutledge}}{{Brown}
  et~al.}{1998}]{browea98}
{Brown} E.~F.,  {Bildsten} L.,   {Rutledge} R.~E.,  1998, \mn@doi [\apjl]
  {10.1086/311578}, \href
  {https://ui-adsabs-harvard-edu.ezp-prod1.hul.harvard.edu/abs/1998ApJ...504L..95B}
  {504, L95}

\bibitem[\protect\citeauthoryear{{Cackett}, {Brown}, {Miller} \& {Wijnands}}{{Cackett} et~al.}{2010}]{cackea10}
{Cackett} E.~M., {Brown} E.~F., {Miller} J.~M., {Wijnands} R.,  2010, \mn@doi [\apj] {10.1088/0004-637X/720/2/1325}, \href
  {https://ui.adsabs.harvard.edu/abs/2010ApJ...720.1325C} {720, 1325}

\bibitem[\protect\citeauthoryear{{Cameron}, {Rutledge}, {Camilo}, {Bildsten},
  {Ransom}  \& {Kulkarni}}{{Cameron} et~al.}{2007}]{cameron07}
{Cameron} P.~B.,  {Rutledge} R.~E.,  {Camilo} F.,  {Bildsten} L.,  {Ransom}
  S.~M.,   {Kulkarni} S.~R.,  2007, \mn@doi [\apj] {10.1086/512229}, \href
  {https://ui.adsabs.harvard.edu/abs/2007ApJ...660..587C} {660, 587}

\bibitem[\protect\citeauthoryear{{Campana}, {Colpi}, {Mereghetti}, {Stella}  \&
  {Tavani}}{{Campana} et~al.}{1998}]{campea98}
{Campana} S.,  {Colpi} M.,  {Mereghetti} S.,  {Stella} L.,   {Tavani} M.,
  1998, \mn@doi [\aapr] {10.1007/s001590050012}, \href
  {https://ui-adsabs-harvard-edu.ezp-prod1.hul.harvard.edu/abs/1998A&ARv...8..279C}
  {8, 279}

\bibitem[\protect\citeauthoryear{{Campana} et~al.,}{{Campana}
  et~al.}{2002}]{Campana02}
{Campana} S.,  et~al., 2002, \mn@doi [\apjl] {10.1086/342505}, \href
  {https://ui.adsabs.harvard.edu/abs/2002ApJ...575L..15C} {575, L15}

\bibitem[\protect\citeauthoryear{{Campana}, {Israel}, {Stella}, {Gastaldello} \& {Mereghetti}}{{Campana} et~al.}{2004}]{campea04}
{Campana} S., {Israel} G.~L., {Stella} L., {Gastaldello} F., {Mereghetti} S. 2004, \mn@doi [\apj] {10.1086/380194}, \href
  {https://ui.adsabs.harvard.edu/abs/2004ApJ...601..474C} {601, 474}

\bibitem[\protect\citeauthoryear{{Chakrabarty} et~al.,}{{Chakrabarty}
  et~al.}{2014}]{chakea14}
{Chakrabarty} D.,  et~al., 2014, \mn@doi [\apj] {10.1088/0004-637X/797/2/92},
  \href {https://ui.adsabs.harvard.edu/abs/2014ApJ...797...92C} {797, 92}

\bibitem[\protect\citeauthoryear{{Chen}, {Richer}, {Caiazzo}  \& {Heyl}}{{Chen}
  et~al.}{2018}]{chenea18}
{Chen} S.,  {Richer} H.,  {Caiazzo} I.,   {Heyl} J.,  2018, \mn@doi [\apj]
  {10.3847/1538-4357/aae089}, \href
  {http://adsabs.harvard.edu/abs/2018ApJ...867..132C} {867, 132}

\bibitem[\protect\citeauthoryear{{Church}, {Strader}, {Davies}  \&
  {Bobrick}}{{Church} et~al.}{2017}]{churea17}
{Church} R.~P.,  {Strader} J.,  {Davies} M.~B.,   {Bobrick} A.,  2017, \mn@doi
  [\apjl] {10.3847/2041-8213/aa9aeb}, \href
  {https://ui-adsabs-harvard-edu.ezp-prod1.hul.harvard.edu/abs/2017ApJ...851L...4C}
  {851, L4}

\bibitem[\protect\citeauthoryear{{Cohn} et~al.,}{{Cohn}
  et~al.}{2010}]{cohnea10}
{Cohn} H.~N.,  et~al., 2010, \mn@doi [\apj] {10.1088/0004-637X/722/1/20}, \href
  {http://adsabs.harvard.edu/abs/2010ApJ...722...20C} {722, 20}

\bibitem[\protect\citeauthoryear{{D'Angelo}, {Fridriksson}, {Messenger}  \&
  {Patruno}}{{D'Angelo} et~al.}{2015}]{dangea15}
{D'Angelo} C.~R.,  {Fridriksson} J.~K.,  {Messenger} C.,   {Patruno} A.,  2015,
  \mn@doi [\mnras] {10.1093/mnras/stv465}, \href
  {https://ui.adsabs.harvard.edu/abs/2015MNRAS.449.2803D} {449, 2803}

\bibitem[\protect\citeauthoryear{{D'Avanzo}, {Campana}, {Casares}, {Israel},
  {Covino}, {Charles}  \& {Stella}}{{D'Avanzo} et~al.}{2005}]{davaea05}
{D'Avanzo} P.,  {Campana} S.,  {Casares} J.,  {Israel} G.~L.,  {Covino} S.,
  {Charles} P.~A.,   {Stella} L.,  2005, \mn@doi [\aap]
  {10.1051/0004-6361:20053517}, \href
  {https://ui.adsabs.harvard.edu/abs/2005A&A...444..905D} {444, 905}

\bibitem[\protect\citeauthoryear{{Deutsch}, {Margon}  \& {Anderson}}{{Deutsch}
  et~al.}{2000}]{deutea00}
{Deutsch} E.~W.,  {Margon} B.,   {Anderson} S.~F.,  2000, \mn@doi [\apjl]
  {10.1086/312486}, \href
  {https://ui.adsabs.harvard.edu/abs/2000ApJ...530L..21D} {530, L21}

\bibitem[\protect\citeauthoryear{{Dolphin}}{{Dolphin}}{2000}]{dolp00}
{Dolphin} A.~E.,  2000, \pasp, \href
  {http://adsabs.harvard.edu/cgi-bin/nph-bib_query?bibcode=2000PASP..112.1383D&db_key=AST}
  {112, 1383}

\bibitem[\protect\citeauthoryear{{Echibur{\'u}}, {Guillot}, {Zhao}, {Heinke},
  {{\"O}zel}  \& {Webb}}{{Echibur{\'u}} et~al.}{2020}]{echiea20}
{Echibur{\'u}} C.~S.,  {Guillot} S.,  {Zhao} Y.,  {Heinke} C.~O.,  {{\"O}zel}
  F.,   {Webb} N.~A.,  2020, \mn@doi [\mnras] {10.1093/mnras/staa1456}, \href
  {https://ui.adsabs.harvard.edu/abs/2020MNRAS.495.4508E} {495, 4508}

\bibitem[\protect\citeauthoryear{{Edmonds}, {Heinke}, {Grindlay}  \&
  {Gilliland}}{{Edmonds} et~al.}{2002a}]{edmoea02a}
{Edmonds} P.~D.,  {Heinke} C.~O.,  {Grindlay} J.~E.,   {Gilliland} R.~L.,
  2002a, \mn@doi [ApJL] {10.1086/338776}, \href
  {http://adsabs.harvard.edu/cgi-bin/nph-bib_query?bibcode=2002ApJ...564L..17E&db_key=AST}
  {564, L17}

\bibitem[\protect\citeauthoryear{{Edmonds}, {Gilliland}, {Camilo}, {Heinke}  \&
  {Grindlay}}{{Edmonds} et~al.}{2002b}]{edmoea02b}
{Edmonds} P.~D.,  {Gilliland} R.~L.,  {Camilo} F.,  {Heinke} C.~O.,
  {Grindlay} J.~E.,  2002b, \mn@doi [ApJ] {10.1086/342985}, \href
  {http://adsabs.harvard.edu/abs/2002ApJ...579..741E} {579, 741}

\bibitem[\protect\citeauthoryear{{Edmonds}, {Gilliland}, {Heinke}  \&
  {Grindlay}}{{Edmonds} et~al.}{2003a}]{edmogillea03a}
{Edmonds} P.~D.,  {Gilliland} R.~L.,  {Heinke} C.~O.,   {Grindlay} J.~E.,
  2003a, \mn@doi [ApJ] {10.1086/378193}, \href
  {http://adsabs.harvard.edu/abs/2003ApJ...596.1177E} {596, 1177}

\bibitem[\protect\citeauthoryear{{Edmonds}, {Gilliland}, {Heinke}  \&
  {Grindlay}}{{Edmonds} et~al.}{2003b}]{edmogillea03b}
{Edmonds} P.~D.,  {Gilliland} R.~L.,  {Heinke} C.~O.,   {Grindlay} J.~E.,
  2003b, \mn@doi [ApJ] {10.1086/378194}, \href
  {http://adsabs.harvard.edu/abs/2003ApJ...596.1197E} {596, 1197}

\bibitem[\protect\citeauthoryear{{Ferraro}, {Pallanca}, {Lanzoni}, {Cadelano},
  {Massari}, {Dalessandro}  \& {Mucciarelli}}{{Ferraro}
  et~al.}{2015}]{ferrea15}
{Ferraro} F.~R.,  {Pallanca} C.,  {Lanzoni} B.,  {Cadelano} M.,  {Massari} D.,
  {Dalessandro} E.,   {Mucciarelli} A.,  2015, \mn@doi [\apjl]
  {10.1088/2041-8205/807/1/L1}, \href
  {https://ui.adsabs.harvard.edu/abs/2015ApJ...807L...1F} {807, L1}

\bibitem[\protect\citeauthoryear{{Gilliland} et~al.,}{{Gilliland}
  et~al.}{2000}]{gillea00}
{Gilliland} R.~L.,  et~al., 2000, \mn@doi [\apjl] {10.1086/317334}, \href
  {http://adsabs.harvard.edu/abs/2000ApJ...545L..47G} {545, L47}

\bibitem[\protect\citeauthoryear{{Gonzaga}, {Hack}, {Fruchter}  \&
  {Mack}}{{Gonzaga} et~al.}{2012}]{gonzea12}
{Gonzaga} S.,  {Hack} W.,  {Fruchter} A.,   {Mack} J.,  eds, 2012, The
  DrizzlePac Handbook

\bibitem[\protect\citeauthoryear{{Grindlay}, {Heinke}, {Edmonds}  \&
  {Murray}}{{Grindlay} et~al.}{2001}]{grinea01a}
{Grindlay} J.~E.,  {Heinke} C.,  {Edmonds} P.~D.,   {Murray} S.~S.,  2001,
  \mn@doi [Science] {10.1126/science.1061135}, \href
  {http://adsabs.harvard.edu/abs/2001Sci...292.2290G} {292, 2290}

\bibitem[\protect\citeauthoryear{{Haggard}, {Cool}, {Anderson}, {Edmonds},
  {Callanan}, {Heinke}, {Grindlay}  \& {Bailyn}}{{Haggard}
  et~al.}{2004}]{haggea04}
{Haggard} D.,  {Cool} A.~M.,  {Anderson} J.,  {Edmonds} P.~D.,  {Callanan}
  P.~J.,  {Heinke} C.~O.,  {Grindlay} J.~E.,   {Bailyn} C.~D.,  2004, \mn@doi
  [\apj] {10.1086/421549}, \href
  {http://adsabs.harvard.edu/cgi-bin/nph-bib_query?bibcode=2004ApJ...613..512H&db_key=AST}
  {613, 512}

\bibitem[\protect\citeauthoryear{{Haggard}, {Cool}  \& {Davies}}{{Haggard}
  et~al.}{2009}]{haggard09}
{Haggard} D.,  {Cool} A.~M.,   {Davies} M.~B.,  2009, \mn@doi [\apj]
  {10.1088/0004-637X/697/1/224}, \href
  {https://ui.adsabs.harvard.edu/abs/2009ApJ...697..224H} {697, 224}

\bibitem[\protect\citeauthoryear{{Hasinger}, {Johnston}  \&
  {Verbunt}}{{Hasinger} et~al.}{1994}]{hasiea94}
{Hasinger} G.,  {Johnston} H.~M.,   {Verbunt} F.,  1994, \aap, \href
  {http://adsabs.harvard.edu/abs/1994A%26A...288..466H} {288, 466}

\bibitem[\protect\citeauthoryear{{Heinke}, {Grindlay}, {Lloyd}  \&
  {Edmonds}}{{Heinke} et~al.}{2003a}]{heinea03a}
{Heinke} C.~O.,  {Grindlay} J.~E.,  {Lloyd} D.~A.,   {Edmonds} P.~D.,  2003a,
  \mn@doi [ApJ] {10.1086/374039}, \href
  {http://adsabs.harvard.edu/cgi-bin/nph-bib_query?bibcode=2003ApJ...588..452H&db_key=AST}
  {588, 452}

\bibitem[\protect\citeauthoryear{{Heinke}, {Grindlay}, {Lugger}, {Cohn},
  {Edmonds}, {Lloyd}  \& {Cool}}{{Heinke} et~al.}{2003b}]{heinea03b}
{Heinke} C.~O.,  {Grindlay} J.~E.,  {Lugger} P.~M.,  {Cohn} H.~N.,  {Edmonds}
  P.~D.,  {Lloyd} D.~A.,   {Cool} A.~M.,  2003b, \mn@doi [ApJ]
  {10.1086/378885}, \href
  {http://adsabs.harvard.edu/cgi-bin/nph-bib_query?bibcode=2003ApJ...598..501H&db_key=AST}
  {598, 501}

\bibitem[\protect\citeauthoryear{{Heinke}, {Grindlay}  \& {Edmonds}}{{Heinke}
  et~al.}{2005a}]{heinea05a}
{Heinke} C.~O.,  {Grindlay} J.~E.,   {Edmonds} P.~D.,  2005a, \mn@doi [ApJ]
  {10.1086/427795}, \href
  {http://adsabs.harvard.edu/cgi-bin/nph-bib_query?bibcode=2005ApJ...622..556H&db_key=AST}
  {622, 556}

\bibitem[\protect\citeauthoryear{{Heinke}, {Grindlay}, {Edmonds}, {Cohn},
  {Lugger}, {Camilo}, {Bogdanov}  \& {Freire}}{{Heinke}
  et~al.}{2005b}]{heinea05b}
{Heinke} C.~O.,  {Grindlay} J.~E.,  {Edmonds} P.~D.,  {Cohn} H.~N.,  {Lugger}
  P.~M.,  {Camilo} F.,  {Bogdanov} S.,   {Freire} P.~C.,  2005b, \mn@doi [ApJ]
  {10.1086/429899}, \href
  {http://adsabs.harvard.edu/cgi-bin/nph-bib_query?bibcode=2005ApJ...625..796H&db_key=AST}
  {625, 796}

\bibitem[\protect\citeauthoryear{{Heinke}, {Rybicki}, {Narayan}  \&
  {Grindlay}}{{Heinke} et~al.}{2006}]{heinea06}
{Heinke} C.~O.,  {Rybicki} G.~B.,  {Narayan} R.,   {Grindlay} J.~E.,  2006,
  \mn@doi [ApJ] {10.1086/503701}, \href
  {http://adsabs.harvard.edu/abs/2006ApJ...644.1090H} {644, 1090}

\bibitem[\protect\citeauthoryear{{Heinke} et~al.,}{{Heinke}
  et~al.}{2014}]{heinea14}
{Heinke} C.~O.,  et~al., 2014, \mn@doi [\mnras] {10.1093/mnras/stu1449}, \href
  {https://ui.adsabs.harvard.edu/abs/2014MNRAS.444..443H} {444, 443}

\bibitem[\protect\citeauthoryear{{Henleywillis}, {Cool}, {Haggard}, {Heinke},
  {Callanan}  \& {Zhao}}{{Henleywillis} et~al.}{2018}]{henlea18}
{Henleywillis} S.,  {Cool} A.~M.,  {Haggard} D.,  {Heinke} C.,  {Callanan} P.,
   {Zhao} Y.,  2018, \mn@doi [\mnras] {10.1093/mnras/sty675}, \href
  {https://ui.adsabs.harvard.edu/abs/2018MNRAS.479.2834H} {479, 2834}

\bibitem[\protect\citeauthoryear{Hong, van~den Berg, Schlegel, Grindlay,
  Koenig, Laycock  \& Zhao}{Hong et~al.}{2005}]{hongvandea05}
Hong J.,  van~den Berg M.,  Schlegel E.,  Grindlay J.,  Koenig X.,  Laycock S.,
    Zhao P.,  2005, ApJ, 635, 907

\bibitem[\protect\citeauthoryear{{Jonker}, {Galloway}, {McClintock}, {Buxton},
  {Garcia}  \& {Murray}}{{Jonker} et~al.}{2004}]{Jonker04}
{Jonker} P.~G.,  {Galloway} D.~K.,  {McClintock} J.~E.,  {Buxton} M.,  {Garcia}
  M.,   {Murray} S.,  2004, \mn@doi [\mnras]
  {10.1111/j.1365-2966.2004.08246.x}, \href
  {https://ui.adsabs.harvard.edu/abs/2004MNRAS.354..666J} {354, 666}

\bibitem[\protect\citeauthoryear{{Katz}}{{Katz}}{1975}]{katz75}
{Katz} J.~I.,  1975, \mn@doi [\nat] {10.1038/253698a0}, \href
  {https://ui-adsabs-harvard-edu.ezp-prod1.hul.harvard.edu/abs/1975Natur.253..698K}
  {253, 698}

\bibitem[\protect\citeauthoryear{{Kim} et~al.,}{{Kim}
  et~al.}{2007}]{kimkimea07}
{Kim} M.,  et~al., 2007, \mn@doi [\apjs] {10.1086/511634}, \href
  {https://ui.adsabs.harvard.edu/abs/2007ApJS..169..401K} {169, 401}

\bibitem[\protect\citeauthoryear{Koekemoer, McLean, McMaster  \&
  Jenkner}{Koekemoer et~al.}{2005}]{koekea05}
Koekemoer A.,  McLean B.,  McMaster M.,   Jenkner H.,  2005, Technical report,
  Instrument Science Report ACS 2005-06.
STScI

\bibitem[\protect\citeauthoryear{{Kononov}, {Lacy}, {Puzin}, {Kozhevnikov},
  {Sytov}  \& {Lyaptsev}}{{Kononov} et~al.}{2017}]{konoea17}
{Kononov} D.,  {Lacy} C.,  {Puzin} V.~B.,  {Kozhevnikov} V.~P.,  {Sytov} A.~Y.,
    {Lyaptsev} A.~P.,  2017, in The Golden Age of Cataclysmic Variables and
  Related Objects IV. p.~31 (\mn@eprint {arXiv} {1607.00265}),
  \mn@doi{10.22323/1.315.0031}

\bibitem[\protect\citeauthoryear{{Li}, {Kastner}, {Prigozhin}, {Schulz},
  {Feigelson}  \& {Getman}}{{Li} et~al.}{2004}]{liea04}
{Li} J.,  {Kastner} J.~H.,  {Prigozhin} G.~Y.,  {Schulz} N.~S.,  {Feigelson}
  E.~D.,   {Getman} K.~V.,  2004, \mn@doi [\apj] {10.1086/421866}, \href
  {https://ui.adsabs.harvard.edu/abs/2004ApJ...610.1204L} {610, 1204}

\bibitem[\protect\citeauthoryear{{Linares}}{{Linares}}{2014}]{lina14}
{Linares} M.,  2014, \mn@doi [\apj] {10.1088/0004-637X/795/1/72}, \href
  {http://adsabs.harvard.edu/abs/2014ApJ...795...72L} {795, 72}

\bibitem[\protect\citeauthoryear{{Lomb}}{{Lomb}}{1976}]{lomb76}
{Lomb} N.~R.,  1976, \mn@doi [\apss] {10.1007/BF00648343}, \href
  {https://ui-adsabs-harvard-edu.ezp-prod1.hul.harvard.edu/abs/1976Ap&SS..39..447L}
  {39, 447}

\bibitem[\protect\citeauthoryear{{Lugger}, {Cohn}, {Cool}, {Heinke}  \&
  {Anderson}}{{Lugger} et~al.}{2017}]{luggea17}
{Lugger} P.~M.,  {Cohn} H.~N.,  {Cool} A.~M.,  {Heinke} C.~O.,   {Anderson} J.,
   2017, \mn@doi [\apj] {10.3847/1538-4357/aa6c56}, \href
  {https://ui-adsabs-harvard-edu.ezp-prod1.hul.harvard.edu/abs/2017ApJ...841...53L}
  {841, 53}

\bibitem[\protect\citeauthoryear{{Miller-Jones} et~al.,}{{Miller-Jones}
  et~al.}{2015}]{millea15}
{Miller-Jones} J.~C.~A.,  et~al., 2015, \mn@doi [\mnras]
  {10.1093/mnras/stv1869}, \href
  {http://adsabs.harvard.edu/abs/2015MNRAS.453.3918M} {453, 3918}

\bibitem[\protect\citeauthoryear{{Nardiello} et~al.,}{{Nardiello}
  et~al.}{2018}]{nardea18}
{Nardiello} D.,  et~al., 2018, \mn@doi [\mnras] {10.1093/mnras/sty2515}, \href
  {https://ui-adsabs-harvard-edu.ezp-prod1.hul.harvard.edu/abs/2018MNRAS.481.3382N}
  {481, 3382}

\bibitem[\protect\citeauthoryear{{{\"O}zel} \& {Freire}}{{{\"O}zel} \&
  {Freire}}{2016}]{ozelfrei16}
{{\"O}zel} F.,  {Freire} P.,  2016, \mn@doi [\araa]
  {10.1146/annurev-astro-081915-023322}, \href
  {https://ui.adsabs.harvard.edu/abs/2016ARA&A..54..401O} {54, 401}

\bibitem[\protect\citeauthoryear{{Pallanca}, {Dalessandro}, {Ferraro},
  {Lanzoni}  \& {Beccari}}{{Pallanca} et~al.}{2013}]{pallea13}
{Pallanca} C.,  {Dalessandro} E.,  {Ferraro} F.~R.,  {Lanzoni} B.,   {Beccari}
  G.,  2013, \mn@doi [\apj] {10.1088/0004-637X/773/2/122}, \href
  {https://ui.adsabs.harvard.edu/abs/2013ApJ...773..122P} {773, 122}

\bibitem[\protect\citeauthoryear{{Pooley} \& {Hut}}{{Pooley} \&
  {Hut}}{2006}]{poolhut06}
{Pooley} D.,  {Hut} P.,  2006, \mn@doi [ApJL] {10.1086/507027}, \href
  {http://adsabs.harvard.edu/abs/2006ApJ...646L.143P} {646, L143}

\bibitem[\protect\citeauthoryear{{Pooley} et~al.,}{{Pooley}
  et~al.}{2003}]{poolea03}
{Pooley} D.,  et~al., 2003, \mn@doi [\apjl] {10.1086/377074}, \href
  {https://ui-adsabs-harvard-edu.ezp-prod1.hul.harvard.edu/abs/2003ApJ...591L.131P}
  {591, L131}

\bibitem[\protect\citeauthoryear{{Qiao} \& {Liu}}{{Qiao} \& {Liu}}{2018}]{qiaoliu2018}
  {Qiao}, E., {Liu}, B.~F., 2018, \mn@doi [\mnras]
  {10.1093/mnras/sty2337}, \href
  {https://ui.adsabs.harvard.edu/abs/2018MNRAS.481..938Q} {481, 938}

\bibitem[\protect\citeauthoryear{{Qiao} \& {Liu}}{{Qiao} \& {Liu}}{2020}]{qiaoliu2020}
  {Qiao}, E., {Liu}, B.~F., 2020, \mn@doi [\mnras]
  {10.1093/mnras/stz3510}, \href
  {https://ui.adsabs.harvard.edu/abs/2020MNRAS.492..615Q} {492, 615}

\bibitem[\protect\citeauthoryear{{Qiao} \& {Liu}}{{Qiao} \& {Liu}}{2021}]{qiaoliu2021}
  {Qiao}, E., {Liu}, B.~F., 2021, \mn@doi [\mnras]
  {10.1093/mnras/stab227}, \href
  {https://ui.adsabs.harvard.edu/abs/2021MNRAS.502.3870Q} {502, 3870}

\bibitem[\protect\citeauthoryear{{Rivera Sandoval} et~al.,}{{Rivera Sandoval}
  et~al.}{2015}]{riveea15}
{Rivera Sandoval} L.~E.,  et~al., 2015, \mn@doi [\mnras]
  {10.1093/mnras/stv1810}, \href
  {http://adsabs.harvard.edu/abs/2015MNRAS.453.2707R} {453, 2707}

\bibitem[\protect\citeauthoryear{{Rivera Sandoval} et~al.,}{{Rivera Sandoval}
  et~al.}{2018}]{riveea18}
{Rivera Sandoval} L.~E.,  et~al., 2018, \mn@doi [\mnras]
  {10.1093/mnras/sty058}, \href
  {http://adsabs.harvard.edu/abs/2018MNRAS.475.4841R} {475, 4841}

\bibitem[\protect\citeauthoryear{{Romani}}{{Romani}}{1987}]{roma87}
{Romani} R.~W.,  1987, \mn@doi [\apj] {10.1086/165010}, \href
  {https://ui.adsabs.harvard.edu/abs/1987ApJ...313..718R} {313, 718}

\bibitem[\protect\citeauthoryear{{Rutledge}, {Bildsten}, {Brown}, {Pavlov}  \&
  {Zavlin}}{{Rutledge} et~al.}{2002}]{rutlea02}
{Rutledge} R.~E.,  {Bildsten} L.,  {Brown} E.~F.,  {Pavlov} G.~G.,   {Zavlin}
  V.~E.,  2002, \mn@doi [\apj] {10.1086/342306}, \href
  {https://ui.adsabs.harvard.edu/abs/2002ApJ...578..405R} {578, 405}

\bibitem[\protect\citeauthoryear{{Salaris}, {Held}, {Ortolani}, {Gullieuszik}
  \& {Momany}}{{Salaris} et~al.}{2007}]{salaea07}
{Salaris} M.,  {Held} E.~V.,  {Ortolani} S.,  {Gullieuszik} M.,   {Momany} Y.,
  2007, \mn@doi [\aap] {10.1051/0004-6361:20078445}, \href
  {http://adsabs.harvard.edu/abs/2007A%26A...476..243S} {476, 243}

\bibitem[\protect\citeauthoryear{{Sarajedini} et~al.,}{{Sarajedini}
  et~al.}{2007}]{saraea07}
{Sarajedini} A.,  et~al., 2007, \mn@doi [\aj] {10.1086/511979}, \href
  {https://ui-adsabs-harvard-edu.ezp-prod1.hul.harvard.edu/abs/2007AJ....133.1658S}
  {133, 1658}

\bibitem[\protect\citeauthoryear{Scargle}{Scargle}{1982}]{scar}
Scargle J.~D.,  1982, ApJ, 263, 835

\bibitem[\protect\citeauthoryear{{Servillat}, {Heinke}, {Ho}, {Grindlay},
  {Hong}, {van den Berg}  \& {Bogdanov}}{{Servillat} et~al.}{2012}]{servea12}
{Servillat} M.,  {Heinke} C.~O.,  {Ho} W.~C.~G.,  {Grindlay} J.~E.,  {Hong} J.,
   {van den Berg} M.,   {Bogdanov} S.,  2012, \mn@doi [\mnras]
  {10.1111/j.1365-2966.2012.20976.x}, \href
  {https://ui.adsabs.harvard.edu/abs/2012MNRAS.423.1556S} {423, 1556}

\bibitem[\protect\citeauthoryear{{Shahbaz}, {Gonz{\'a}lez-Hern{\'a}ndez}, {Breton}, {Kennedy}, {Mata S{\'a}nchez} \& {Linares}}{{Shahbaz} et~al.}{2022}]{shahea22}
  {Shahbaz}, T.,  {Gonz{\'a}lez-Hern{\'a}ndez}, J.~I., {Breton}, R.~P., {Kennedy}, M.~R., {Mata S{\'a}nchez}, D., {Linares}, M., 2022, \mn@doi [\mnras]
  {10.1093/mnras/stac492}, \href
  {https://ui.adsabs.harvard.edu/abs/2022MNRAS.513...71S} {513, 71}

\bibitem[\protect\citeauthoryear{{Sirianni} et~al.,}{{Sirianni}
  et~al.}{2005}]{siriea05}
{Sirianni} M.,  et~al., 2005, \mn@doi [\pasp] {10.1086/444553}, \href
  {http://adsabs.harvard.edu/abs/2005PASP..117.1049S} {117, 1049}

\bibitem[\protect\citeauthoryear{{Steiner}, {Heinke}, {Bogdanov}, {Li}, {Ho},
  {Bahramian}  \& {Han}}{{Steiner} et~al.}{2018}]{steiea18}
{Steiner} A.~W.,  {Heinke} C.~O.,  {Bogdanov} S.,  {Li} C.~K.,  {Ho} W.~C.~G.,
  {Bahramian} A.,   {Han} S.,  2018, \mn@doi [\mnras] {10.1093/mnras/sty215},
  \href
  {https://ui-adsabs-harvard-edu.ezp-prod1.hul.harvard.edu/abs/2018MNRAS.476..421S}
  {476, 421}

\bibitem[\protect\citeauthoryear{{Testa} et~al.,}{{Testa}
  et~al.}{2012}]{testea12}
{Testa} V.,  et~al., 2012, \mn@doi [\aap] {10.1051/0004-6361/201219904}, \href
  {https://ui.adsabs.harvard.edu/abs/2012A&A...547A..28T} {547, A28}

\bibitem[\protect\citeauthoryear{{Tudor} et~al.,}{{Tudor}
  et~al.}{2018}]{tudoea18}
{Tudor} V.,  et~al., 2018, \mn@doi [\mnras] {10.1093/mnras/sty284}, \href
  {http://adsabs.harvard.edu/abs/2018MNRAS.476.1889T} {476, 1889}

\bibitem[\protect\citeauthoryear{{Verbunt}, {van Kerkwijk}, {in't Zand}  \&
  {Heise}}{{Verbunt} et~al.}{2000}]{verbea00}
{Verbunt} F.,  {van Kerkwijk} M.~H.,  {in't Zand} J.~J.~M.,   {Heise} J.,
  2000, \mn@doi [\aap] {10.48550/arXiv.astro-ph/0005337}, \href
  {https://ui.adsabs.harvard.edu/abs/2000A&A...359..960V} {359, 960}

\bibitem[\protect\citeauthoryear{{Verbunt}, {Pooley}  \& {Bassa}}{{Verbunt}
  et~al.}{2008}]{verbea08}
{Verbunt} F.,  {Pooley} D.,   {Bassa} C.,  2008, in {Vesperini} E.,  {Giersz}
  M.,   {Sills} A.,  eds,  International Astronomical Union Symposium no. 246
  Vol. 246, Dynamical Evolution of Dense Stellar Systems. pp 301--310
  (\mn@eprint {arXiv} {0710.1804}), \mn@doi{10.1017/S1743921308015822}

\bibitem[\protect\citeauthoryear{{Walsh}, {Cackett}  \& {Bernardini}}{{Walsh}
  et~al.}{2015}]{walsea15}
{Walsh} A.~R.,  {Cackett} E.~M.,   {Bernardini} F.,  2015, \mn@doi [\mnras]
  {10.1093/mnras/stv315}, \href
  {https://ui-adsabs-harvard-edu.ezp-prod1.hul.harvard.edu/abs/2015MNRAS.449.1238W}
  {449, 1238}

\bibitem[\protect\citeauthoryear{{Wijnands}, {Heinke}, {Pooley}, {Edmonds},
  {Lewin}, {Grindlay}, {Jonker}  \& {Miller}}{{Wijnands}
  et~al.}{2005}]{Wijnands05}
{Wijnands} R.,  {Heinke} C.~O.,  {Pooley} D.,  {Edmonds} P.~D.,  {Lewin} W.
  H.~G.,  {Grindlay} J.~E.,  {Jonker} P.~G.,   {Miller} J.~M.,  2005, \mn@doi
  [\apj] {10.1086/426127}, \href
  {https://ui.adsabs.harvard.edu/abs/2005ApJ...618..883W} {618, 883}

\bibitem[\protect\citeauthoryear{{Wijnands}, {Degenaar}, {Armas Padilla}, {Altamirano}, {Cavecchi}, {Linares}, {Bahramian} \& {Heinke}}{{Wijnands}
  et~al.}{2015}]{wijnea15}
{Wijnands}, R., {Degenaar}, N., {Armas Padilla}, M., {Altamirano}, D., {Cavecchi}, Y., {Linares}, M., {Bahramian}, A., {Heinke}, C.~O., 2015, \mn@doi [\mnras]
  {10.1093/mnras/stv1974}, \href
  {https://ui.adsabs.harvard.edu/abs/2015MNRAS.454.1371W}
  {454, 1371}

\bibitem[\protect\citeauthoryear{{Zacharias}, {Urban}, {Zacharias}, {Wycoff},
  {Hall}, {Monet}  \& {Rafferty}}{{Zacharias} et~al.}{2004}]{zachea04}
{Zacharias} N.,  {Urban} S.~E.,  {Zacharias} M.~I.,  {Wycoff} G.~L.,  {Hall}
  D.~M.,  {Monet} D.~G.,   {Rafferty} T.~J.,  2004, \mn@doi [\aj]
  {10.1086/386353}, \href {http://adsabs.harvard.edu/abs/2004AJ....127.3043Z}
  {127, 3043}

\bibitem[\protect\citeauthoryear{{in't Zand}, {van Kerkwijk}, {Pooley},
  {Verbunt}, {Wijnands}  \& {Lewin}}{{in't Zand} et~al.}{2001}]{intzea01}
{in't Zand} J.~J.~M.,  {van Kerkwijk} M.~H.,  {Pooley} D.,  {Verbunt} F.,
  {Wijnands} R.,   {Lewin} W.~H.~G.,  2001, \mn@doi [\apjl] {10.1086/338361},
  \href {https://ui.adsabs.harvard.edu/abs/2001ApJ...563L..41I} {563, L41}

\bibitem[\protect\citeauthoryear{{in't Zand} et~al.,}{{in't Zand}
  et~al.}{2003}]{intzea03}
{in't Zand} J.~J.~M.,  et~al., 2003, \mn@doi [\aap]
  {10.1051/0004-6361:20030681}, \href
  {https://ui-adsabs-harvard-edu.ezp-prod1.hul.harvard.edu/abs/2003A&A...406..233I}
  {406, 233}

\makeatother
\end{thebibliography}

\appendix
\section{Refining the orbital period of X5} \label{app_x5period}

The eclipses of X5 are total eclipses, averaging 2500 s long, which
can be explained by the X-rays coming exclusively from the neutron
star surface. Only occasional background photons fall within the
eclipse.  To provide the photon list, we extract lightcurves at the
instrument time resolution (for ACIS, or 2.0 s for HRC) from
barycentred event files. We use the 0.5--2 keV energy range for ACIS
data (which includes most source photons, while reducing background)
and a 1.0\arcsec\,extraction region to create the lightcurves. We use
ACIS observations from 2000 (ObsIDs 953, 955; \citealt{heinea03a}),
2002 (ObsIDs 2735--3386 in Table~\ref{tab_xray_eclipses}, described in
\S 2.1), 2014 (ObsIDs 16527--15748, \citealt{bogdea16}), and 2022
(ObsID 26229, PI Paduano), along with HRC observations in 2005-2006
(ObsIDs 5542--6240, \citealt{cameron07}).

The countrate from X5 is low enough (typically 0.1 counts/second, or
lower) that we choose not to try to fit the eclipse with a specified
function, nor to fit the ingresses and egresses, as when studying
eclipses of bright LMXBs \citep[e.g.][]{intzea03}. Instead, we find
the best approach is to find the midpoint of each eclipse using the
last photon recorded before the eclipse, and the first photon after
it.  The uncertainty in the eclipse midpoint is thus $\sigma_{mid}$ =
$\sqrt{\text{(}\sigma^2_{in} \text{+} \sigma^2_{out}\text{)/2}}$,
where e.g. $\sigma_{in}$ is the inverse of the measured counts in the
100 seconds up to the last photon.  This requires some iteration, to
identify likely background photons interrupting eclipses (we find only
14 such interlopers). We omit using eclipses if the continuum
countrate is so low that the eclipse is substantially longer than
average ($>$2700 s), or if we measure only an ingress or an
egress. This leaves us with 35 complete and high-quality eclipses,
spread from 2000 to 2022, with 20 observed during the 2005--2006 HRC
data. The eclipse midpoints, and their associated errors, are given in
Table~\ref{tab_xray_eclipses}, including three not used for our
calculation, in units of TDB seconds since MJD\_REF=50814.0
(barycentred).

We calculate the orbital period iteratively, using eclipses close in
time to find the orbital period and its error, and then predict more
distant eclipses, which then refine the orbital period and its error.
If the error in the orbital period for the next eclipse is less than
half an orbital period, we maintain cycle count.

We begin with ObsID 6238 in 2005, which has two well-measured
eclipses, which give $P$=$31195\pm30$ s. We then extrapolate this
solution 3 cycles to ObsID 5546, refining the period to
$P$=$31204\pm7.5$ s, and continue similarly.  We include the cycle
count (from the 2nd eclipse of ObsID 6238) in
Table~\ref{tab_xray_eclipses}.  Using the full range of eclipses, we
arrive at an orbital period solution for X5's X-ray eclipse midpoints,
measured from MJD\_REF=50814.0, of TDB=251975445($\pm16$)+$n \times
31200.192$($\pm0.002$) s, where $n$ is an integer.

\begin{table}
  \begin{tabular}{l@{\hskip0.3cm}l@{\hskip0.3cm}l@{\hskip0.18cm}l@{\hskip0.18cm}l}
    \hline
    ObsID    & Midpoint$^a$ & Offset & Error & Cycle  \\
             & s            &  s     &  s    & Count    \\
   \hline
953 &	69610285 &	  -38 &     26  &	-5845 \\
955 &	69641492.5 &	 -30 &    17 &	-5844 \\		
2735 &	149732465.5 &	50 &   17 &	-3277 \\
2735 &	149763664* &	48 &    33 &	-3276 \\
2736 &	149794986.5* &	170 &   119	 & -3275 \\
2736 &	149826053 &	    37 &     20 &	-3274 \\
2737 &	149982022.5 &	5 &   15 &	-3269 \\
2737 &	150013441* &	223 &    376 &	-3268 \\
3386 &	150044439 &	  21 &   15 &	-3267 \\
2738 &	150699598 &	 -24 &    15 &	-3246 \\
5542 & 251382640 &    -1 &   15 & -19 \\
5543 & 251507389 &    -53 &   17 & -15 \\
5544 &	251601060 &	   17 &  15 &	-12 \\
5544 &	251632237 &	 -6 &    19 &	-11 \\
5545 &	251725832 &	  -11 &   21 &	-8 \\
6237 &	251850633 &	  -11 &   21 &	-4 \\
6238 &	251944250 &	   5 &  26 &	-1 \\
6238 &	251975445 &	  0 &   16 &	0 \\
5546 &	252069066 &	  20 &   15 &	3 \\
6230 &	252193857 &	  11 &   21 &	7 \\
6231 &	252287482 &	  35 &   38 &	10 \\
6231 &	252318628 &	  -19 &   20 &	11 \\
6232 &	252412263 &	  15 &   19 &	14 \\
6233 &	252599456 &	  7 &   25 &	20 \\
6233 &	252630661 &	 12 &    25 &	21 \\
6233 &	252661850 &	  1 &   20 &	22 \\
6235 &	252755438 &	 -12 &    21 &	25 \\
6236 &	252880266 &	  15 &   17 &	29 \\
6239 &	253005058 &	  7 &   13 &	33 \\
6240 &	253098648 &	  -4 &   13 &	36 \\
16527 &	526287538 &	   5 &    38 &	8792 \\
16527 &	526318725 &	   -8 &    17 &	8793 \\
15747 &	526693116 &	   -20 &   18 &	8805 \\
15747 &	526724336 &	    0 &    20 &	8806 \\
16529 &	527691504.5 & -37 &	  19 &	8837 \\
15748 &	528627523.5 & 	-24 &  38 &	8867 \\
16528 &	539297989.5 &	-24 &  20 &	9209 \\
26229 &	759602543.5	&   -25 &   40 & 16270 \\
    \hline    
  \end{tabular}
  
  \caption{X-ray eclipse midpoints. Offsets are the differences
    between the observed eclipse midpoint and our calculated ephemeris
    (observed-calculated times). Errors are the estimated uncertainty
    in the eclipse midpoint, based on the countrate in the
    earlier/later 100 seconds.  $^a$ TDB barycentred time, in seconds,
    since MJD\_REF=50814.0. *-Not used in our
    calculations. } \label{tab_xray_eclipses}

  \end{table}



\bsp	
\label{lastpage}
\end{document}